\documentclass[a4paper,fleqn,usenatbib]{mnras}

\usepackage{newtxtext,newtxmath}

\usepackage[T1]{fontenc}
\usepackage{ae,aecompl}

\usepackage{graphicx,longtable}	
\usepackage{amsmath}	
\usepackage{amssymb}	
\usepackage{xcolor}

\title[Doppler tomography of exoplanet atmospheres]{Doppler tomography as a tool for detecting exoplanet atmospheres}

\author[C. A. Watson et al.]{C.\,A.\ Watson,$^{1}$\thanks{E-mail:
    c.a.watson@qub.ac.uk.uk}
E. J. W. de Mooij,$^{1,2}$
D. Steeghs $^{3}$, T. R. Marsh$^{3}$, M. Brogi$^{3}$, \newauthor
N. P. Gibson$^{1,4}$, and S. Matthews$^{1}$ \\
$^{1}$Astrophysics Research Centre, Queen's University Belfast, Belfast BT7 1NN, UK\\
$^{2}$School of Physical Sciences and Centre for Astrophysics \& Relativity, Dublin City University, Glasnevin, Dublin 9, Ireland\\
$^{3}$Department of Physics, University of Warwick, Coventry CV4 7AL, UK \\
$^{4}$School of Physics, Trinity College Dublin, The University of Dublin, Dublin 2, Ireland \\
}

\date{Accepted XXX. Received YYY; in original form ZZZ}

\pubyear{2016}

\begin{document}
\label{firstpage}
\pagerange{\pageref{firstpage}--\pageref{lastpage}}
\maketitle

\begin{abstract}

  High-resolution Doppler spectroscopy is a powerful tool for
  identifying molecular species in the atmospheres of both transiting
  and non-transiting exoplanets.  Currently, such data is analysed
  using cross-correlation techniques to detect the Doppler shifting
  signal from the orbiting planet. In this paper we demonstrate that,
  compared to cross-correlation methods currently used, the technique
  of Doppler tomography has improved sensitivity in detecting the
  subtle signatures expected from exoplanet atmospheres. This is
  partly due to the use of a regularizing statistic, which acts to
  suppress noise, coupled to the fact that all the data is fit
  simultaneously. In addition, we show that the technique can also
  effectively suppress contanimating spectral features that may arise
  due to overlapping lines, repeating line patterns,
  or the use of incorrect linelists. These
  issues can confuse conventional cross-correlation approaches,
  primarily due to aliasing issues inherent in such techniques,
  whereas Doppler tomography is less susceptible to such effects.  In
  particular, Doppler tomography shows exceptional promise for
  simultaneously detecting multiple line species (e.g. isotopologues),
  even when there are high contrasts between such species -- and far
  outperforms current CCF analyses in this respect.  Finally, we
  demonstrate that Doppler tomography is capable of recovering
  molecular signals from exoplanets using real data, by confirming the
  strong detection of CO in the atmosphere of $\tau$ Boo b. We recover
  a signal with a planetary radial velocity semi-amplitude K$_p$ =
  109.6 $\pm$ 2.2 km s$^{-1}$, in excellent agreement with the
  previously reported value of 110.0 $\pm$ 3.2 km s$^{-1}$.

\end{abstract}




\section{Introduction}

Ground-based high-resolution Doppler spectroscopy has proven itself
to be a powerful tool for probing the atmospheres of both transiting and
non-transiting hot-Jupiters. Initial attempts to use time-series
high-resolution Doppler spectroscopy to provide direct detections of hot-Jupiter
atmospheres concentrated on trying to measure reflected starlight from the
planets in the optical (e.g. \citealt{cameron99};
\citealt{leigh03a, leigh03b}). Unfortunately,
the low optical albedos of hot-Jupiters meant that these earlier attempts were
unsuccessful. More recently, \cite{martins15} have claimed evidence for
detected reflected starlight from 51 Peg b using Doppler spectroscopy, but this
has since been challenged by \cite{hoeijmakers18}.

\cite{snellen10} were the first to successfully use high resolution
Doppler spectroscopy to detect the atmosphere of an exoplanet.  In
this case they used the CRyogenic Infra-Red Echelle Spectrograph
(CRIRES) Spectrograph on the Very Large Telescope (VLT) to detect CO
absorption during the transit of HD209458b. This was followed by the
detection of CO molecular absorption from the dayside of the
non-transiting hot-Jupiter $\tau$ Boo b by both \cite{brogi12} and
\cite{rodler12} (with water vapour later found for the same planet
by \citealt{lockwood14}). Since these discoveries, additional detections
(including for other molecular species such  H$_2$O and TiO) have also been reported for
51 Peg b (\citealt{brogi13}; \citealt{birkby17}), HD189733b (\citealt{dekok13};
\citealt{birkby13}; \citealt{brogi16}; \citealt{brogi18}), HD179949b (\citealt{brogi14}),
upsilon Andromedae b (\citealt{piskorz17}), HD88133b (\citealt{Piskorz16}), and WASP-33b
(\citealt{nugroho17}).

In addition to probing exoplanet atmospheres, high resolution Doppler
spectroscopy also enables the radial velocity motion of the planet to
be determined, effectively reducing the system to a double-lined
spectroscopic binary.  As a result, the absolute mass of the planet
can be determined even for non-transiting planets. With new
instrumentation such as CRIRES+ soon to come online, the capability of
high-resolution spectroscopy to drive forward our understanding of the
fundamental parameters and atmospheres of short-period exoplanets will
greatly increase. For example, the wider wavelength range coverage of
CRIRES+ ($\sim$ 6 times wider than that of CRIRES) will enable a far
larger number of spectral lines from molecular species such as H$_2$O
and CH$_4$ to be captured. Since the signal scales roughly as
$\sqrt{n}$, where $n$ is the number of lines observed (different line
strengths and signal-to-noise as a function of wavelength complicate
this assumption), such instruments promise far greater sensitivity to
atmospheric signatures in the near-future.

Despite this, the application of high-resolution Doppler spectroscopy to
exoplanet atmospheres is still limited, and the analyses still revolve
around the use of cross-correlation functions. In this paper, we
demonstrate that the application of Doppler tomography has a number of
clear advantages over such cross-correlation techniques. We begin by
briefly describing the technique in Section~\ref{sec:doptom} and
in Section~\ref{sec:tauboo} we apply Doppler tomography to detect
CO from $\tau$ Boo b using the data of \cite{brogi12}. We then test
the ability of Doppler tomography relative to more conventional
cross-correlation techniques in Section~\ref{sec:sims}. We present our
conclusions in Section~\ref{sec:conclusions}. Finally, we look
at further adaptations and potential for novel uses of Doppler tomography
for exoplanet atmosphere studies in Section~\ref{sec:future}.

\section{Doppler Tomography}
\label{sec:doptom}

The technique of Doppler tomography was developed by \cite{marsh88b}
and is normally used to study close-binary systems. As applied to
binaries, the technique aims to recover a model-independent map in
velocity space that resolves the distribution of line emission and/or
absorption within the binary using a time-series of spectra.
In essence, Doppler tomography assumes that orbiting material
(whether that be a parcel of material in an accretion disk, or emission from a
planet) traces out a sinusoidal radial velocity curve of the form,
\begin{equation}
  \mathrm{v}_R\left(\phi\right) = \gamma - \mathrm{v}_x\cos2\pi\phi + \mathrm{v}_y\sin2\pi\phi.
\label{eq:vr}
\end{equation}
\noindent Here, $\phi$ is the orbital phase, $\gamma$ is the
systemic velocity of the system, and v$_x$ and v$_y$ are the radial velocity
semi-amplitudes of the cosine and sine terms, respectively. One can then use
the information encoded in line-profiles observed as a function of orbital
phase to calculate the strength of the line emission/absorption as a function
of velocity. This, in turn, can  be used to construct a velocity `image' or
map of the orbiting material in velocity space, defined as the strength
of the line emission/absorption as a function of velocity, $I(\mathrm{v}_x,\mathrm{v}_y)$.

\begin{figure}
\includegraphics[width=9.0cm,keepaspectratio]{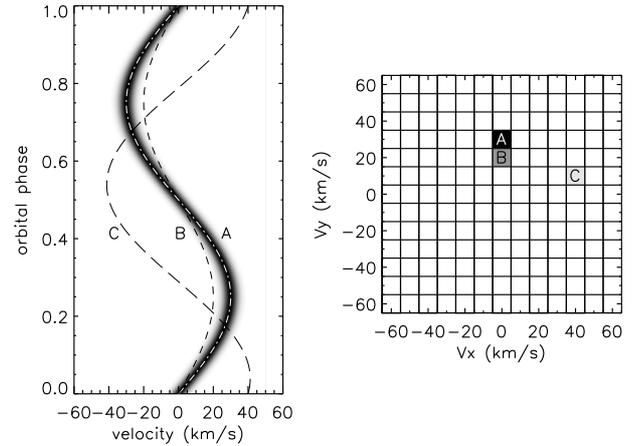}
\caption{Left-hand panel: schematic showing an observed signal following a
  radial-velocity path described by Equation~\ref{eq:vr} with v$_x$ = 0 km
  s$^{-1}$ and v$_y$ = 30 km s$^{-1}$. The dot-dashed line, `A', and
  dashed lines `B' and `C' show different radial-velocity paths -- with path A
  matching the observed signal. The right-hand panel depicts a velocity, or
  Doppler, map where the flux in each pixel represents a different integral
  along a path in the data given by Equation~\ref{eq:vr} and the respective
  v$_x$ and v$_y$ values. Where such a path is close to the true one, the
  result is a strong signal in the corresponding pixel (in this case pixel
  A).}
\label{fig:doppler_plot}
\end{figure}

Figure~\ref{fig:doppler_plot} presents a simple schematic
  outlining this process. Here we have a `real' signal with v$_x$ =
  0 km s$^{-1}$ and v$_y$ = 30 km s$^{-1}$. Integrating the observed flux
  along the correct radial-velocity curve (given by the dashed line `A' in
  Figure~\ref{fig:doppler_plot}) results in a strong signal at the corresponding
  pixel in the velocity map (right-hand panel of Figure~\ref{fig:doppler_plot}).
  Integrating the flux along a radial-velocity curve that is slightly offset
  from the true signal (e.g. path `B') would produce a reduced signal, since
  there is a reduced overlap with the real signal. Finally, path `C' traces
  a radial-velocity path with minimal overlap with the real signal, resulting in
  little signal appearing in the corresponding pixel in the velocity map.

Strictly speaking, the above describes the process of
  back-projection, where each pixel in the velocity map represents the integral
  of the signal along each individual radial velocity path described by
  Equation~\ref{eq:vr}.
  However, unlike back-projection, Doppler tomography fits the observed data
  simultaneously across all the observed lines and over all
  observed phases subject to assumptions regarding the expected modulation of
  the amplitude of the signal.
  In its most basic form, Doppler tomography assumes that a point is equally
  visible at all phases, and thus attributing a signal to paths B and C
  in Figure~\ref{fig:doppler_plot} would yield a poorer fit to the data than
  just assigning all the signal to path A. As a
  result, Doppler tomography returns a much weaker (or no) signal
  at incorrect velocities in the resultant Doppler map relative to
  standard back-projection techniques. For the same reason, aliasing effects
  common to
  back-projection approaches are also greatly reduced (see
  Section~\ref{sec:cont}). We note that Doppler tomography is not just
  constrained to assuming that the signal strength is constant at all orbital
  phases, but that it can also be modified to deal with a modulating signal.
  In the context of exoplanet atmospheres, this means Doppler tomography
  is able to deal with modulating signals that may arise from day-night
  phase variations, for example.

Although Doppler tomography has mainly been applied to
  cataclysmic variable binaries, primarily as a tool to probe the
  accretion regions in such systems, it has shown itself to be
  remarkably versatile -- and its adaptation to a star-planet system
  is straight-forward. For this work, we assume that the planet is on
a circular orbit, and thus the planetary radial-velocity
  variation can be described by Equation~\ref{eq:vr} with v$_x = 0$
  km s$^{-1}$ and v$_y$ = K$_p$, where K$_p$ is the radial-velocity
  semi-amplitude of the planet. The system can then be represented as an
image in velocity space, which is the line emission/absorption strength as a
function of velocity, I(v$_x$, v$_y$) -- as depicted in the
  right-hand panel of Figure~\ref{fig:doppler_plot}.  If each pixel in velocity
space has a width dv$_x$ and dv$_y$, then the contribution from that
pixel is given by,
\begin{equation}
  I\left(\mathrm{v}_x,\mathrm{v}_y\right)\left[g\left(\mathrm{v} - \mathrm{v}_R \right)\right]\mathrm{dv}_x\mathrm{dv}_y,
\end{equation}
\noindent where $g$ represents the line-profile shape from any point in
the image and includes effects such as instrumental broadening as well as line
broadening mechanisms.

The contribution of the lines from each position in our velocity map
between v and v+dv at orbital phase $\phi$ can then be determined
by carrying out the integral,
\begin{equation}
  f(\mathrm{v},\phi) = \int_{-\infty}^{\infty} \int_{-\infty}^{\infty} I\left(\mathrm{v}_x, \mathrm{v}_y\right)g\left(\mathrm{v}-\mathrm{v}_R\right)\mathrm{d}v_x\mathrm{d}v_y,
\end{equation}
\noindent where the integral limits are, in practice, set to cover an
appropriate range in radial velocity.

The line profile at any orbital phase can be thought of as a projection
of the velocity-space image along the direction determined by the orbital
phase. In Doppler tomography, the reverse process is performed and a
time-series of line profiles taken at different phases are inverted to
construct a velocity image. From Equation~\ref{eq:vr},
different values of v$_R$ define a whole set of parallel straight lines
across the Doppler map, with the direction dependent upon the orbital phase.
For example, orbital phase 0 corresponds to a projection along the
positive v$_y$ axis, while phase 0.25 corresponds to a projection along
the positive v$_x$ axis. Thus the formation of the synthetic line profile
for comparison to the dataset at a particular phase can be thought of as a
projection of the Doppler map along a direction defined by the orbital phase.
Technically, all projections can be covered in half-an-orbit, since
projections separated by half-an-orbit are merely mirror images of
each other.

Broadly speaking, there are 2 primary
techniques employed to perform the inversion: Fourier-filtered back-projection
or the use of the Maximum Entropy Method (MEM). We do not discuss the
back-projection method here and refer the reader to numerous reviews on the
topic (e.g. \citealt{marsh01a}), other than to note that it is limited in
practice as it cannot take into account effects such as high line optical
depths, blended lines, and any bad data needs to be interpolated over. The
MEM approach is more flexible and has greater tolerance to inconsistencies
within the dataset, and it is this version that we implement.

In the MEM reconstruction method, the intensities of the pixels in the
velocity image are adjusted in order to fit the observed data to a target
reduced $\chi^2$. The $\chi^2$ is not minimised, however, as this results in
a map dominated by noise. Instead, the data are fit until the predicted
and observed data are consistent. Since there are a number of different
images that could satisfy this situation, a regularisation statistic
is implemented to select one, and the image of maximum entropy (the
map with least information content relative to some comparison map --
called the `default map') is selected. The default map may encode
prior information on the expected line distribution, or a smoothed
(e.g. Gaussian smoothed) version of itself which helps preserve
narrow features -- but in its least informative form is simply a uniform map.
Thus, Doppler tomography can be thought of as returning a velocity map
of the system that contains the least amount of information required
to describe the data to the required $\chi^2$.

The power of the MEM is that it requires no special
provision to deal with missing data, it takes into account the error
bars on the data, and the maximum entropy regularisation
statistic tends to suppress the growth of noise (which acts to increase the
information content of the velocity map) in the final results.
For more
details on Doppler tomography, we refer the reader to
the many extensive reviews of the technique undertaken in the past
(e.g. \citealt{marsh01a};
\citealt{marsh04}; \citealt{morales-rueda04}; \citealt{richards04};
\citealt{schwope04}; \citealt{steeghs04}; \citealt{vrtilek04}). We now
turn to outlining ts application to the high-resolution study of exoplanet
atmospheres. In order to demonstrate the applicability of Doppler tomography to
studying the Doppler shifted signatures of exoplanet atmospheres, we first
apply it in the next Section to the exoplanet $\tau$ Boo b.

\section{Doppler tomography of tau Boo b}
\label{sec:tauboo}

To demonstrate Doppler tomography's capability to recover the signal
from molecules in real exoplanet data, we used CRIRES (\citealt{kaeufl04})
data for $\tau$ Boo b from \cite{brogi12}, who used it to succesfully detect CO
in the atmosphere of this non-transiting planet. This dataset consists
of 452 spectra obtained over 3 epochs amounting to a total of $\sim$18
hours of observing time. We have implemented exactly the same data reduction
and pre-processing steps as \cite{brogi12}, including systematics removal
and adoption of their orbital phases corrected from the original
\cite{butler06} orbital solution, and we refer the reader to that paper for details.

For the Doppler tomography analysis we used a CO linelist for the
$^{12}$C$^{16}$O isotopologue obtained from the HITRAN database
(\citealt{rothman13}) using HAPI (\citealt{kochanov16}).
HAPI was also
used to calculate the expected strengths of the lines for a
temperature of 1700K and a pressure of 0.1bar. The results from this
were subsequently fitted in IDL to generate a linelist file to use in
the analysis, a total of 122 of the strongest CO lines in the
wavelength range covered were used, with the strength of the faintest line
at $\sim$2.5\% that of the strongest line. This cut was made
in order to retain a sufficient number of lines for use in Doppler
tomography while removing the weakest lines that will have
increasingly smaller contributions to the final reconstruction. A
uniform default map was employed
during the Doppler tomography reconstruction, and the velocity image
consisted of 300 by 300 pixels with a 1.5 km s$^{-1}$ stepsize.
The same orbital phases as those in the \cite{brogi12} analysis
were adopted, and the known systemic velocity of $\tau$ Boo was removed from the
data. Hence any planetary signal should lie on the v$_x$ = 0 km s$^{-1}$ line
in the Doppler tomogram.

\begin{figure}
\includegraphics[width=7.8cm,keepaspectratio]{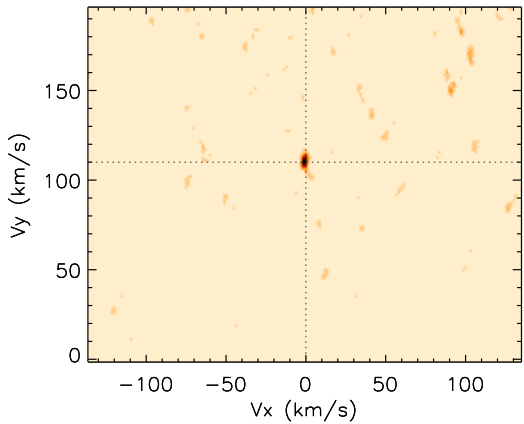}
\caption{Doppler tomogram of the $\tau$ Boo b system using a CO linelist.
  The intersection of  the dashed lines indicates the expected location
  of the planetary signature based on the detection of Brogi et al. (2012).
  CO is clearly detected in the Doppler tomogram at the expected location.}
\label{fig:tauboo}
\end{figure}

We ran the Doppler tomogram to a reduced $\chi^2$=0.96346, and the resulting map
can be found in Figure~\ref{fig:tauboo}. The CO signal from the planet
can clearly be seen at the expected position with a planetary radial
velocity semi-amplitude (K$_p$ = v$_y$) of 109.6 $\pm$ 2.2 km s$^{-1}$.
The uncertainty on K$_p$ was determined by fitting a two-dimensional
Gaussian to the planet signal in the Doppler map and adopting the
$\sigma-$width of the Gaussian along the v$_y$ axis.  Our measured
K$_p$ is in excellent agreement with the value of 110.0 $\pm$ 3.2 km
s$^{-1}$ reported by \cite{brogi12}, and demonstrates that Doppler
tomography can be used to directly detect exoplanet atmospheres.

We note that the
  fact we were able to fit to a reduced $\chi^2$ of less than 1 indicates
  that we have slightly over-estimated the uncertainties on the data. In the
  case of weak signals (as expected in the study of exoplanet atmospheres)
  there is normally a relatively narrow $\chi^2$ window over which a
  satisfactory Doppler tomogram can be recovered. Fitting to higher
  aim $\chi^2$'s results in a blank featureless map, while conversely fitting
  to too low a $\chi^2$ results in a map dominated by noise. Thus, while the
  exact choice of aim $\chi^2$ is somewhat subjective, it is obvious when
  one lies outside a reasonable range. In the case of $\tau$ Boo b, if we fit
  to a reduced $\chi^2$ = 1 then, while the planet signal is still present,
  the Doppler map is poorly constrained by the data and does not converge.
  However, it is evident in the iterative process that Doppler tomography
  wants to push towards a lower  $\chi^2$.

\section{Doppler tomography comparison to CCF approach}
\label{sec:sims}

In this section, we compare the ability of
Doppler tomography to the conventional CCF analysis approach
for the purposes of analysing time-series high-resolution Doppler observations
of exoplanet atmospheres. This was done through tests using a series of
simulated datasets. Each synthetic dataset consists of 201 evenly spaced
orbital phases from 0.25 to 0.75. Lines were injected into the simulations
assuming a systemic velocity (v$_{\mathrm{sys}}$) of 0 km s$^{-1}$,
a planetary radial
velocity semi-amplitude K$_p$ = 150 km s$^{-1}$, and a
full-width-half-maximum (FWHM) of 4.7 km s$^{-1}$ unless otherwise stated in
the relevant section. The wavelength
coverage varies between each simulation (indicated where relevant), and is
sampled by a pixel grid of 1 km s$^{-1}$. Each simulation assumes
white-noise (of varying levels between simulations), and the
same noise map is applied to each dataset. The Doppler tomography
reconstructions were all run using a uniform default map.

We note that we have not considered the impact that temperature
inversions can have on the line-shapes themselves within our linelist
(e.g. \citealt{parmentier18}). This would require a full model
atmosphere in order to produce the full complement of emission,
absorption, and `nulled' lines. This could be incorporated into
Doppler tomography by using multiple line-lists capturing the main
morphological line-shapes, and then fitting for these lines simultaneously
(e.g. see Section~\ref{sec:isotop}).

In this paper we compare the Doppler tomograms
  (in terms of v$_x$ and v$_y$, as shown in Figure~\ref{fig:tauboo}) to
  the conventional way of showing the CCF analyses as a plot of the CCF
  strength as a function of v$_{\mathrm{sys}}$ and K$_p$ (e.g. see right-hand
  panel of Figure~\ref{fig:dotest}). We note that,
while v$_y$ and K$_p$ are comparable on the y-axis of both the Doppler
tomogram and CCF maps, v$_x \neq$ v$_{\mathrm{sys}}$. However, we maintain
this format for presenting the CCF maps in this work as it allows for direct
comparison to extrasolar planet atmosphere detections presented in the
literature to date. We also note here that taking a cut along v$_x$ = 0 km
s$^{-1}$ in a Doppler map and a cut at v$_{\mathrm{sys}}$ =  0 km
s$^{-1}$ in the CCFs (after removing the system's systemic velocity,
as later presented in the first 2 columns of Figure~\ref{fig:simulations2})
are directly comparable.

In Appendix~\ref{sec:appendix} we also compare a sample
  of the Doppler tomography results versus those obtained by back-projecting
the CCFs in Figure~\ref{fig:appsims}. This provides a direct comparison with Doppler
tomography as it uses an identical v$_x$ and v$_y$ grid. We also present an alternative
method of phase-folding the CCFs as a function of the orbital phase-offset and K$_P$
(see Figure~\ref{fig:appsims2}).

\subsection{Noise properties}
\label{sec:dotest}

Our first simulations investigated the relative performance of the
techniques at different signal-to-noise levels. The signal-to-noise
ratios reported in these simulations are per frame and are such that
the values quoted represents the signal-to-noise of the planetary signal
when combining all the lines present -- as measured from the peak of the
lines. However, we should note that since the lines are not confined to a
single pixel (but span several), the
actual signal-to-noise ratio of the simulations will be higher (by a
factor of 2-3 as estimated using the FWHM and pixel scale).  For the
tests described in this Section, we have generated synthetic spectra
using the same CO linelist as applied in our analysis of $\tau$ Boo
data presented in Section~\ref{sec:tauboo}. 

\begin{figure*}
\begin{tabular}{ccc}
\includegraphics[width=5.6cm,keepaspectratio]{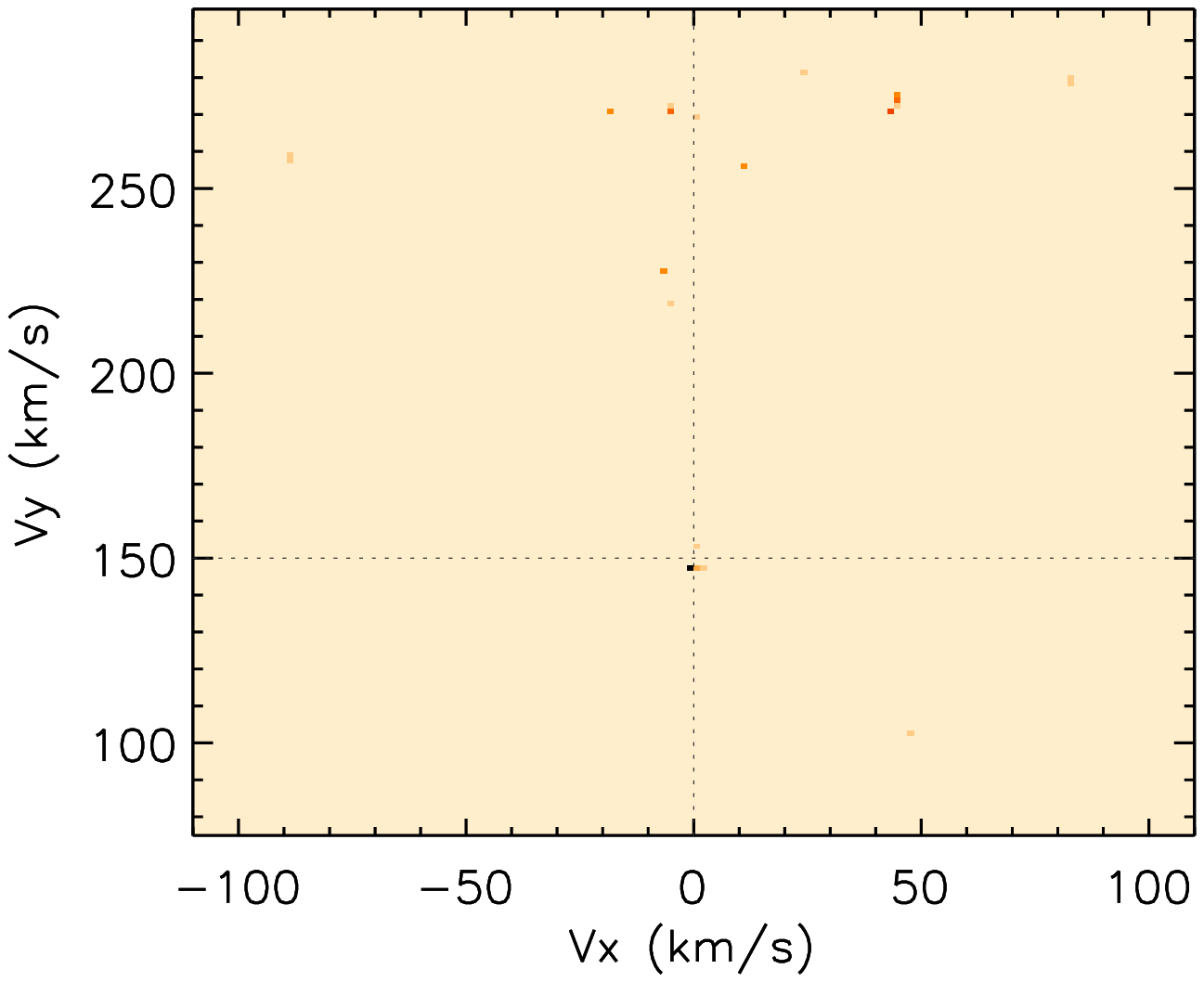} &
\includegraphics[width=5.6cm,keepaspectratio]{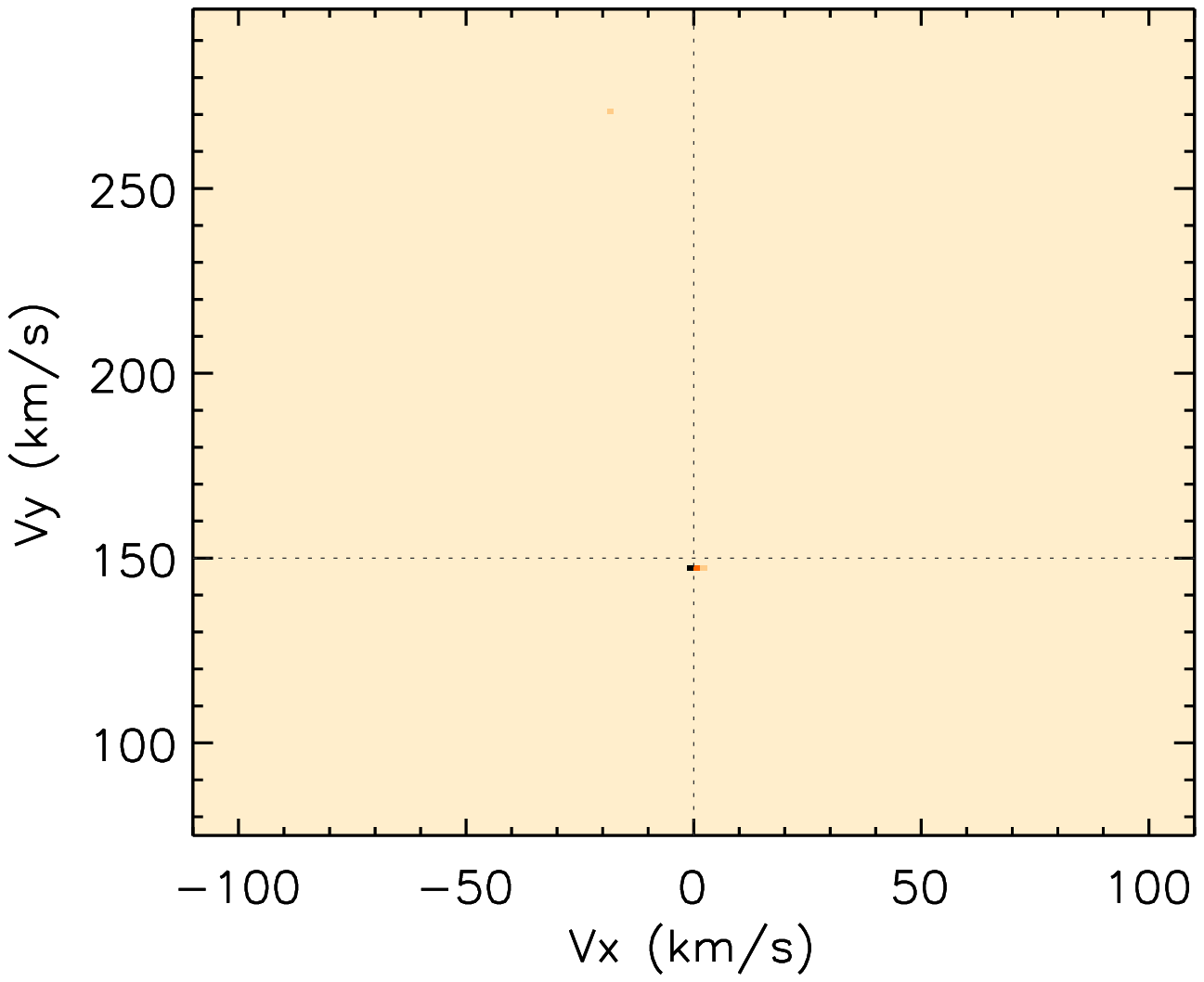} &
\includegraphics[width=5.6cm,keepaspectratio]{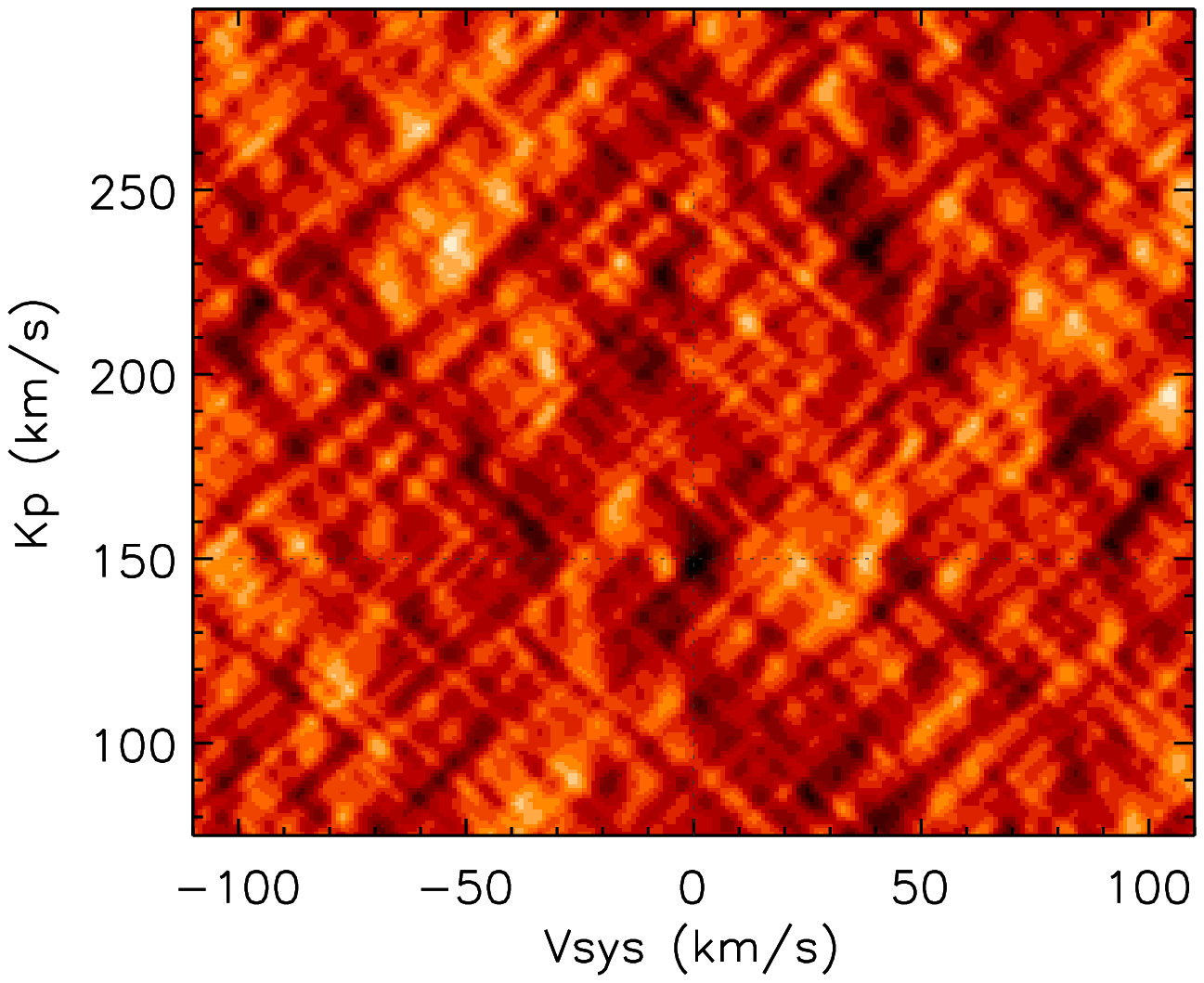} \\
\end{tabular}
\caption{Left panel: Doppler tomogram of the synthetic dataset as outlined
  in Section~\ref{sec:dotest}. The intersection of the dashed lines indicates
  the location of the injected planetary signal, which Doppler tomography
  recovers. Noise manifests itself as localised intensity spikes in the Doppler
  map and in this case is particularly evident at higher v$_y$ values.
  Central panel: The results of the signal
  coherency test whereby each pixel represents the minimum intensity value of
  the respective pixel in either the Doppler tomogram constructed using only
  odd-numbered spectra, or using only even-numbered spectra. Since random noise
  is not a truly coherent signal then most of the spikes are filtered out in
  this process, leaving only the true (injected) planet signal. Right panel:
  The CCF map for the same dataset used in the Doppler tomography
  reconstruction shown on the left panel.}
\label{fig:dotest}
\end{figure*}

Figure~\ref{fig:dotest} shows the results of the Doppler tomogram
(left-panel) and the CCF analysis (right panel), respectively, for a
signal-to-noise ratio of 1/13 (this S/N was adopted as it gives a
realistically challenging signal to recover in the CCF analysis). The
intersection of the dashed lines indicate the location of the injected
planet. In all cases the images are plotted on a linear scale running
from the minimum to the maximum value within the image. This scaling
allows the noise properties within each map relative to the recovered
line profile contributions to be visually compared.

While both the Doppler tomogram and CCF map recover the planetary signal,
the first thing to note is the reduced noise levels in the Doppler
tomograms. This is due to the fact that
Doppler tomography fits the entire dataset coupled to the use of
the maximum entropy regularisation statistic, which seeks to find the image of
least information content. This has the effect of suppressing
noise features that do not follow a radial velocity trajectory through the
data as described by Equation~\ref{eq:vr}. 

\begin{figure*}
\begin{tabular}{cc}
\includegraphics[width=7.8cm,keepaspectratio]{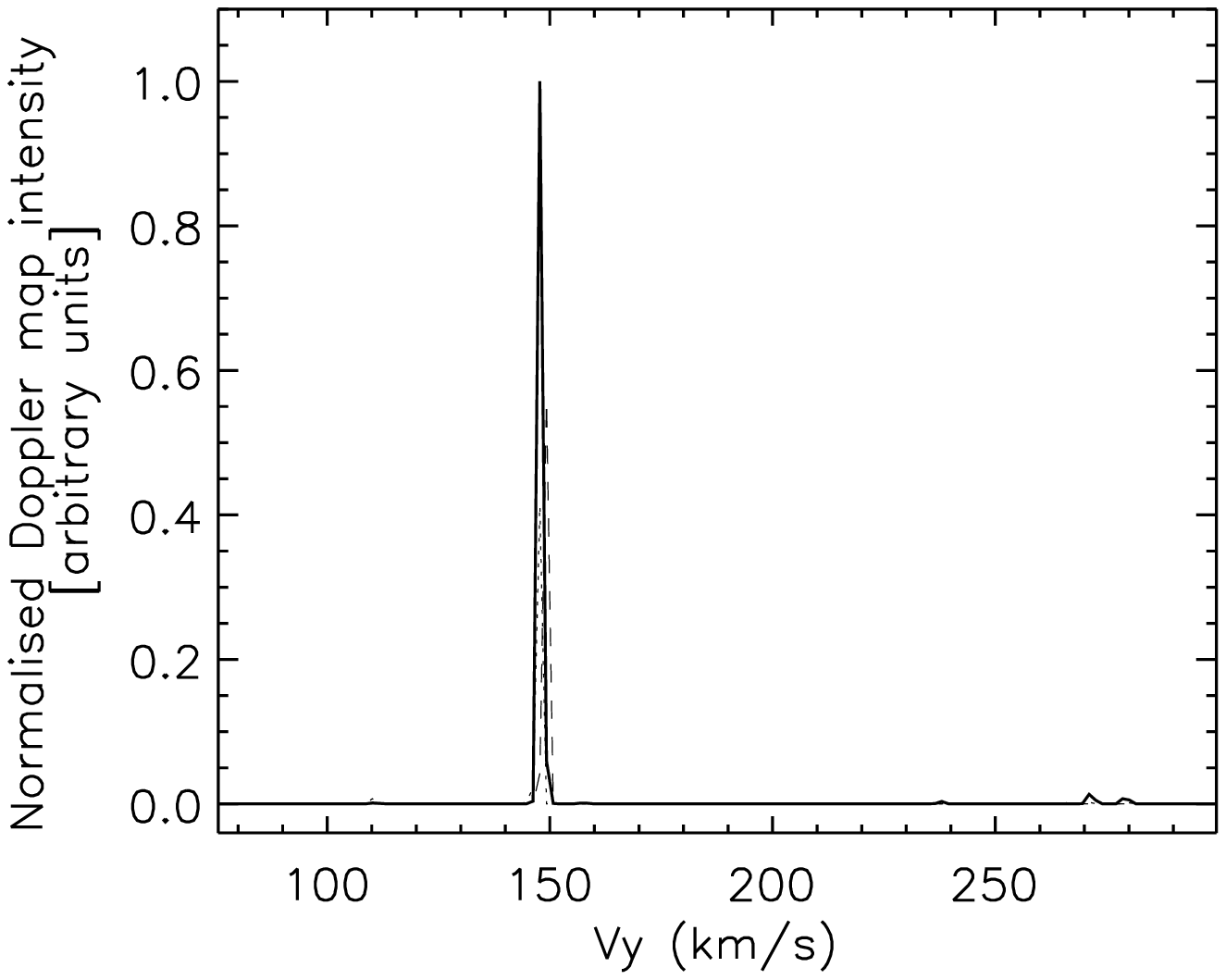} &
\includegraphics[width=7.8cm,keepaspectratio]{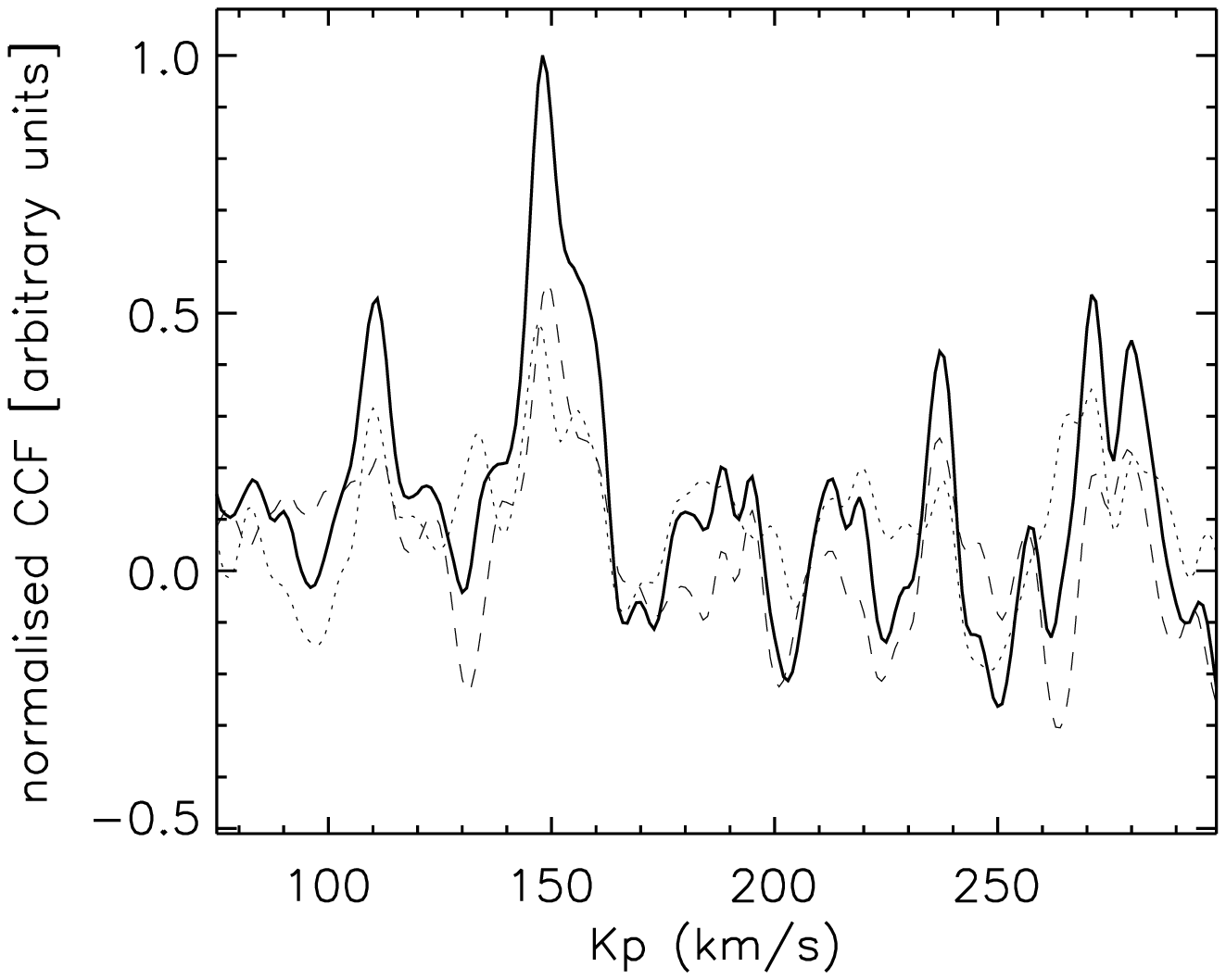} \\
\vspace{0.4cm} \\
\includegraphics[width=7.8cm,keepaspectratio]{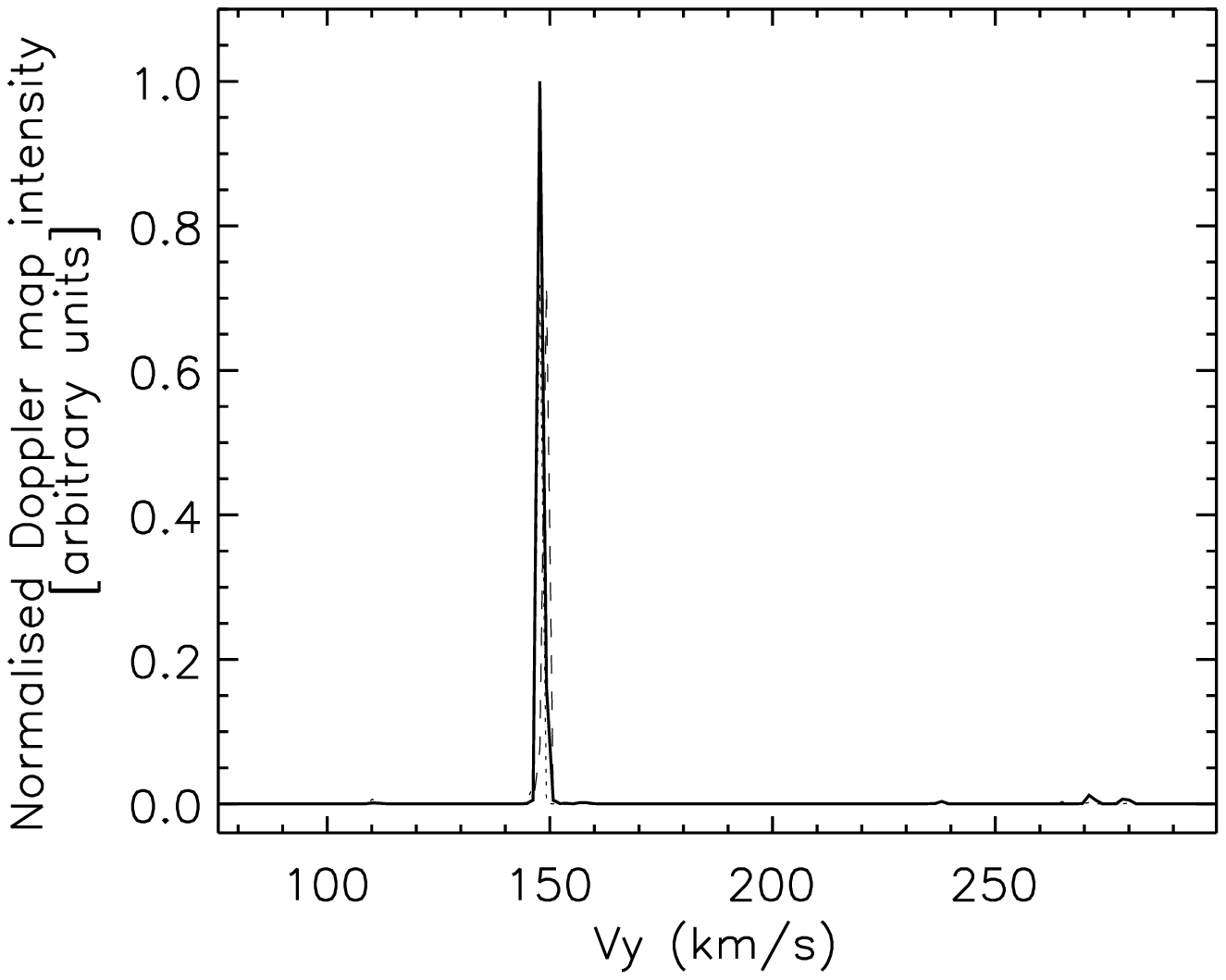} &
\includegraphics[width=7.8cm,keepaspectratio]{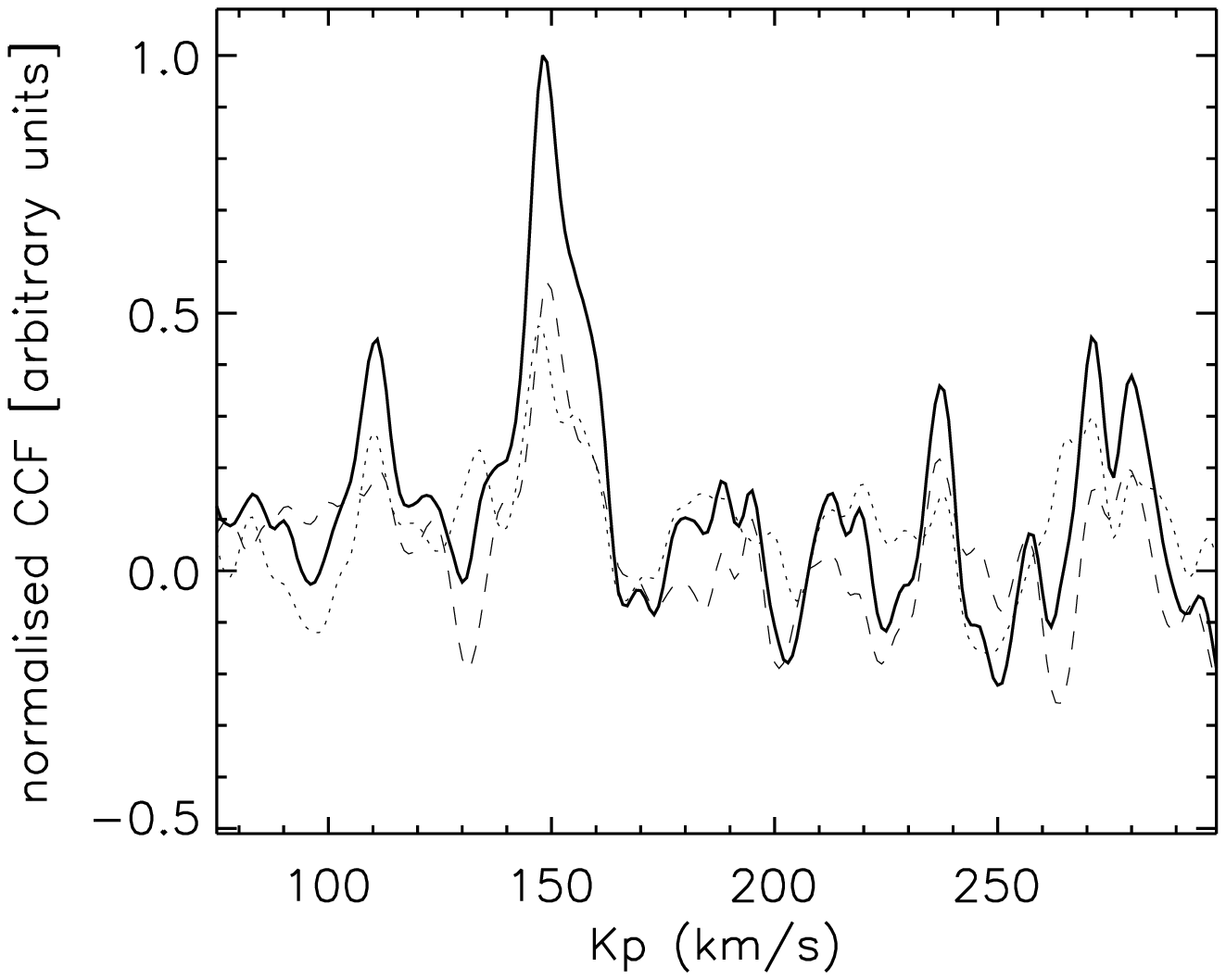} \\
\end{tabular}
\caption{Left panels: Cuts through the Doppler tomograms at v$_{x}$ = 0 km
  s$^{-1}$. Right panel: Equivalent cut through the CCF maps at v$_{\mathrm{sys}}$ = 0 km
  s$^{-1}$. The top panels are for the maps shown in Figure~\ref{fig:dotest},
  while the bottom panels are for simulations with the same noise map but
  with a 1.44 times higher signal-to-noise ratio for comparison. Over-plotted
  are the results when only even- (dashed lines) or odd-numbered (dotted lines)
  spectra are considered. The strengths of the odd/even tests are
  scaled relative to the signals generated by the full dataset results.}
\label{fig:dotest2}
\end{figure*}

This can be further demonstrated in Figure~\ref{fig:dotest2}, which
shows 1-dimensional cuts across regions of physical meaning in both
the Doppler tomograms and the CCF maps. The top panel shows these cuts
for the simulations used for Figure~\ref{fig:dotest}, while the bottom
panels show the same for a simulation where the signal-to-noise is a
factor of 1.44 times higher.  In the case of the Doppler tomogram, the
only cut across the map of physical significance is along the v$_x = 0$
line (vertical dashed line in the Doppler tomograms), where v$_y$ maps
along the direction of the planet's radial velocity semi-amplitude,
K$_p$ (though see Section~\ref{sec:phaseoff}). For the CCF maps, we
present a cut along the K$_p$ axis, which provides a direct
  comparison between the two methods. Each plot is
normalised by the maximum value of the corresponding map presented in
Figure~\ref{fig:dotest}, thus a value of 1 means that it is the
strongest signal present in the entire map. As can be seen in
Figure~\ref{fig:dotest2}, as well as the reduced noise, the Doppler
tomogram provides a much sharper line recovery compared to the CCF,
despite the fact that a uniform default map has been implemented
(generally a Gaussian smoothed default map would be better at
recovering point-like line contributions to the Doppler tomogram).

Naturally, it is important to establish the robustness of any potential
planetary signal. For CCF analysis, the significance of any detection is
often estimated by measuring the signal peak-to-RMS ratio, where the
RMS is measured from the `background' noise in the CCF
maps. Unfortunately, such an analysis cannot be performed for Doppler
tomograms as the maximum-entropy regularisation statistic suppresses
noise, leading to a highly non-Gaussian noise distribution in the
resultant maps. As can be seen in the left-hand panel of
Figure~\ref{fig:dotest}, while the injected planetary signal is
clearly recovered, noise in Doppler tomograms manifest themselves as
localised spots while the majority of the tomogram is noise free due
to the regularisation statistic.  These noise `spikes' could easily be
mis-identified as a bona-fide planet signal if they lie close to the
v$_x$ = 0 km s$^{-1}$ line along which any planet should lie.

In order to test the veracity of the recovered signal we examined the
coherency of features in the map by constructing two additional
Doppler tomograms.  These tomograms were formed using only the
odd-numbered spectra for one and the even-numbered spectra for the
other, effectively forming two independent datasets. Only coherent
signals in both datsets will appear in both tomograms, while the noise
spikes will (typically) appear at different locations in the two
tomograms. This odd-even approach is applied as standard to assess the
reality of features produced by Doppler-imaging techniques
(e.g. \citealt{barnes04}; \citealt{watson06}; \citealt{watson07};
\citealt{xiang15}). By then constructing a Doppler map formed by
selecting the minimum intensity from either the even- or odd-phased
Doppler tomogram at each pixel position, we can examine what features
remain constant.

The central panel of Figure~\ref{fig:dotest} displays
the results of this coherency test and, in comparison to the left-hand panel
of Figure~\ref{fig:dotest}, shows that this is an effective means
for filtering out incoherent (noise) features. Naturally, it is
possible that random noise could produce a broadly coherent signal in
the Doppler tomogram that would not filter out, but under such
circumstances a CCF analysis would suffer from the same effect.
To show that the odd/even peaks are coherent in the Doppler tomogram,
these are over-plotted in Figure~\ref{fig:dotest2}. In this figure we also
show the same for the CCF analysis for comparison.

In summary, at the signal-to-noise levels in the simulations we have
conducted, Doppler tomography gives clearer signal detections.
There are a number of reasons why one might expect Doppler tomography,
in principle, to perform better than cross-correlation approaches.
For example, aliasing of signals is a problem inherent to CCF analyses (as
explored later in this section). This projects additional structures
in the resulting CCF maps, which may both confuse the identification
of the true molecular signal as well as act to reduce its significance.
Since Doppler tomography actually fits to the lines themselves, aliasing
issues are not so prominent (as demonstrated later), and the use of a
regularising statistic also helps to suppress noise by minimising the
information content of the final Doppler map.

There are, however, a number of possible caveats that should be considered.
One is that Doppler tomography has the additional complication that a target
$\chi^2$ has to be selected, and this is somewhat subjective -- there is no
such issue for CCF analysis. In addition, the maximum entopy regularisation
enforces a strict positivity criterion to the data. This effectively removes
any negative signals (i.e. removes any positive noise lying above the
continuum when an absorption signal is being searched for). This comes with
the potential expense that a poor continuum fit can result in a planetary
signal being lost, while the CCF analysis will be robust against this.
We should note, however, that the results from our $\tau$ Boo analysis
(Section~\ref{sec:tauboo}) indicates that this is not a particular issue
in reality as implemented here. One should also note that the
strict positivity criterion does not mean that absorption lines cannot be
traced as this simply requires flipping the sign of the dataset. In addition,
it is possible to map both absorption and emission lines simultaneously by
adopting a `virtual' pixel in the Doppler map that contributes to all
wavelengths. This could then be used to accommodate any
(deliberate or otherwise) continuum offests to ensure absorption lines obeyed
the positivity criterion, and the use of such a 'virtual' pixel has
previously been adopted in Roche tomography reconstructions \cite{watson03}.

\subsection{Impact of contaminating lines}
\label{sec:cont}

Here we probe the relative performance of the techniques
when there is a nearby, strong contaminating line with the same
radial velocity variation as the line of interest. This
also provides a simple demonstration of the impact of errors
(e.g. missing lines, incorrect line positions) in linelists used for such
work (see, for example, \citealt{hoeijmakers15}).

In the first 2 simulations, the synthetic dataset consists of two
equal strength lines separated by 50 km s$^{-1}$. In the first
instance (see Figure~\ref{fig:simulations}, Images A1 \& A2) the noise was set
such that the RMS of the map was 3.98 times greater than the line-depth, and
in the second instance the RMS was 6.31 times greater than the line-depth
(see Figure~\ref{fig:simulations}, Images B1 \& B2). The results of the Doppler
tomogram are shown in the left panel, with the comparison of the CCF
results shown in the right panel. The intersection of the dashed
lines indicate the location of the injected planet.

In Figure~\ref{fig:simulations2} we show 1-dimensional cuts across regions
of physical meaning in both the Doppler tomograms and the CCF maps.
These cuts are similar to those presented in Figure~\ref{fig:dotest2}, however
for the CCF maps we also includes cuts along the line of constant K$_p$
(i.e. at the velocity of the planet but at different systemic velocites,
v$_{\mathrm{sys}}$). Again, as before, each plot is normalised by the maximum
value of the corresponding map presented in Figure~\ref{fig:simulations}.

In both signal-to-noise cases, the true planet
signature is the strongest feature present in the Doppler tomogram.
This is not the case for the CCF map, which shows the most power arising
from the contaminating line not included in the linelist,
seen as a strong feature shifted by v$_{\mathrm{sys}}$ = +50 km s$^{-1}$ from the
actual planet position in Figure~\ref{fig:simulations} A2 \&~B2
and Figure~\ref{fig:simulations2} A3 \&~B3. Indeed, this could potentially
yield confusion as to the nature of the signal and/or systemic velocity.
This is in contrast to the Doppler tomogram, which has efficiently
filtered out the contanimating line. The explanation for this can
be seen in the Image C rows of Figures~\ref{fig:simulations}
\&~\ref{fig:simulations2}, which is the same dataset used for Image A,
but with the contanimating line now offset by 20 km s$^{-1}$. The Doppler
tomogram now shows a semi-circular arc-structure to the right of the true
planet position. This arises due to the fact that Doppler tomography
sees the contanimating line as having a non-zero systemic velocity and
therefore, from Equation~\ref{eq:vr}, it is forced to map the contribution
of the contaminating line across a range of v$_x$ and v$_y$ pixels on
the map. This essentially dilutes the signal in the Doppler map by
spreading it over a larger area in velocity space -- which is then
suppressed by the maximum entropy regularisation. Thus, Doppler
tomography is less affected by missing lines in the linelist,
or potential contamination from other lines or species
(see section~\ref{sec:lineerrors} for a further example of this). Indeed,
in this particular case, careful phase selection can prevent any power
from contaminating lines leaking onto the v$_x$ = 0 km s$^{-1}$ line in
the Doppler tomogram, along which any planet signal is expected to lie if the
systemic velocity and orbital phases are known for the system.

In Appendix~\ref{sec:appendix} we also present the back-projections of the CCFs
  in Figure~\ref{fig:appsims}
  (Images A and C). In this case, the CCFs also show a similar semi-circular
  artefact due to the nearby contaminating line, but also exhibits much higher
  noise levels than the Doppler tomogram. We note that, in the case where the
  contaminating line source pattern is much more complicated than simply a
  nearby single line, Doppler tomography performs much better than the
  CCF back-projection method (see Sections~\ref{sec:lineerrors}
  and~\ref{sec:isotop}) due to the fact that it fits all the data
  simultaneously.

\subsection{Line-list confusion}
\label{sec:lineerrors}

In our next set of simulations, we downloaded a CO linelist for both the
$^{12}$C$^{16}$O and $^{12}$C$^{17}$O isotopologues obtained from
the HITRAN database (\citealt{rothman13}) using HAPI (\citealt{kochanov16}).
Both linelists were used to generate synthetic spectra covering a
wavelength range from 22,861.833$\AA$~to 23503.069$\AA$~in 0.077$\AA$~steps.
The $^{12}$C$^{16}$O contained 107 lines within this wavelength range, whereas
the $^{12}$C$^{17}$O contained 53 lines. An unrealistic case of equal
abundance of both isotopologues was assumed, and the signal-to-noise in
each individual spectrum was set such that 
the resultant CCF signal for $^{12}$C$^{17}$O in each individual spectrum
approximately equaled the noise level (i.e. an approximately 1-$\sigma$
detection of $^{12}$C$^{17}$O in 1 spectrum). Since $^{12}$C$^{16}$O has more
lines in the wavelength range under consideration, it gives a stronger signal.

\begin{figure*}
\begin{tabular}{cc}
\includegraphics[width=7.8cm,keepaspectratio]{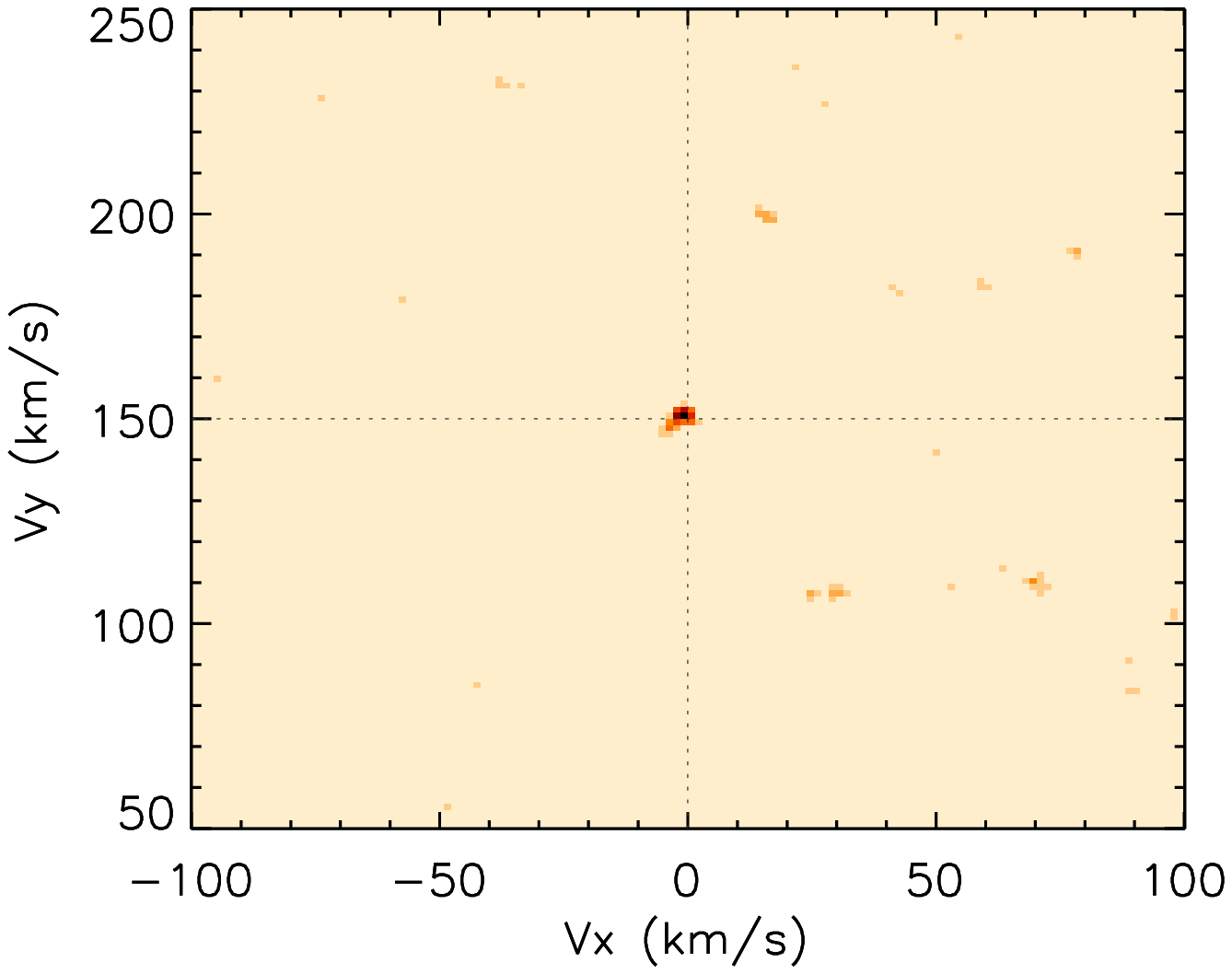} &
\includegraphics[width=7.8cm,keepaspectratio]{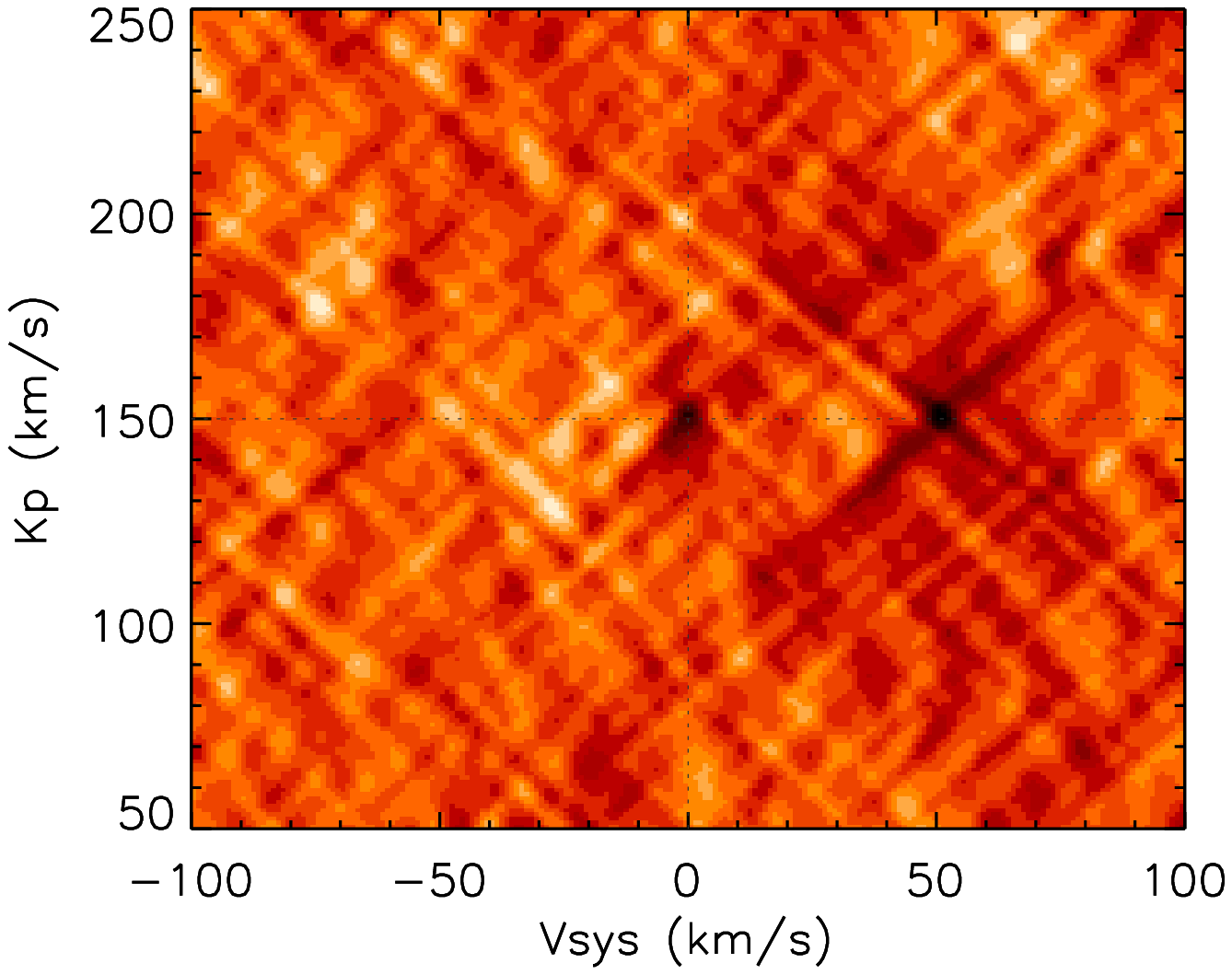} \\
Image A1: Doppler tomogram, contaminating line at +50 km s$^{-1}$ & Image A2:
CCF, contaminating line at +50 km s$^{-1}$
\vspace{0.4cm}\\
\includegraphics[width=7.8cm,keepaspectratio]{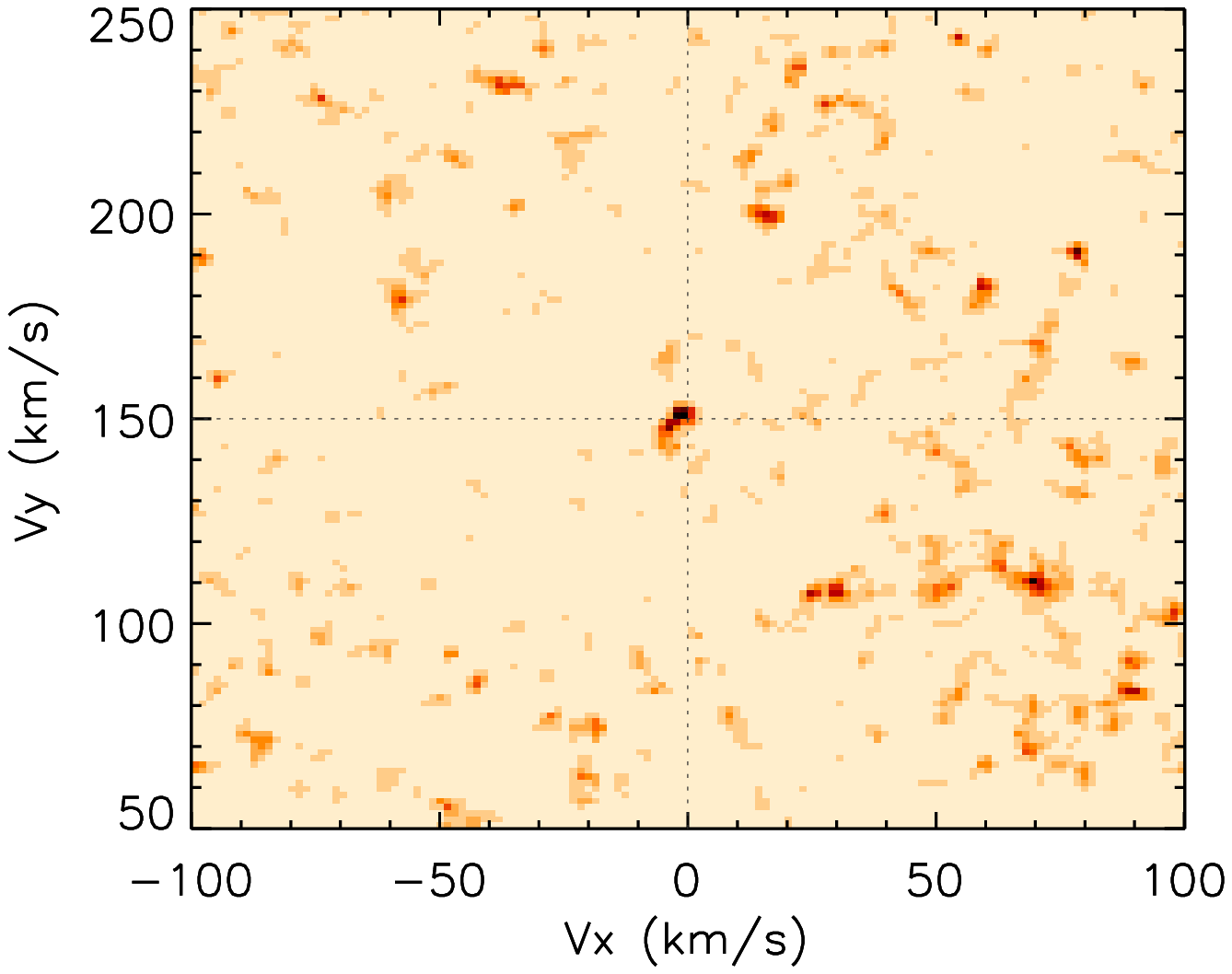} &
\includegraphics[width=7.8cm,keepaspectratio]{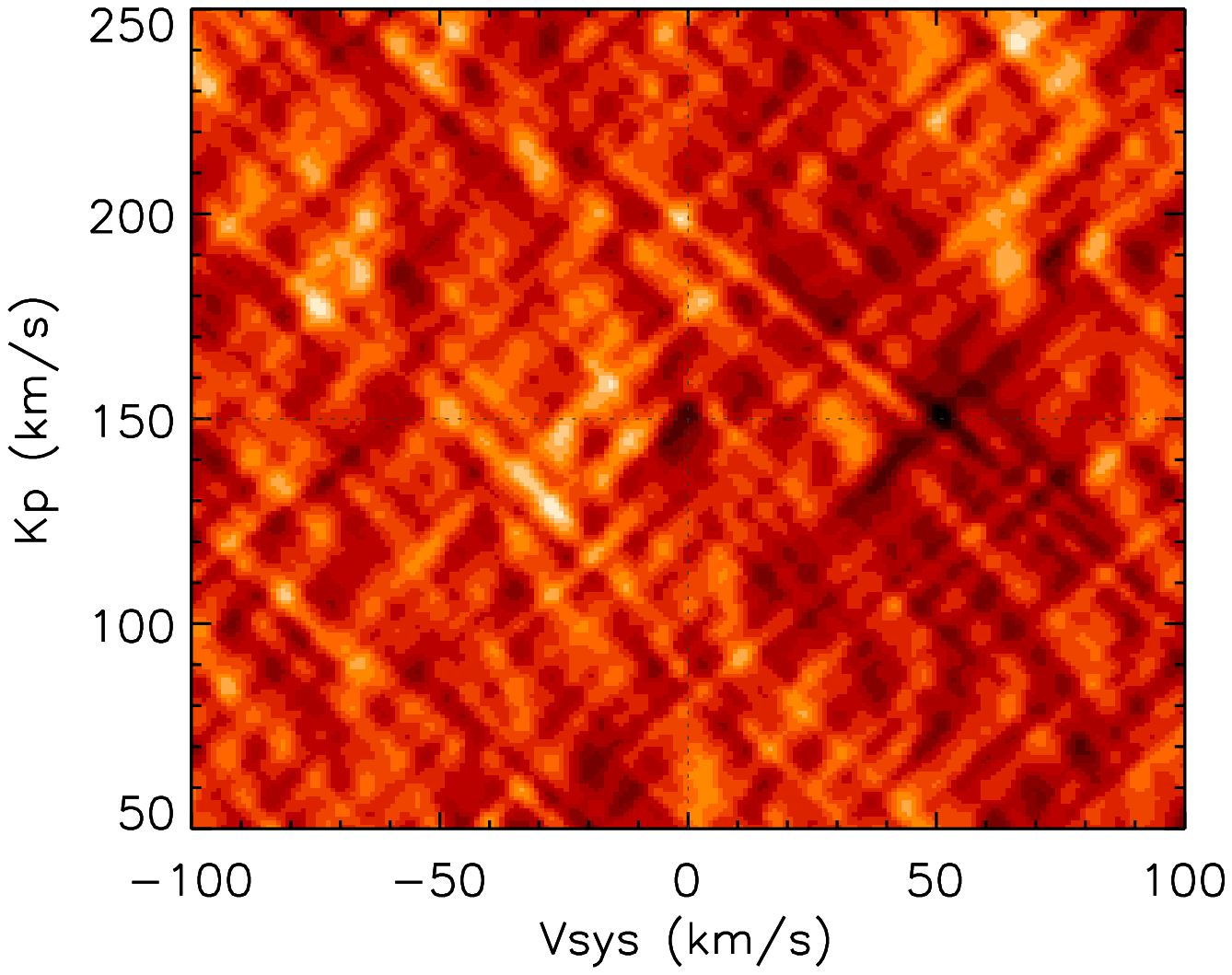} \\
Image B1: As Image A1, but with a noise-level $\sim$1.6 times higher &
Image B2: As Image A2, but with a noise-level $\sim$1.6 times higher
\vspace{0.4cm}\\
\includegraphics[width=7.8cm,keepaspectratio]{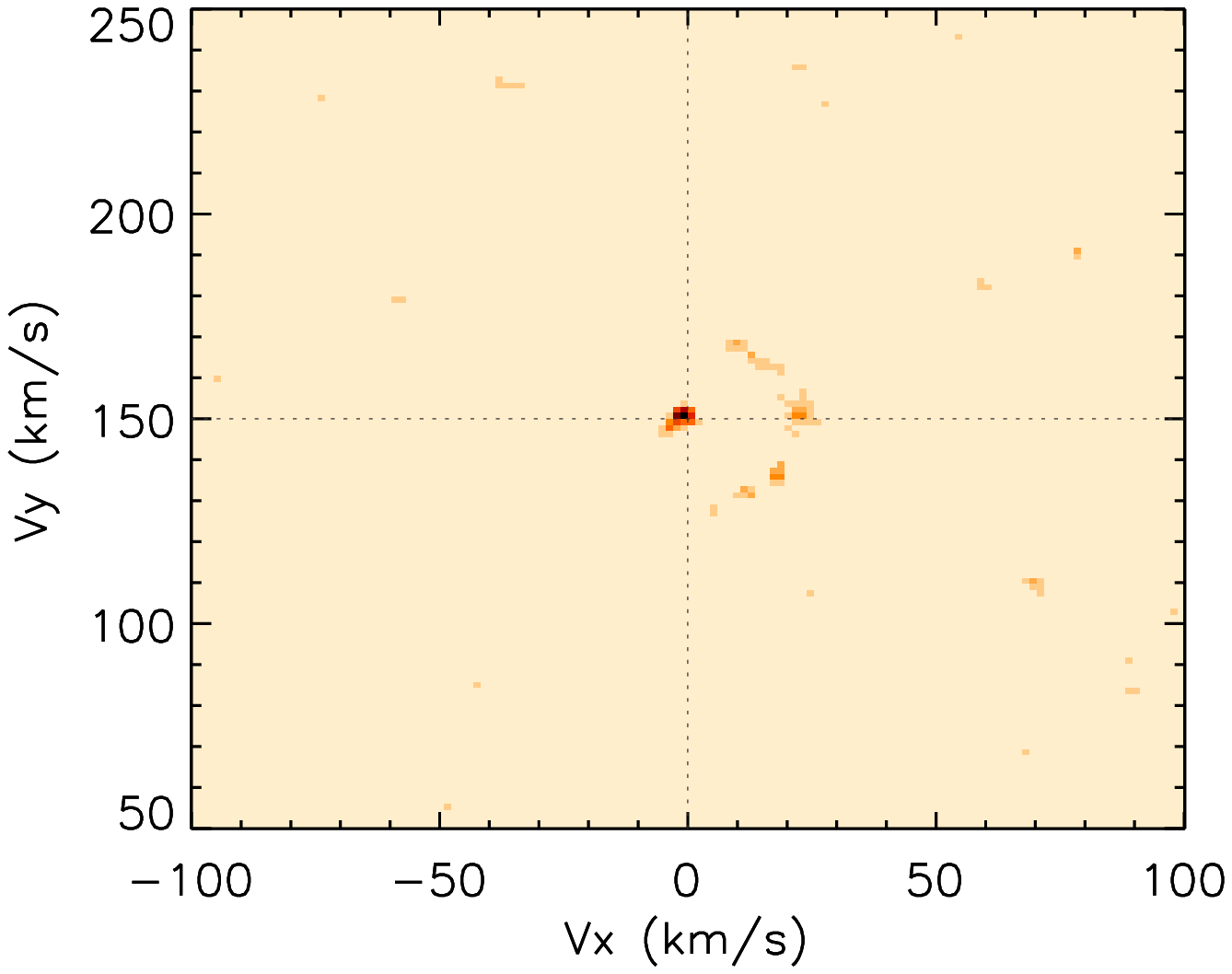} &
\includegraphics[width=7.8cm,keepaspectratio]{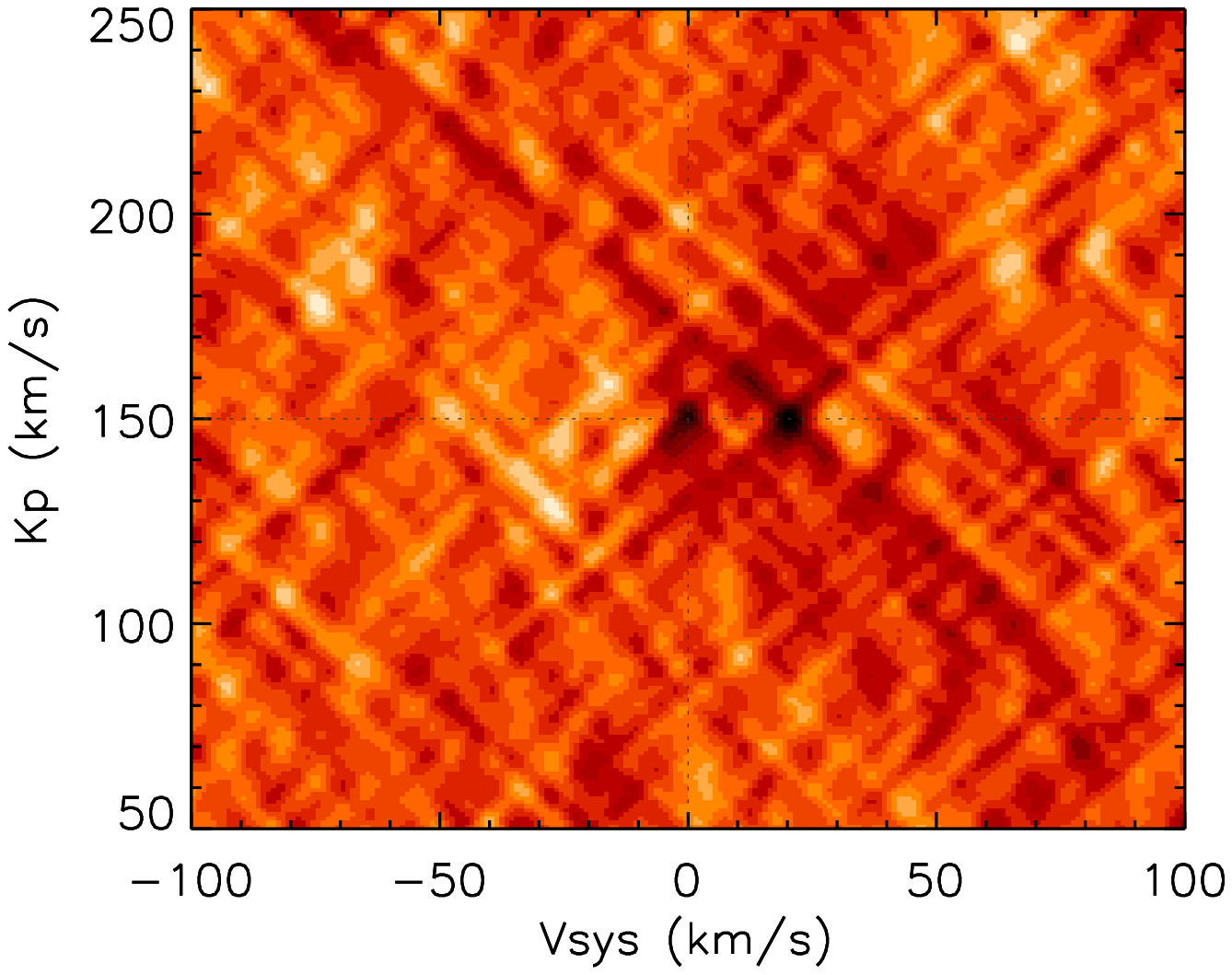} \\
Image C1: Doppler tomogram, contaminating line at +20 km s$^{-1}$ &
Image C2: CCF, contaminating line at +20 km s$^{-1}$
\vspace{0.4cm}\\
\end{tabular}
\caption{Comparison between Doppler tomography (left panels) and the
  CCF method (right panels) for different simulations. Images A and B
  show the effects for a contaminating line that is not included in
  the linelist, at +50km s$^{-1}$. For Image A the noise level was
  set such that the RMS in the data was $\sim$4 times the expected
  line strength, while in Image B the noise-level is set to 6.31
  times the line strength. Image C shows a similar simulation, but now
  with the contaminating line offset by only 20 km s$^{-1}$ and a
  noise level that is $\sim$4 times that of the targeted line. All images
  are linearly scaled from the minimum to the maximum values in the
  respective map. The intersection of the dashed lines indicate the location of
  the true injected signal.}
\label{fig:simulations}
\end{figure*}

\begin{figure*}
\begin{tabular}{ccc}
\includegraphics[width=5.5cm,keepaspectratio]{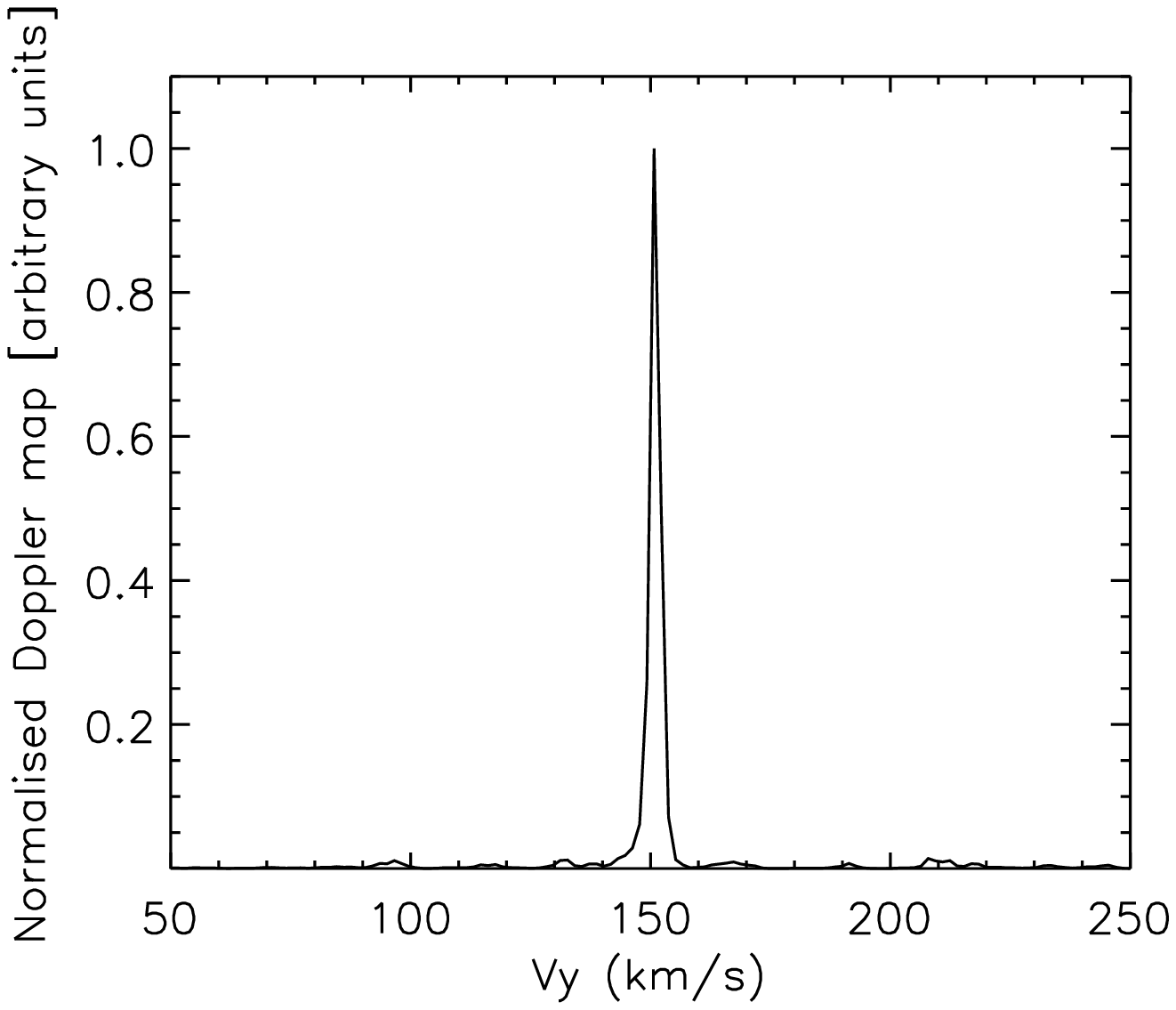} &
\includegraphics[width=5.5cm,keepaspectratio]{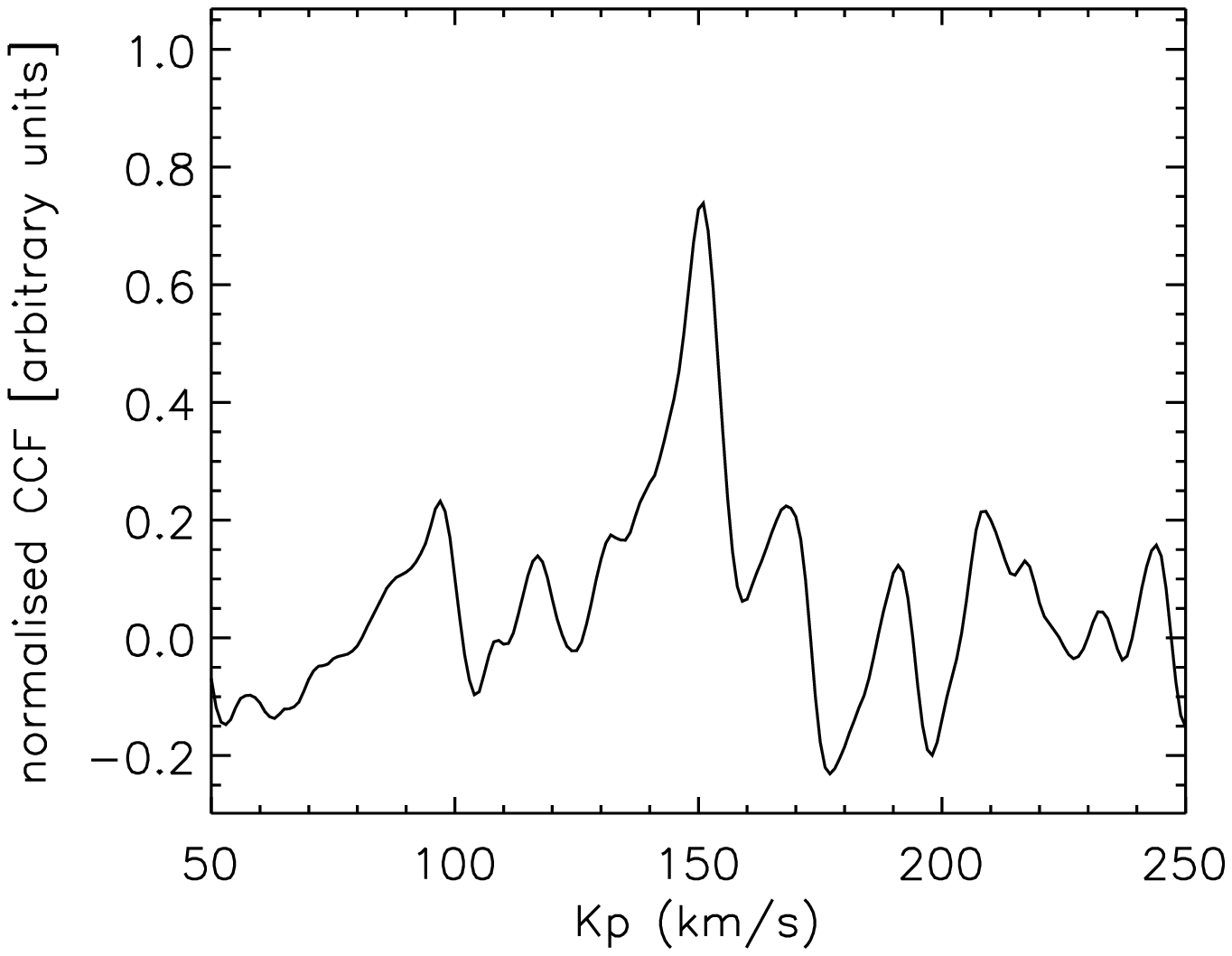} &
\includegraphics[width=5.5cm,keepaspectratio]{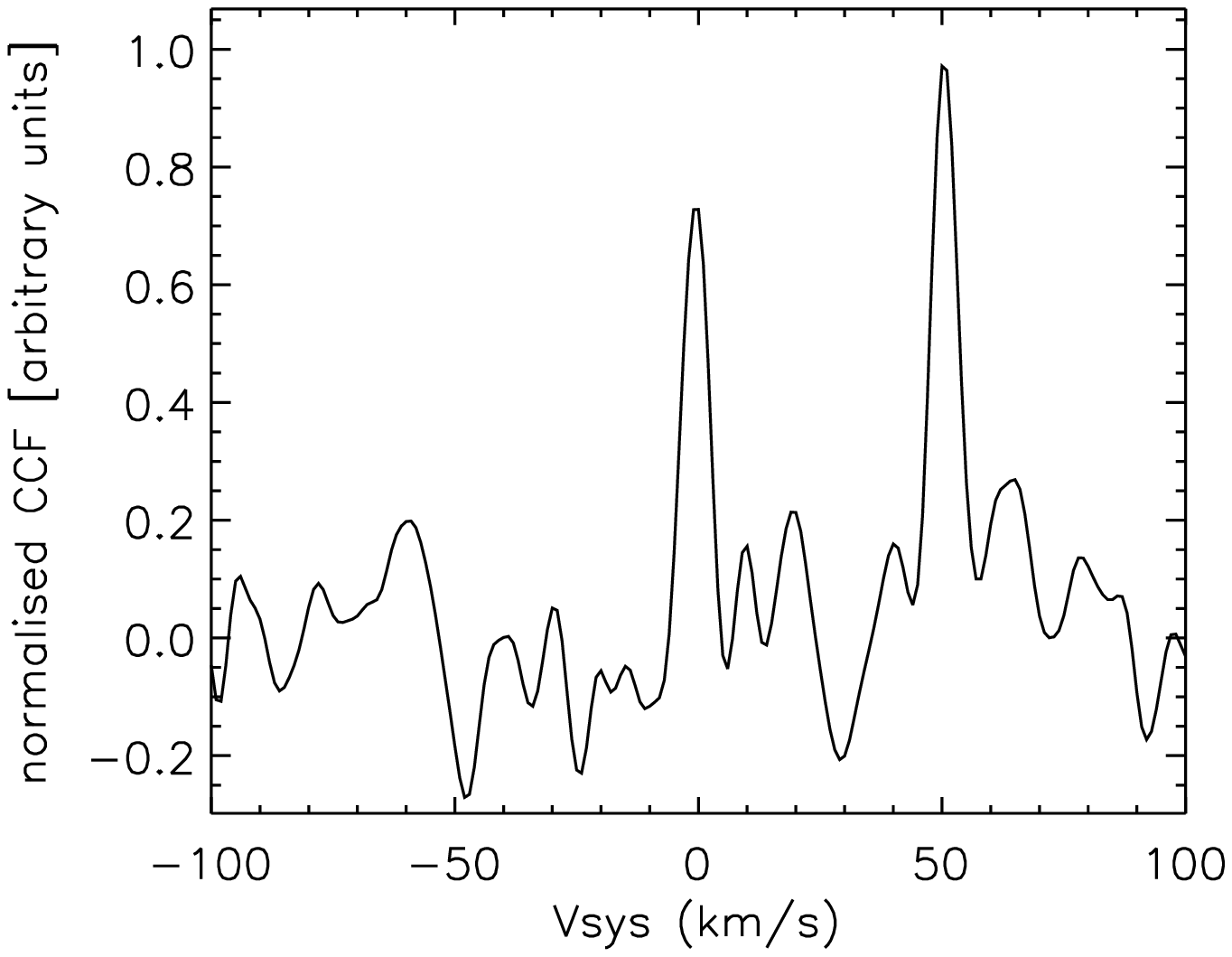} \\
Image A1 & Image A2 & Image A3
\vspace{0.4cm}\\
\includegraphics[width=5.5cm,keepaspectratio]{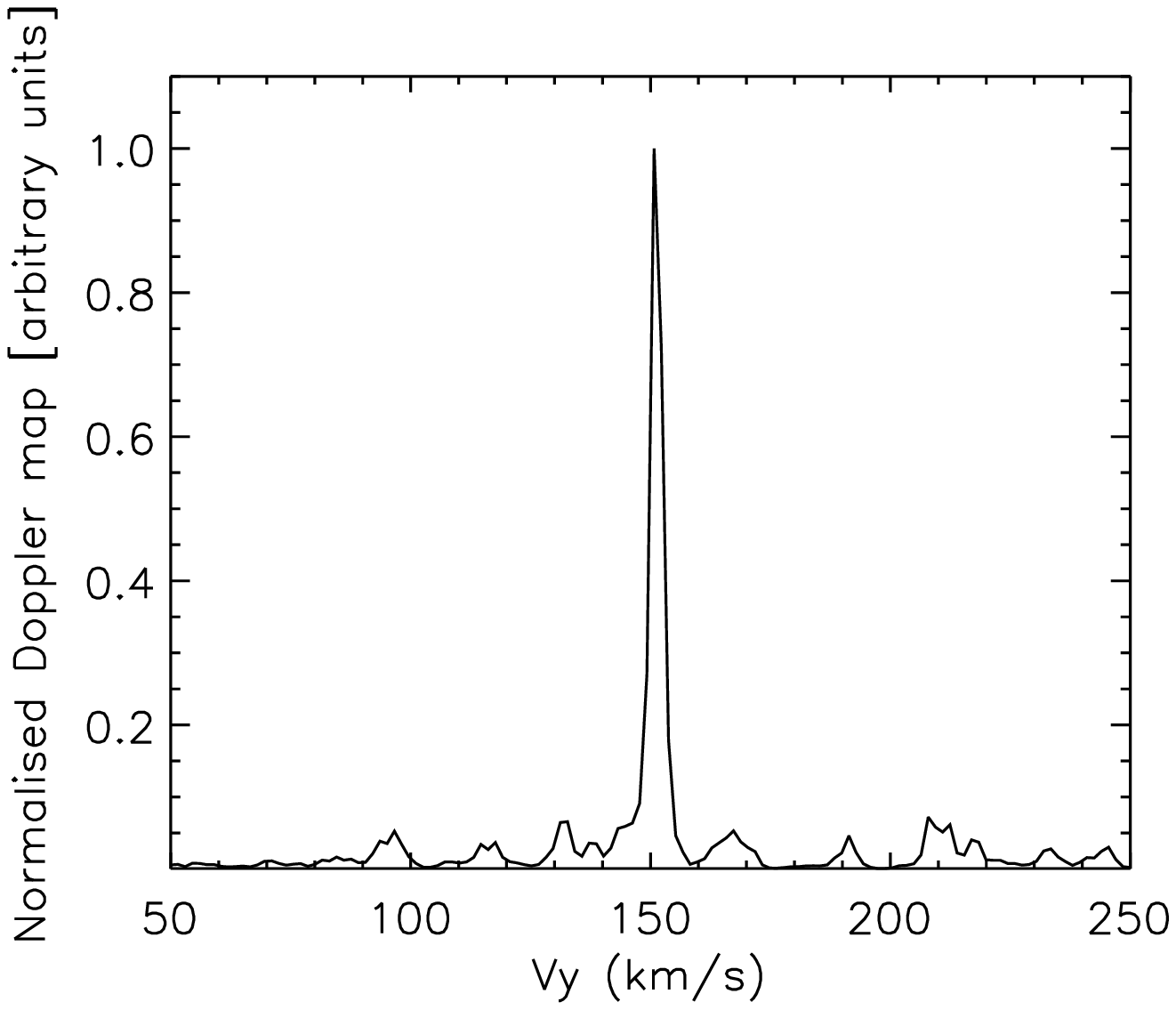} &
\includegraphics[width=5.5cm,keepaspectratio]{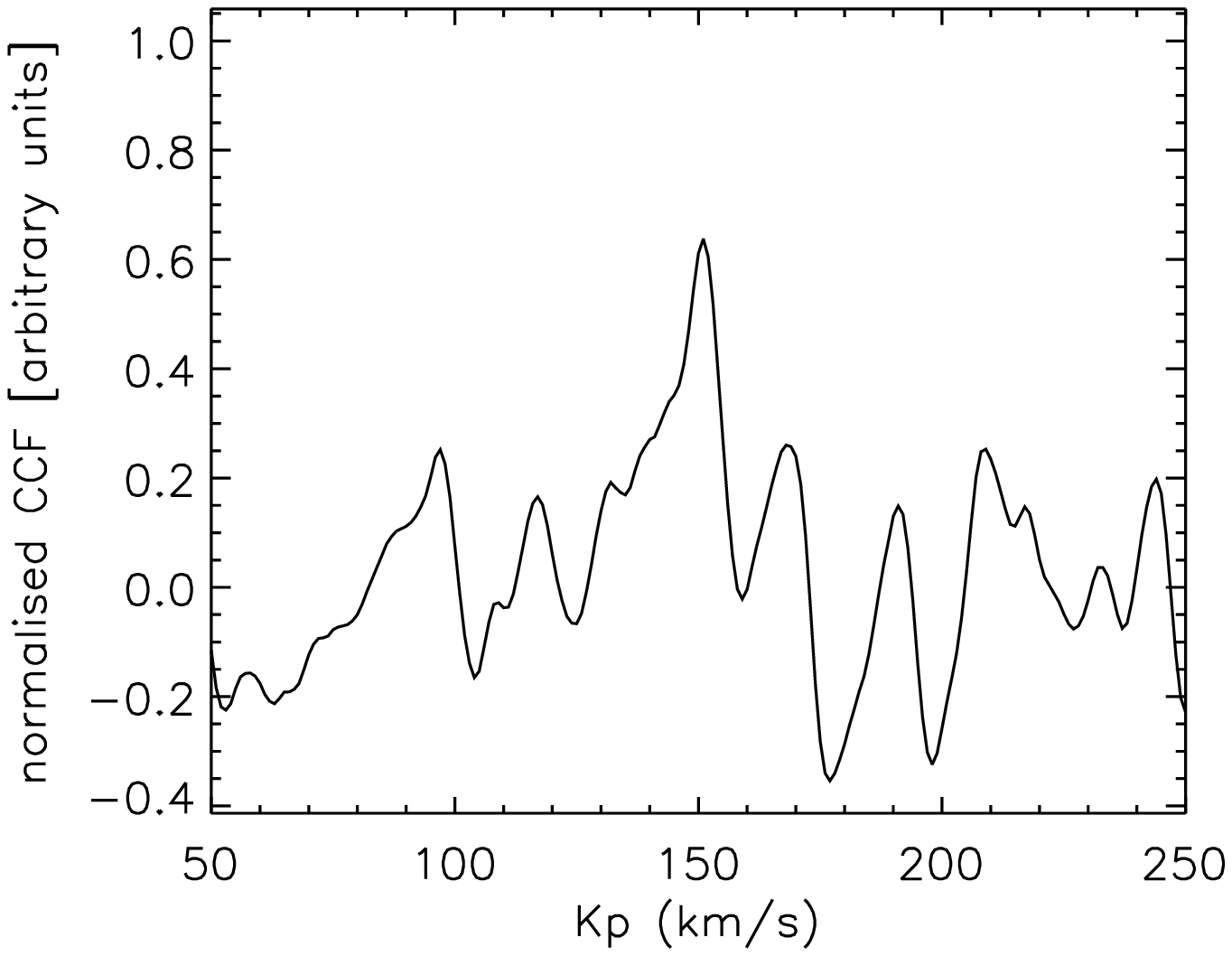} &
\includegraphics[width=5.5cm,keepaspectratio]{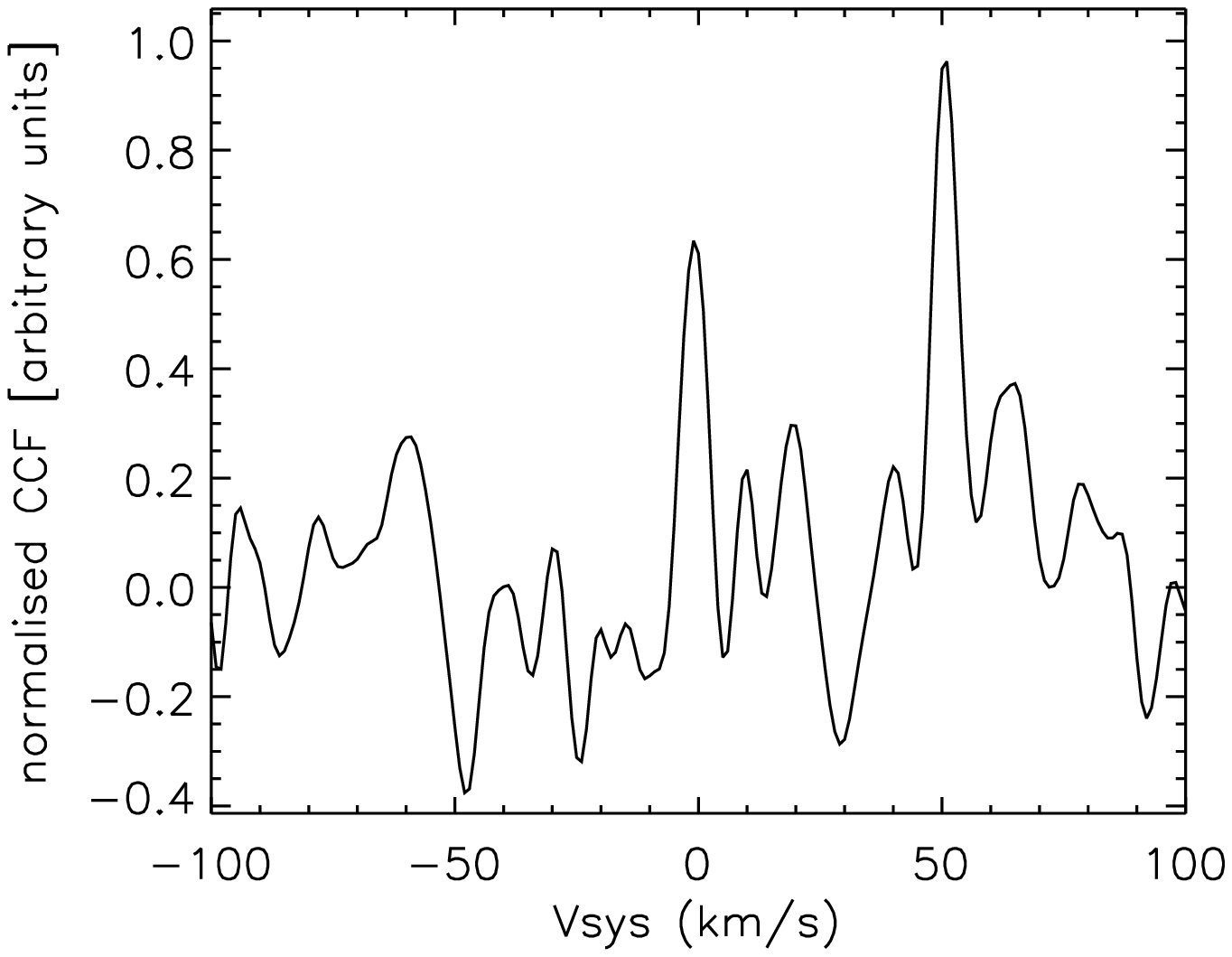} \\
Image B1 & Image B2 & Image B3
\vspace{0.4cm}\\
\includegraphics[width=5.5cm,keepaspectratio]{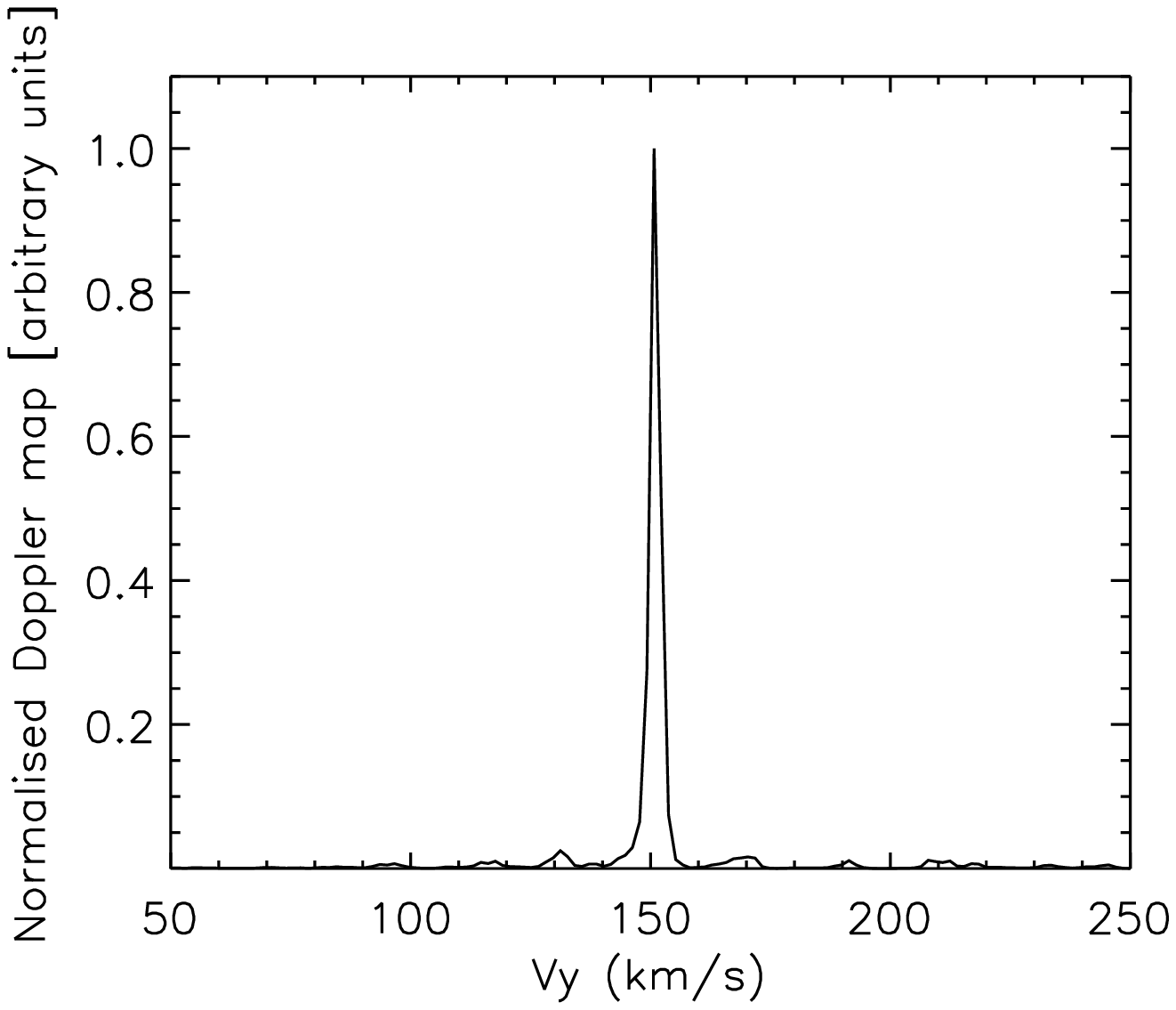} &
\includegraphics[width=5.5cm,keepaspectratio]{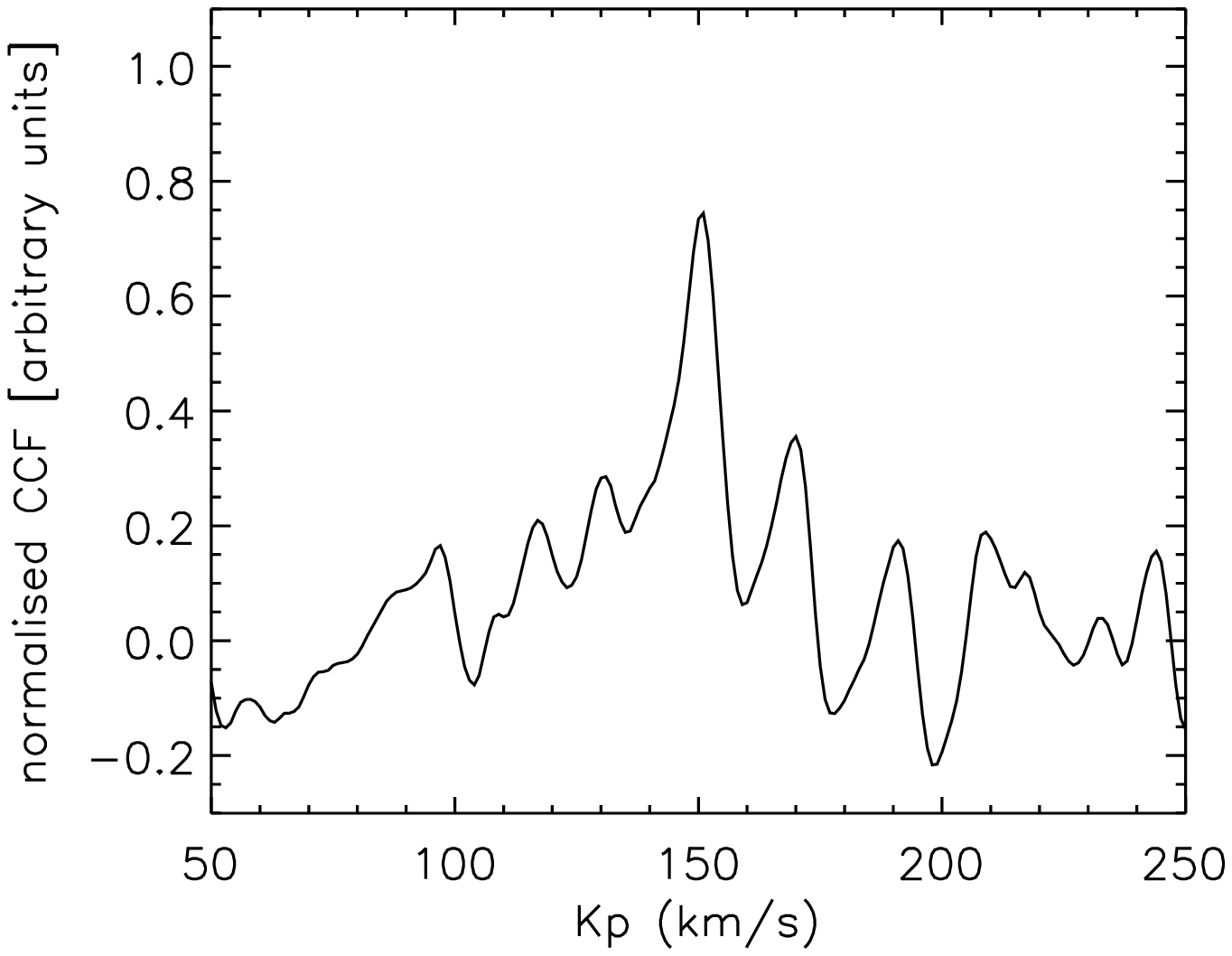} &
\includegraphics[width=5.5cm,keepaspectratio]{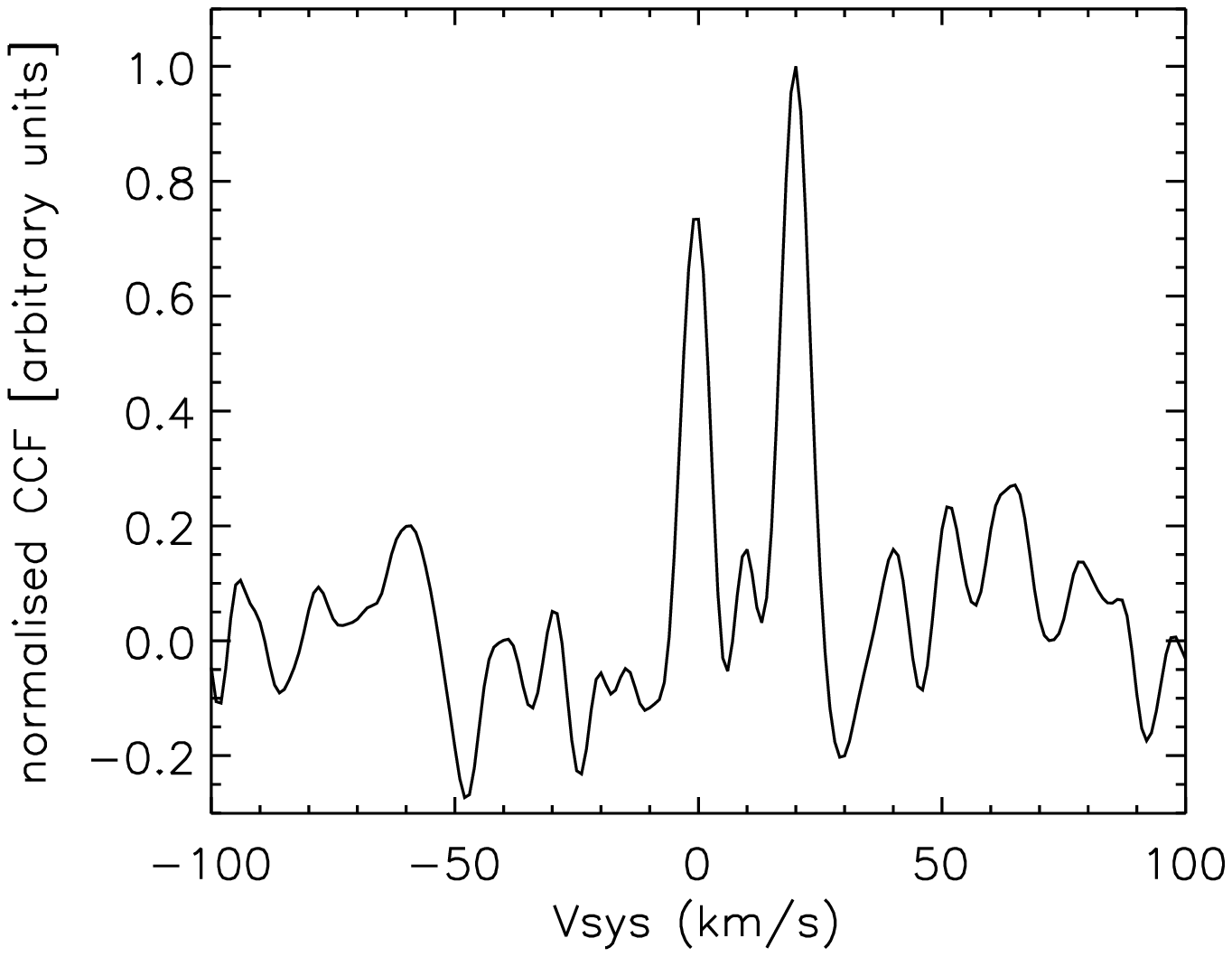} \\
Image C1 & Image C2 & Image C3
\vspace{0.4cm}\\
\includegraphics[width=5.5cm,keepaspectratio]{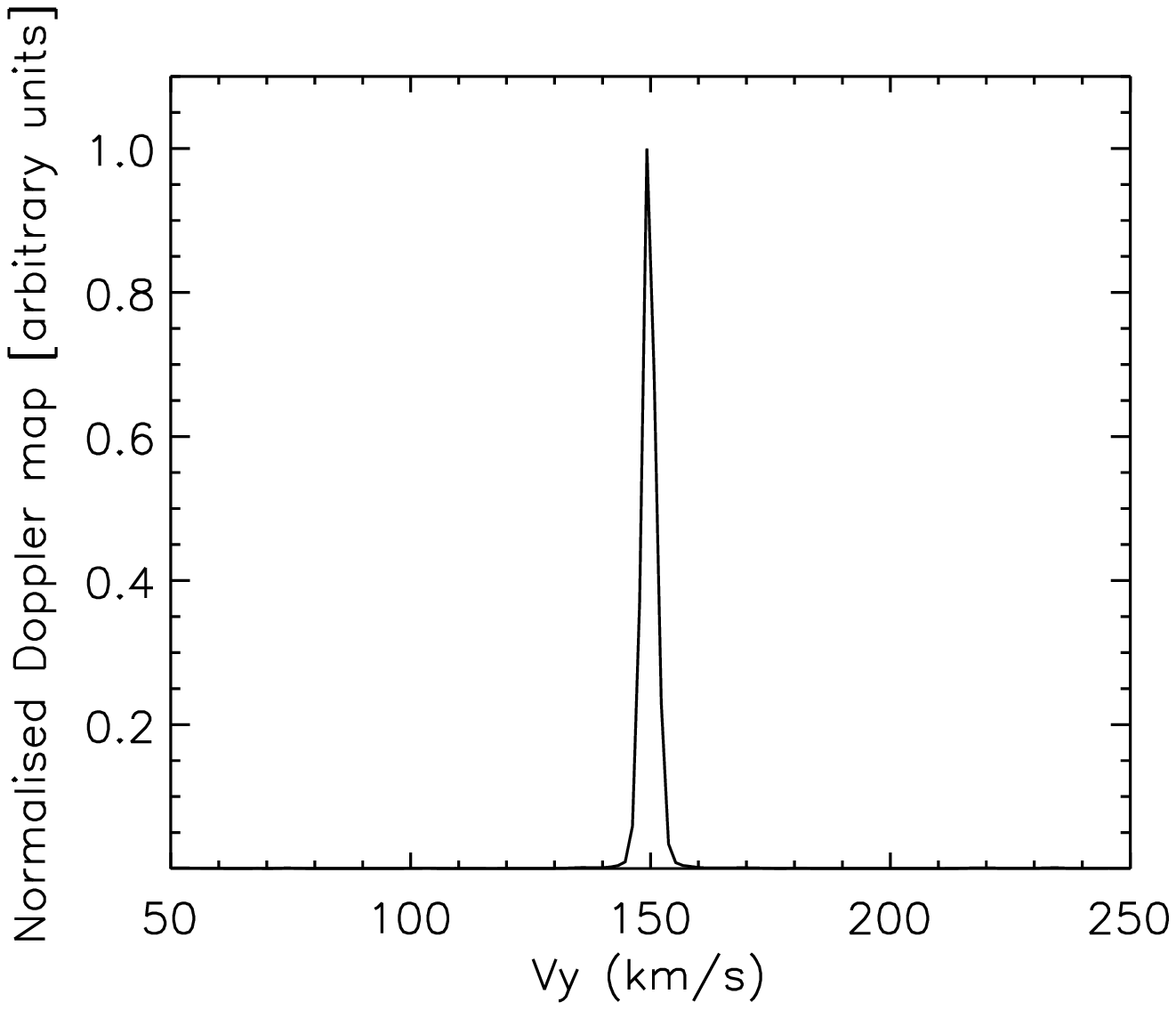} &
\includegraphics[width=5.5cm,keepaspectratio]{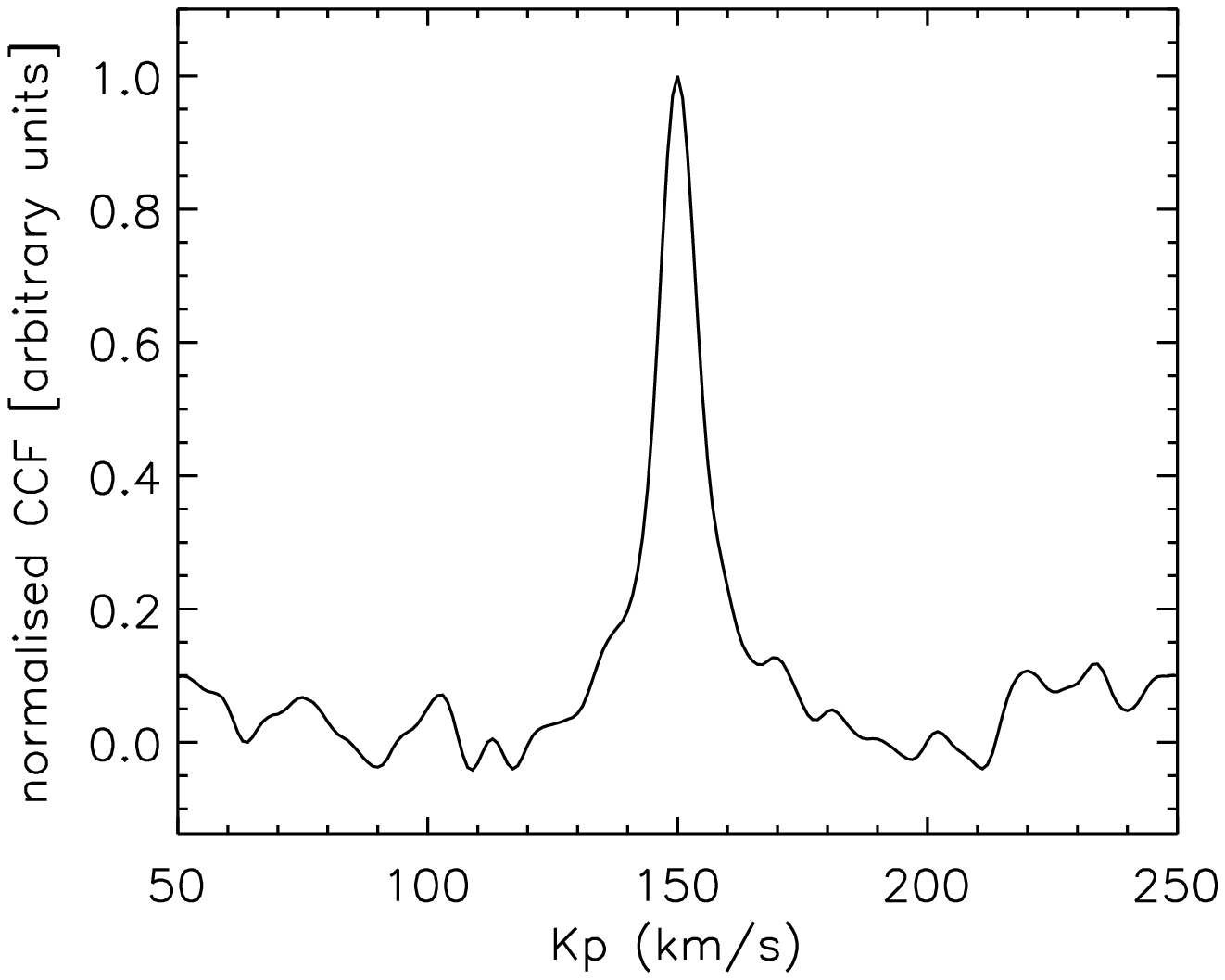} &
\includegraphics[width=5.5cm,keepaspectratio]{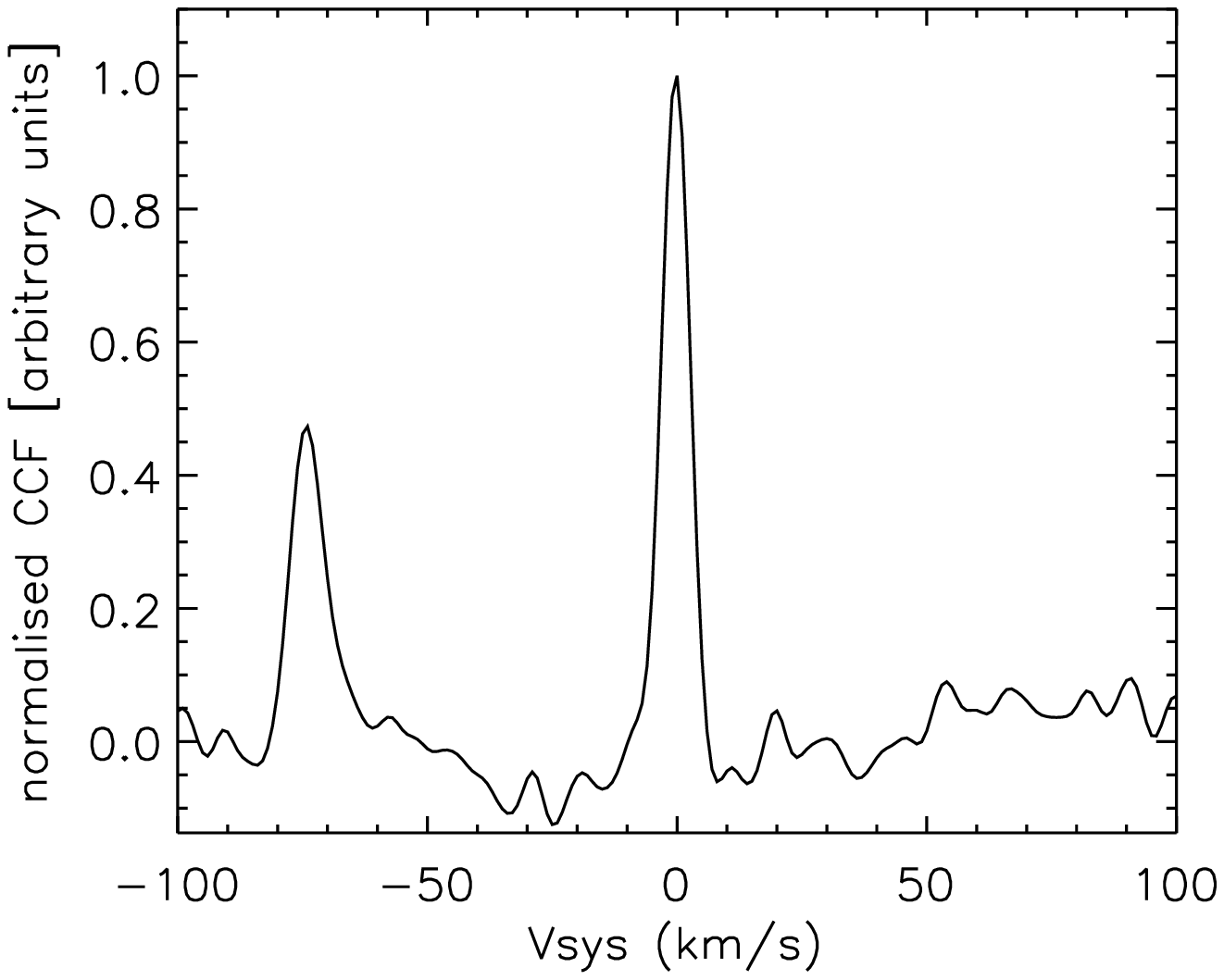} \\
Image D1 & Image D2 & Image D3
\vspace{0.4cm}\\
\end{tabular}
\caption{Cuts through the Doppler tomograms and CCF maps from Figure
  4. The left panels (column 1) show a cut through the Doppler tomograms at
  v$_x$=0 (where we expect the planet's signal if the phase is
  correct). For the CCFs we show cuts along both K$_P$ for
  v$_{\mathrm{sys}}$=0 (middle panel, column 2) and
  along v$_{\mathrm{sys}}$ at the injected planet velocity K$_P$=150 km
  s$^{-1}$ (right panels, column 3). The image identifier (A, B, C, etc.)
  correspond
  to those used in Figure~\ref{fig:simulations}. Note that cuts across Image G are
  not shown, as the planet signal is no longer at v$_x$=0 in the
  Doppler tomogram and there is significant spurious structure in the CCF map.}
\label{fig:simulations2}
\end{figure*}

Both CCF maps and Doppler tomograms of the combined $^{12}$C$^{16}$O and
$^{12}$C$^{17}$O synthetic spectra were reconstructed by first
applying the $^{12}$C$^{17}$O linelist {\em only} (Figure~\ref{fig:simulations}
Image D1 \&~D2), followed by applying the $^{12}$C$^{16}$O linelist {\em only}
(Figure~\ref{fig:simulations} Image E1 \&~E2). The respective cuts
through these images are presented in panels D and E in
Figure~\ref{fig:simulations2}.

While the planet signature is successfully recovered at the correct location
in both the Doppler tomograms and the CCF maps, the CCF maps show strong
evidence of a spurious signal at the correct K$_p$, but with a systemic
velocity of approximately -75 km s$^{-1}$ in Figure~\ref{fig:simulations}
Image D2 (also see Figure~\ref{fig:simulations2} Image D3) when the
$^{12}$C$^{17}$O linelist is used. A similar, albeit weaker, signal
is also seen at a systemic velocity of approximately +75 km s$^{-1}$ when
the $^{12}$C$^{16}$O linelist is applied (Figure~\ref{fig:simulations} Image E2
\&~Figure~\ref{fig:simulations2} Image E3).

This is caused by cross-talk in the CCF process between the linelist
being used, and the lines from the isotopologue present in the spectra that
are not accounted for in the linelist. Since the $^{12}$C$^{16}$O isotopologue
contains more lines, it shows up more prominently when the simulation is
carried out using the $^{12}$C$^{17}$O linelist (Figure~\ref{fig:simulations}
Image D2). While close isotopologues of the same molecular species tend to
appear (very loosely) as shifted versions of one other, the typical
wavelength shifts of the lines are of the order of 100's km s$^{-1}$. Thus,
it is not this overall `shift' between the 2 isotopologues that results
in the cross-talk. Rather, the cause is slightly more subtle.
For molecules such as CO, the line pattern tends to repeat fairly
regularly -- thus even if there is a large wavelength shift between
isotopologues, there is still a high possibility of a similar
pattern of lines appearing near the lines of interest, and it is this that
confuses the CCFs. It is clear from the Doppler tomograms, however, that there
is no such confusion -- as once more the erroneous lines are greatly
smeared out in velocity space and suppressed by the regularisation process
as outlined in Section~\ref{sec:lineerrors}.

Again, in Appendix~\ref{sec:appendix}, we present the back-projections of the CCFs in
  Figure~\ref{fig:appsims} (Image D). In this case, the CCF
  back-projection yields a semi-circular artefact similar to those seen
  in the single contaminating line case presented in
  Section~\ref{sec:lineerrors}. As in that case, the CCF sees the nearly
  repeating line-pattern again, whereas Doppler tomography is able to
  suppress this erroneous signal.

We provide another example of this, where once more we have analysed
the combined $^{12}$C$^{16}$O + $^{12}$C$^{17}$O synthetic spectra assuming
only the presence of the $^{12}$C$^{17}$O isotopologue. On this
occasion the noise level with respect to the $^{12}$C$^{17}$O lines has been
increased by a factor of 18, and the line strengths of the $^{12}$C$^{16}$O
isotopologue has been enhanced by a factor of 3. The correct planet
signature in the CCFs (Figures~\ref{fig:simulations} \&~\ref{fig:simulations2}
Image F2) is now becoming difficult to discern, while it is still
clearly picked up in the corresponding Doppler tomogram. We note that
the peak in the Doppler tomogram signal at the planet location
in Figure~\ref{fig:simulations2} Image F1 is at 0.47. In actual fact, the
planet signature is still the strongest peak in the Doppler tomogram,
but happens to land on a pixel adjacent to the correct location.
Again, there is no sign of confusion due to cross-talk in the Doppler maps.
The same cannot be said for the CCF map, which is dominated by the erroneous
power injected into the signal by the presence of an unacconted isotopologue.
The ability of Doppler tomography to cleanly separate multiple
molecular species/isotopologues is further examined in
Section~\ref{sec:isotop}.

\subsection{Phase offsets}
\label{sec:phaseoff}

The results of both CCF analyses and Doppler tomography are affected
in different ways by adopting incorrect values for the systemic
velocity, $\gamma$, and the orbital phasing. In the case of CCF analysis,
the systemic velocity is one of the parameters that are determined, and
therefore it is robust against errors in $\gamma$.
Doppler tomography is, in theory, sensitive to errors in the systemic
velocity. However, in practice, the systemic velocity of planetary systems
is usually known
to extremely high precision for planets studied with the radial velocity
technique. Even a reflex Doppler wobble of $\sim$100 m s$^{-1}$ induced
on the host star by a hot-Jupiter is insufficient to inject enough
uncertainty into $\gamma$ to have any discernable impact on the Doppler
tomogram. In addition, it can be easily measured directly from the
observed spectra in order to identify any instrumental velocity shifts
or calibration errors that may have arisen.

The orbital phasing, on the otherhand, may be uncertain in the
case of non-transiting planets where phase slippage over many orbits
can mount up over time to yield a substantial phase offset. In the case
of Doppler tomography, a phase offset ($\Delta\phi$) merely results in a
rotation of the planet signature by an angle of $\Delta\phi$ radians
from the origin, while otherwise completely preserving the signal
strength. This is simulated using a synthetic dataset comprising of
100 lines evenly spaced by 5.6$\AA$~starting at a wavelength of 22,900$\AA$.
In this case, an error in the phasing of the planetary orbit of
$\Delta\phi$ = +0.03 was injected, where the assumed phase
$\phi_{\mathrm{ass}}$ is related to the true phase
$\phi_{\mathrm{true}}$ by $\phi_{\mathrm{ass}} = \phi_{\mathrm{true}} + \Delta\phi$.

The resulting Doppler tomogram
is shown in Image G1 of Figure~\ref{fig:simulations}. As can be seen, the
planet signal is preserved but rotated about the origin by
$\Delta\phi \times 360^{\circ} = 10.8^{\circ}$. The value of
$\sqrt{\mathrm{v}_x^2 + \mathrm{v}_y^2}$ is also preserved, leading to its new
location in the velocity map of v$_x =$ 147 km s$^{-1}$,
v$_y$ = 28.1 km s$^{-1}$. Thus Doppler tomography
behaves in an entirely predictable way and preserves the
planetary signature and radial velocity amplitude of the planet -- and
can be used to determine the correct phase offset. The same simulated dataset
was then analysed using the CCF analysis, and is shown in Image G2
of Figure~\ref{fig:simulations}.
This marks an important difference between currently implemented CCF
techniques, which are sensitive to phase offsets (though see
Appendix~\ref{sec:appendix} for an alternative method of analysing CCFs)
that may not be well constrained in many non-transiting planetary systems
compared to Doppler tomography that
is more sensitive to errors in $\gamma$ (which is normally one of the best
known system parameters). While trial phase offsets can be searched using
the CCF technique, as Image G2 of Figure~\ref{fig:simulations} shows, even small
phase offsets can destroy a planetary signal, and thus may result in an
erroneous non-detection. This is particularly pertinent for the recovery
of very weak signals, and we note that all Doppler spectroscopy studies of
non-transiting planets with CRIRES have so far required phase-shift corrections
to be applied.

We note that carrying out a back-projection of the CCFs can
  largely mitigate against the impact of a phase offset (see
  Figure~\ref{fig:appsims}, Image G), resulting in the same rotation of the
  signal about the origin as seen in Doppler maps. However, both the
  complete elimination of any aliasing (ringing) from the repeating linelist
  pattern and the superior noise reduction in the Doppler map are clearly
  evident.

\subsection{Separation of species and isotopologues}
\label{sec:isotop}

The identification of different species and isotopologues
  in exoplanet atmospheres is an important goal. \cite{molliere19} have
  disussed isotopologue detection using the CCF technique in some detail,
  and approach the problem by removing the strongest planetary contributions
  in an iterative fashion. One of the advantages of Doppler tomography
over conventional CCF analyses
is that it can {\em simultaneously} fit multiple different linelists
of different species, and iteratively adjust the relative strengths
between the different species. To demonstrate this, the same synthetic
spectrum containing both $^{12}$C$^{16}$O and $^{12}$C$^{17}$O isotopologues
was analysed, but on this occasion the strength of the $^{12}$C$^{16}$O
lines was enhanced to be $\sim$100 times stronger that those of the
$^{12}$C$^{17}$O isotopologue. The signal-to-noise was adjusted such that 
the resultant signal for $^{12}$C$^{17}$O in each individual spectrum
approximately equaled the noise level (i.e. an approximately 1-$\sigma$
detection of $^{12}$C$^{17}$O in 1 spectrum).

Doppler tomography was run as before, except on this occasion it was
supplied linelists for both isotopologues to fit simultaneously. At
the start of the Doppler tomography reconstruction it was assumed that
the strength of the lines from both isotopologues was roughly equal, but
the relative weighting between the two linelists was allowed to vary.
The results of this simultaneous dual-linelist fitting is shown
in Figure~\ref{fig:simulations} Image H1 (for the $^{12}$C$^{16}$O map)
and Image I1 (for the $^{12}$C$^{17}$O map). As can be clearly seen,
Doppler tomography is able to cleanly disentangle the signals from
both isotopologue at the correct velocity, despite a contrast of 2 orders
of magnitude. Indeed, at this signal-to-noise there is no
discernable cross-talk between the Doppler tomograms. In addition, the peak
intensity in the two Doppler maps differ by a factor of 107 -- close to the
injected difference in weighting between the isotopologues. The robustness
of this process can also be seen in Figure~\ref{fig:simulations2}
(Image H1 and I1). By comparison, running the CCF using each linelist in
turn fails to detect the weaker $^{12}$C$^{17}$O isotopologue, and is
dominated by the contribution from $^{12}$C$^{16}$O (see
Figure~\ref{fig:simulations} Image H2 \&~I2). The same
  result is found if one back-projects the CCFs (see
  Figure~\ref{fig:appsims}, Images H \&~J).

\section{Conclusions}
\label{sec:conclusions}

In all of the simulations conducted, Doppler tomography has proven itself
to yield either comparable or better results than the standard CCF approach,
or the additional CCF back-projection approach presented
  in Appendix~\ref{sec:appendix}.
We have so far yet to create a scenario where Doppler tomography yields
poorer results, except in cases where the systemic velocity of the system
is poorly constrained (leading to a ringing pattern in the Doppler tomogram).
However, the systemic velocity is one of the best constrained parameters of
exoplanets with radial-velocity measurements, and can also be easily
measured with the same dataset used for the exoplanet atmosphere
detection itself. The technique is robust against orbital phase offsets
(and can be used to measure them),
can handle contaminating lines, and of particular interest is its capability
of simultaneously mapping different line species of unknown relative strengths.
With instruments such as CRIRES+ soon to come online at the time of writing,
this ability may prove immensely useful for measuring multiple
molecular species across its wider wavelength coverage. By doing so,
it may be possible to robustly determine
the relative abundances of the major C- and O-bearing molecules such as
H$_2$O, CO$_2$, and CH$_4$. This would be crucial for determining the
planetary C/O abundance ratio (\citealt{brogi14}), which has been
cited as having a potentially critical influence on the properties of (and hence
our understanding of) hot-hydrogen-dominated atmospheres
(e.g. \citealt{madhusudhan12}).

\section{Future Work}
\label{sec:future}

The aim of this paper was mainly to demonstrate some of
the advantages of applying Doppler tomography to the task of
detecting exoplanet atmospheres using time-series high-resolution
spectroscopy. There are, however, a number of extensions and further
considerations that should be be investigated, some of which are
outlined below.

One is to develop robust methods for determining planetary atmosphere detection
significancies with Doppler tomography. Possible routes for doing this include
the use of a bootstrapping approach in order to assess the impact of noise on
the Doppler tomograms (as implemented in other tomographical
image reconstructions, e.g. \citealt{watson01a}, and also in
  Doppler tomography analysis of binary systems by \citealt{wang17} and
  \citealt{wang18}), or phase scrambling. In order
to implement such methods, it may be necessary to reconstruct the Doppler
tomograms to a target entropy value, rather than to a target $\chi^2$ as
currently implemented. This comes about due to the fact that both approaches
will change the noise properties slightly or, in the latter option,
scrambling the phase order of the data would render any bona-fide signal
indiscernable by Doppler tomography. As a result, it would no longer be
possible to fit as closely to those data-points. Therefore, fitting to the
same $\chi^2$ would force the algorithm to fit more closely to noise in the
data, thereby altering the noise properties of the resulting Doppler tomogram.
While this would lead to a systematic underestimation of the significance
of any detected signals, in the worst case scenarios it may result in the
rejection of a real feature. Iterating to a constant entropy value would
largely preserve the noise properties of the Doppler map, thereby
helping to reveal subtle signatures. Adoption of a more sophisticated
default map or regularisation statistic may also deliver tangible gains
in sensitivity.

Finally, Doppler tomography can be adapted to account for any
phase variation in the changing day/night aspect of the planet.
Such Modulated Doppler Tomography codes have been applied to other
objects, and \cite{steeghs03} has demonstrated that such adaptations
delivers improved fits. This also leads to the prospect of constraining
planetary phase curves directly from time-series high-resolution spectroscopy.

\begin{figure*}
\begin{tabular}{cc}
\includegraphics[width=7.8cm,keepaspectratio]{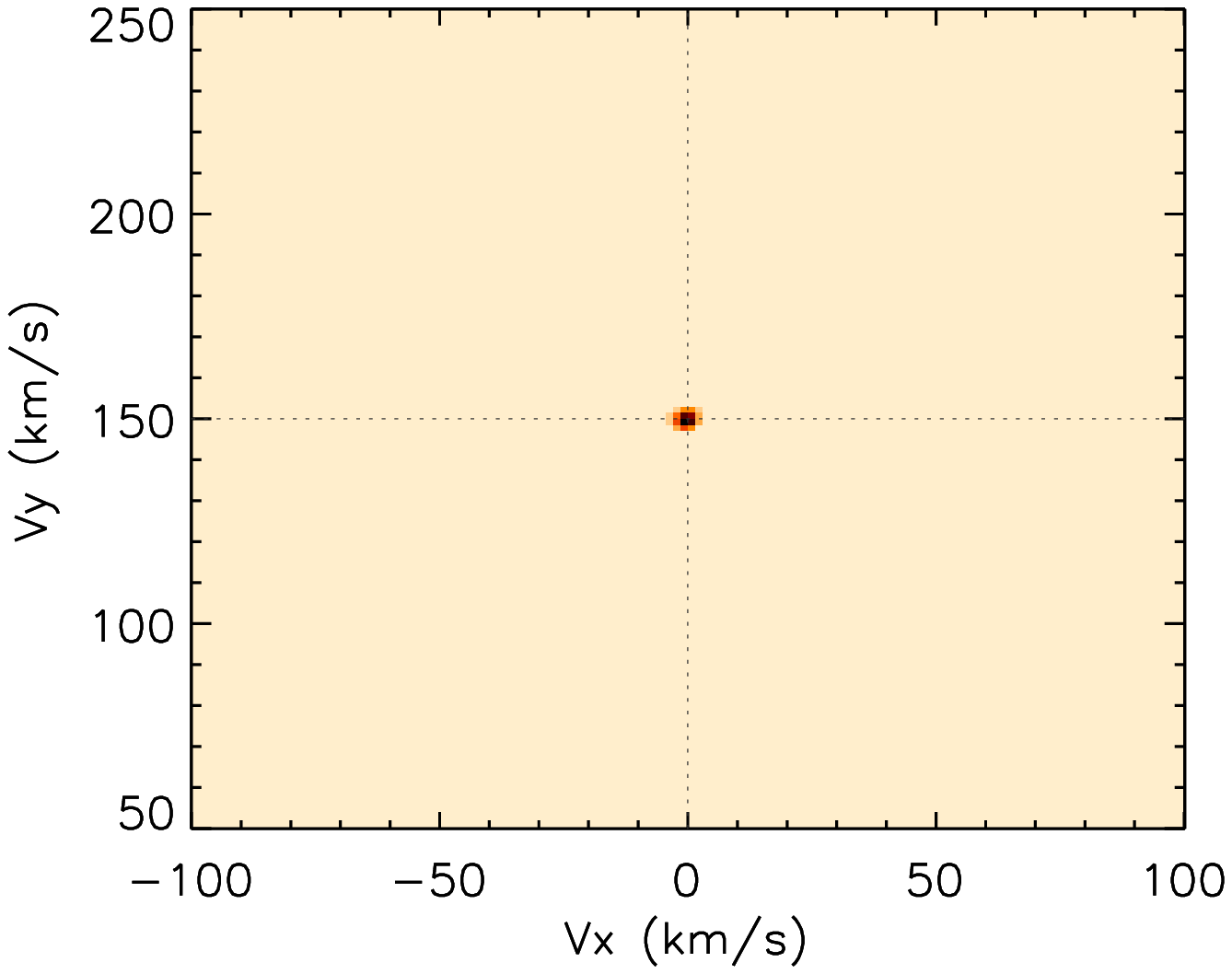} &
\includegraphics[width=7.8cm,keepaspectratio]{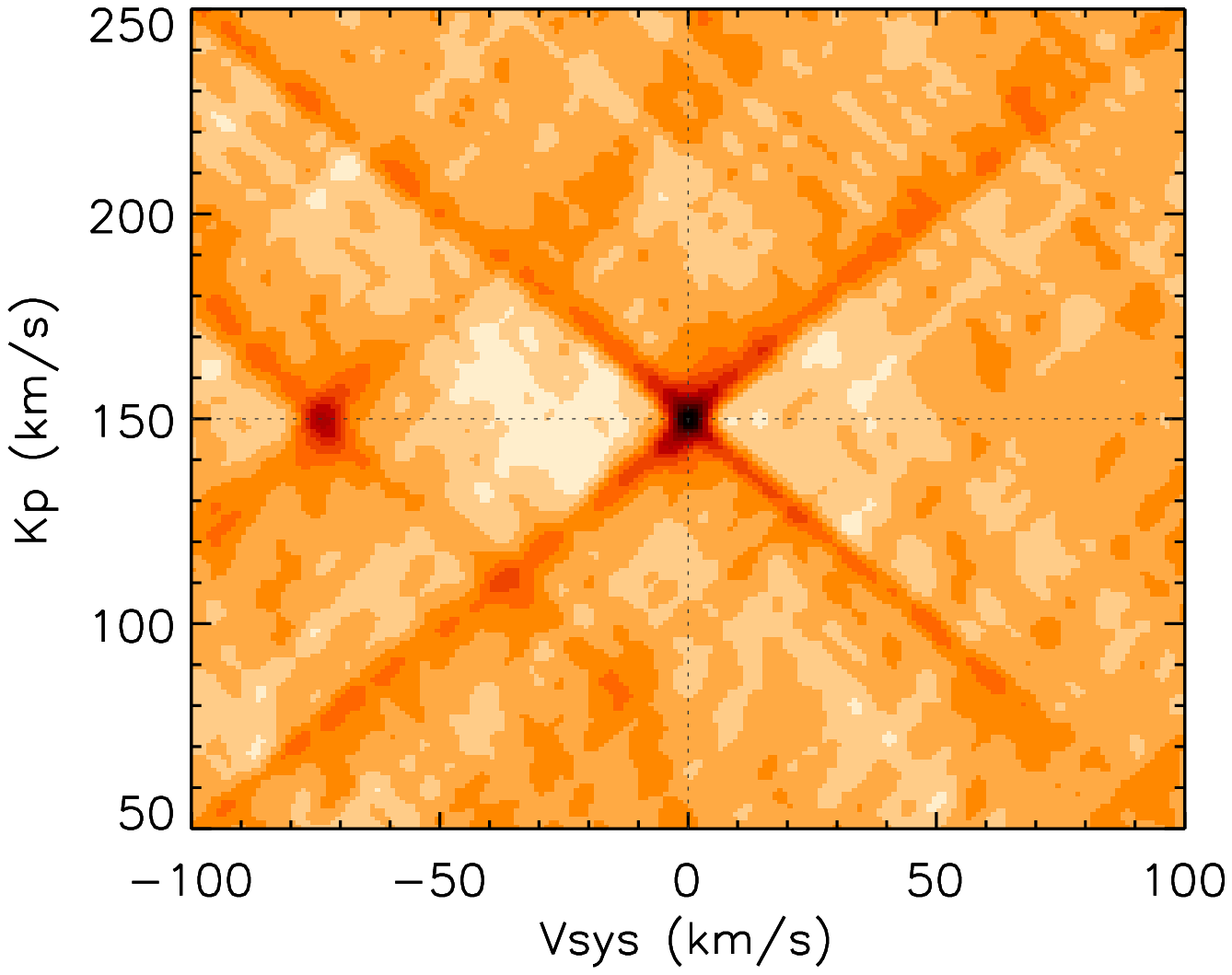} \\
Image D1: Doppler tomogram for $^{12}$C$^{17}$O &
Image D2: CCF for $^{12}$C$^{17}$O
\vspace{0.4cm}\\
\includegraphics[width=7.8cm,keepaspectratio]{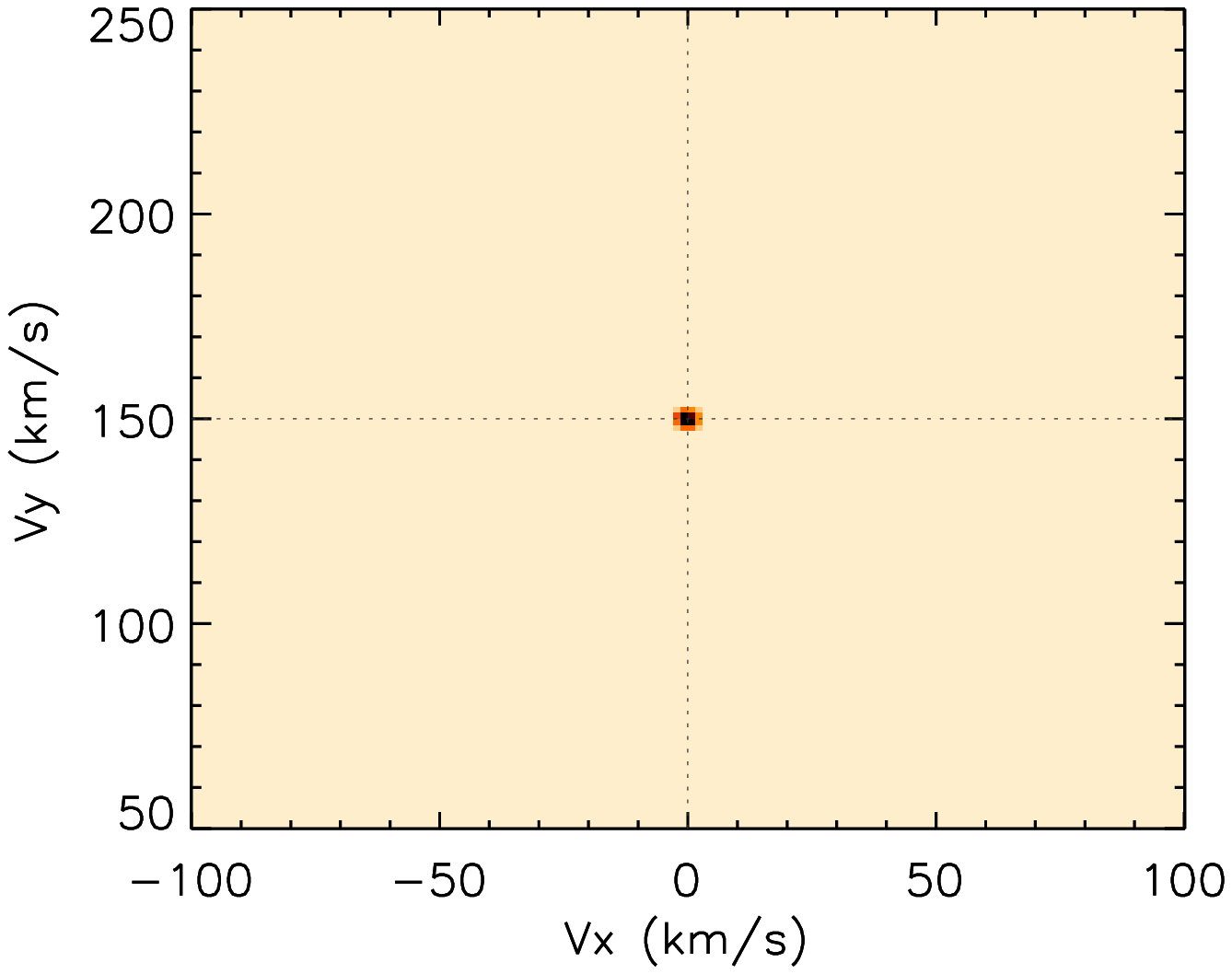} &
\includegraphics[width=7.8cm,keepaspectratio]{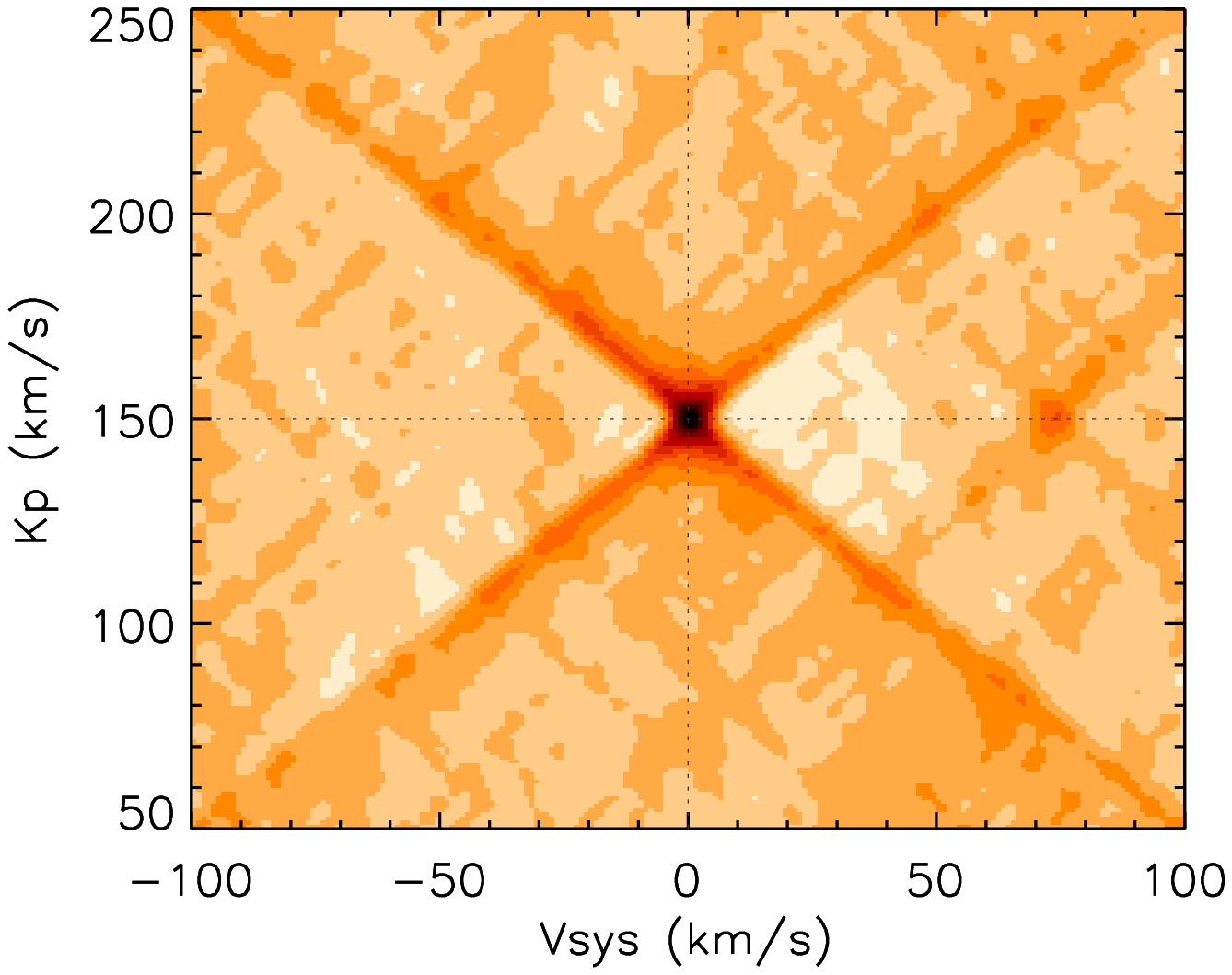} \\
Image E1: Doppler tomogram for $^{12}$C$^{16}$O &
Image E2: CCF for $^{12}$C$^{16}$O
\vspace{0.4cm}\\
\includegraphics[width=7.8cm,keepaspectratio]{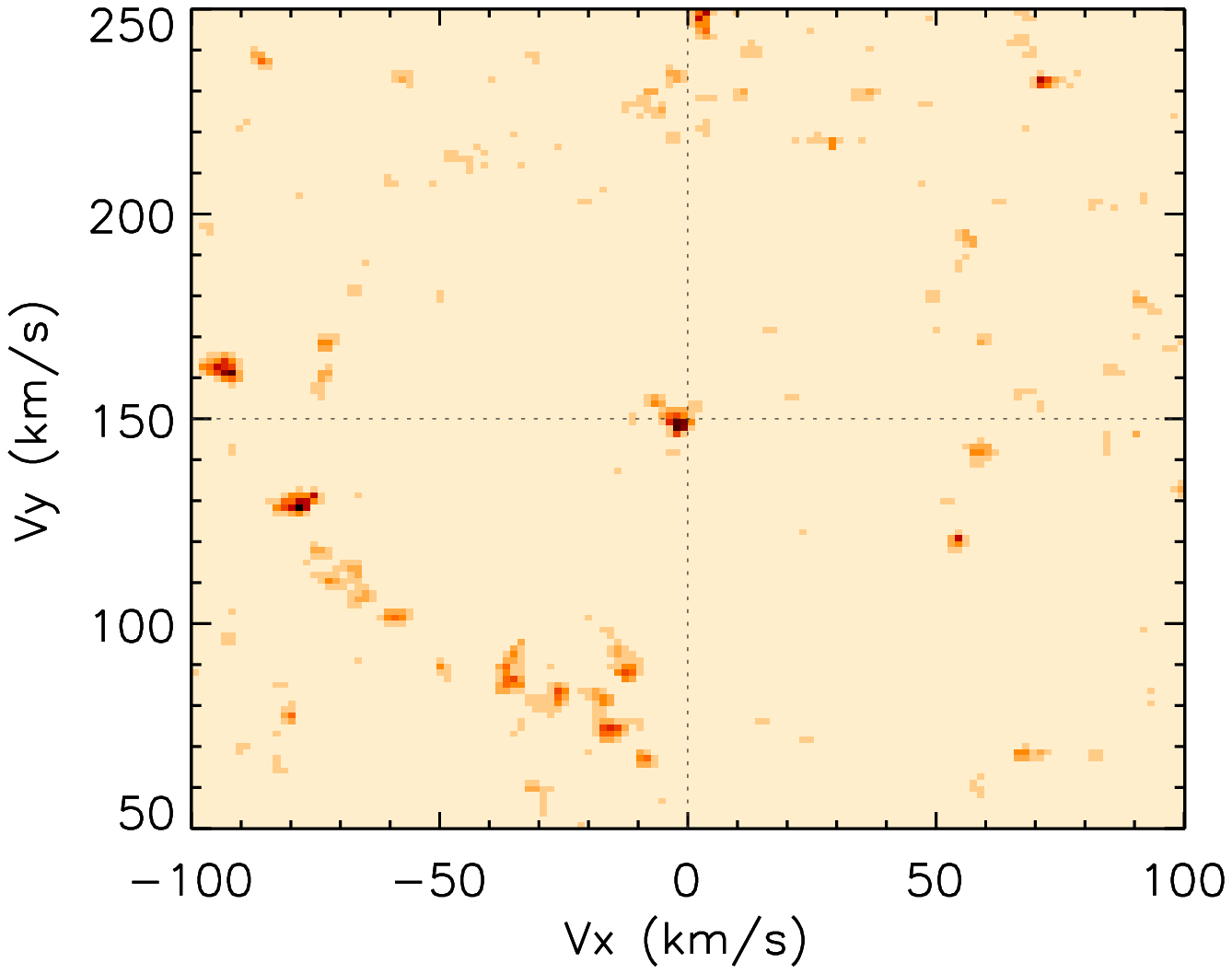} &
\includegraphics[width=7.8cm,keepaspectratio]{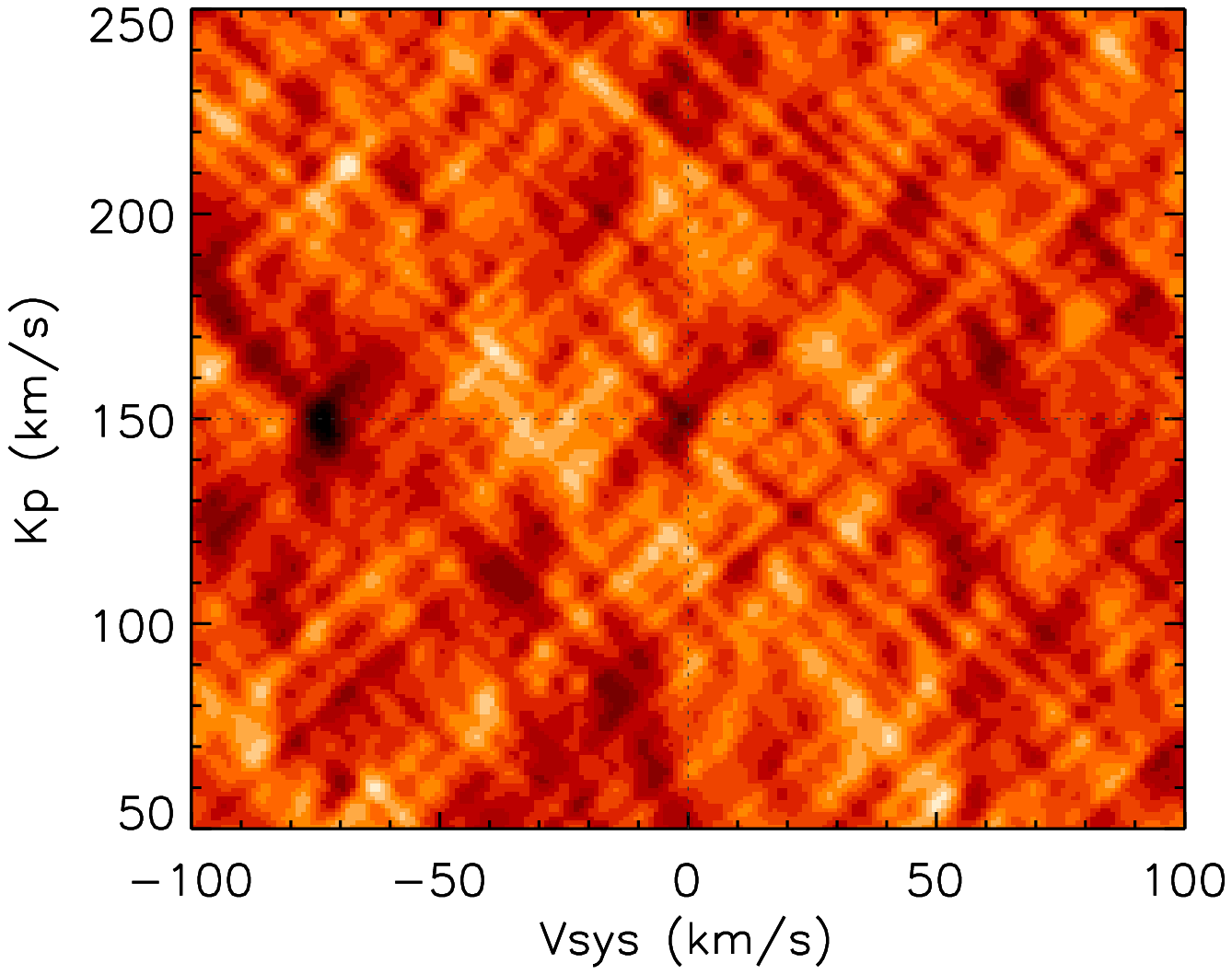} \\
Image F1: Doppler tomogram for $^{12}$C$^{17}$O &
Image F2: CCF for $^{12}$C$^{17}$O
\vspace{0.4cm}\\
\end{tabular}
\parbox{17.6cm}{{\bf Figure~\ref{fig:simulations}.} -- continued. Images D and E show the
  impact of contaminating lines when targeting specific isotopologues
  of CO. The data for both maps contain $^{12}$C$^{16}$O and
  $^{12}$C$^{17}$O at equal strength. In Image D the data is analysed
  using {\it only} the $^{12}$C$^{17}$O linelist, while in Image E
  only a $^{12}$C$^{16}$O linelist is used. Image F show a simulation
  containing both $^{12}$C$^{16}$O and $^{12}$C$^{17}$O, but now the
  SNR is reduced significantly, and the abundance of $^{12}$C$^{16}$O
  is enhanced by a factor of 3. The results here show an analysis
  targeting only the weaker $^{12}$C$^{17}$O.}
\end{figure*}

\begin{figure*}
\begin{tabular}{cc}
\includegraphics[width=7.8cm,keepaspectratio]{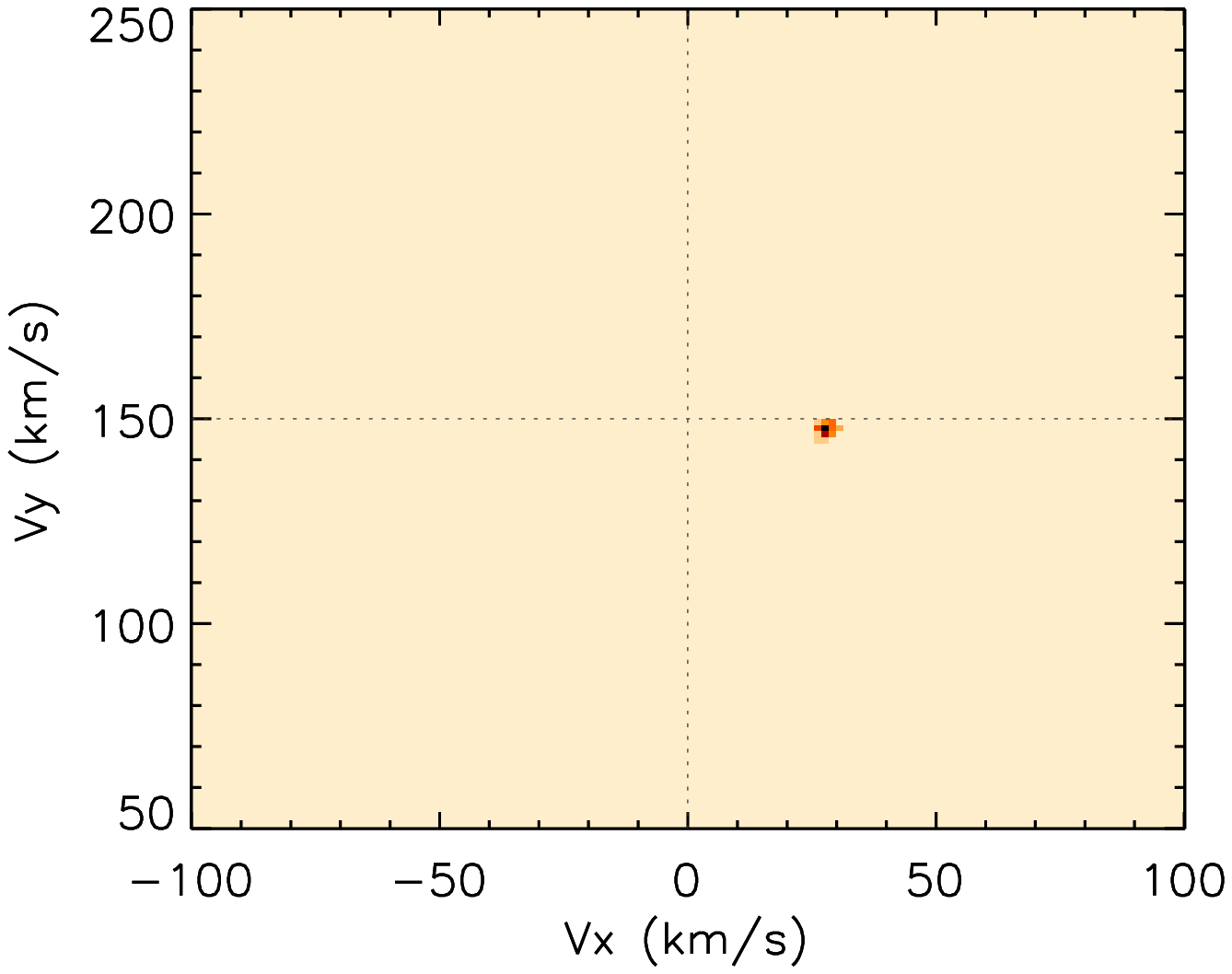} &
\includegraphics[width=7.8cm,keepaspectratio]{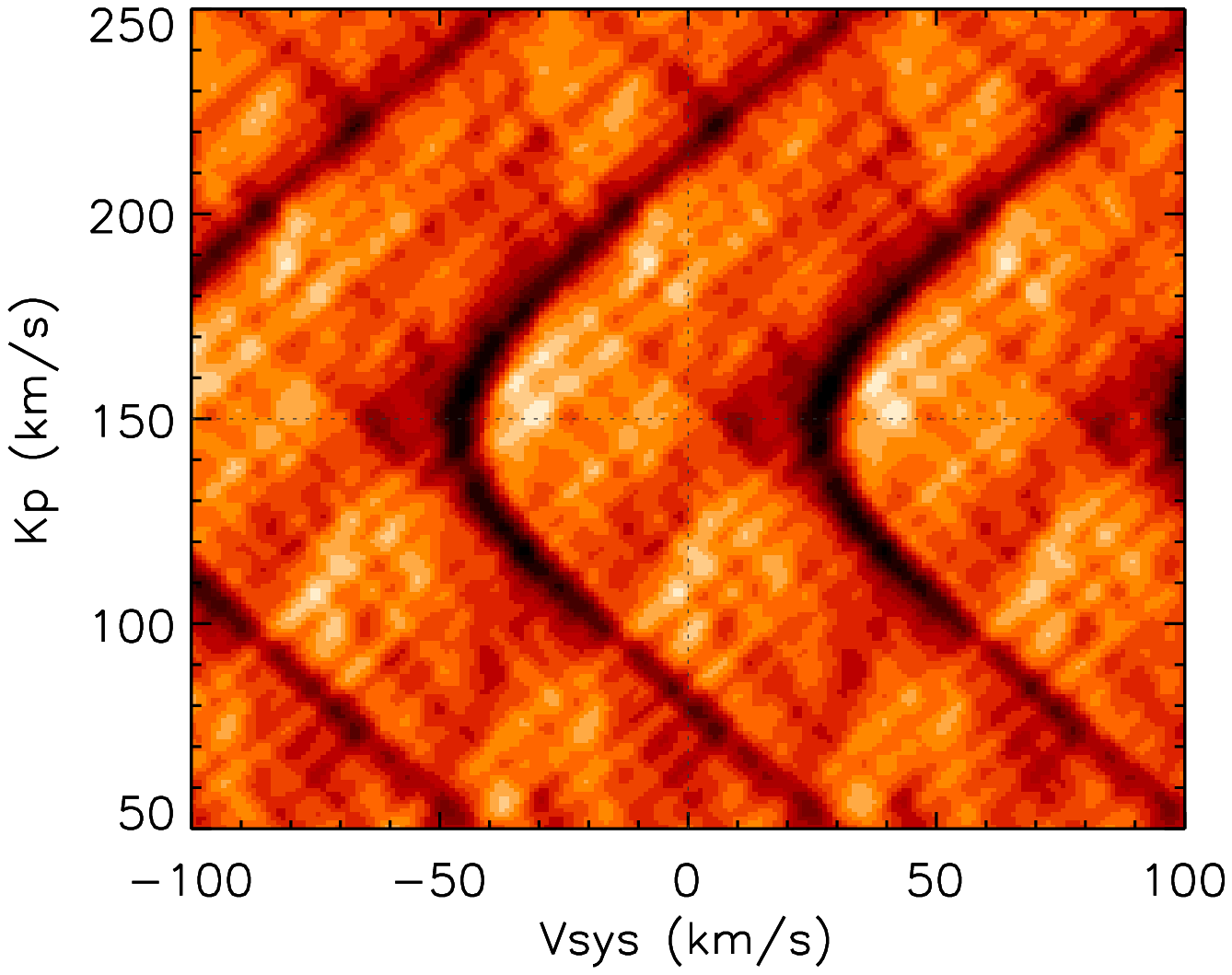} \\
Image G1: Doppler tomogram for data with $\Delta\phi$=+0.03 &
Image G2: CCF for data with $\Delta\phi$=+0.03
\vspace{0.4cm}\\
\includegraphics[width=7.8cm,keepaspectratio]{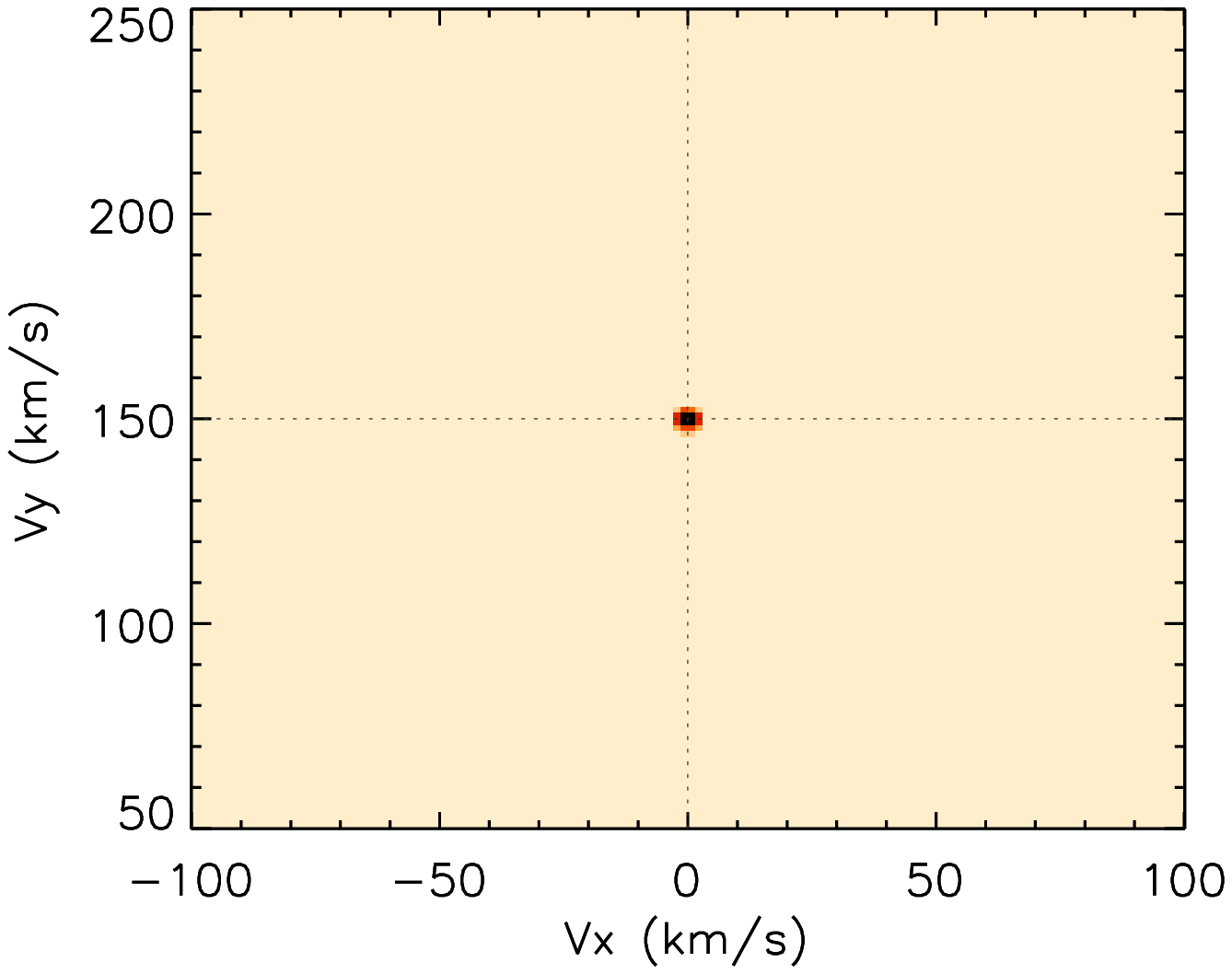} &
\includegraphics[width=7.8cm,keepaspectratio]{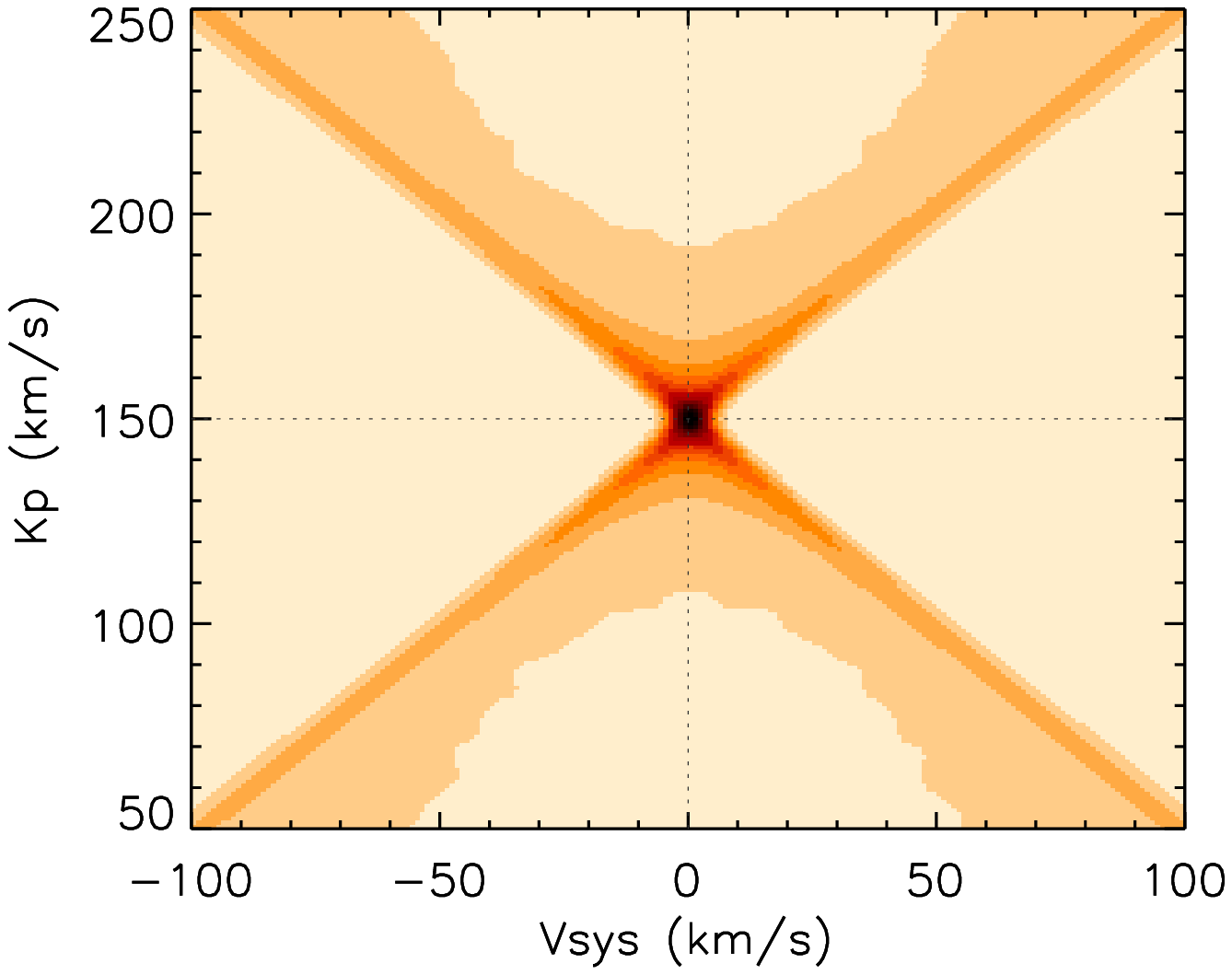} \\
Image H1: Isotopologue test $^{12}$C$^{16}$O Doppler map &
Image H2: Isotopologue test $^{12}$C$^{16}$O CCF
\vspace{0.4cm}\\
\includegraphics[width=7.8cm,keepaspectratio]{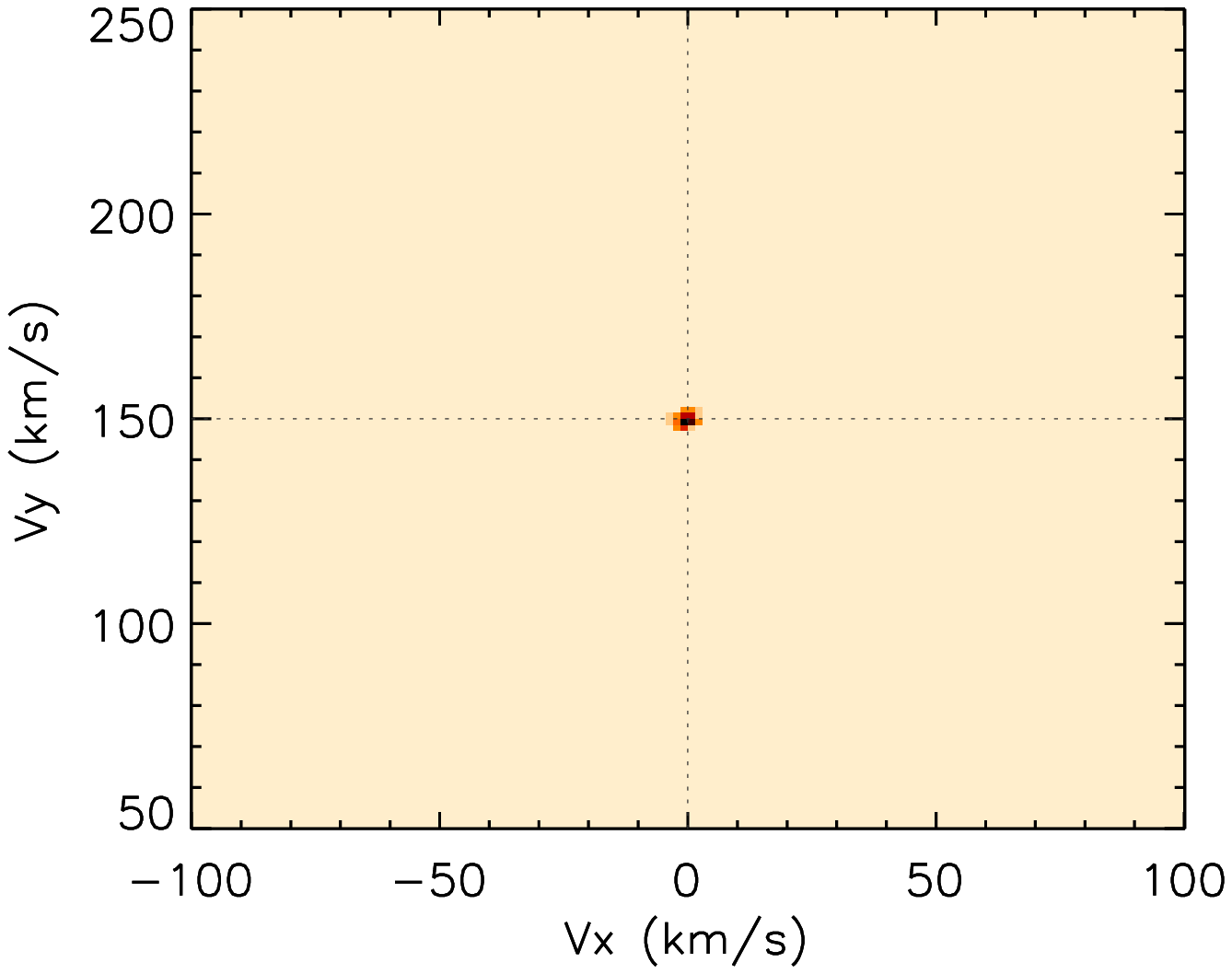} &
\includegraphics[width=7.8cm,keepaspectratio]{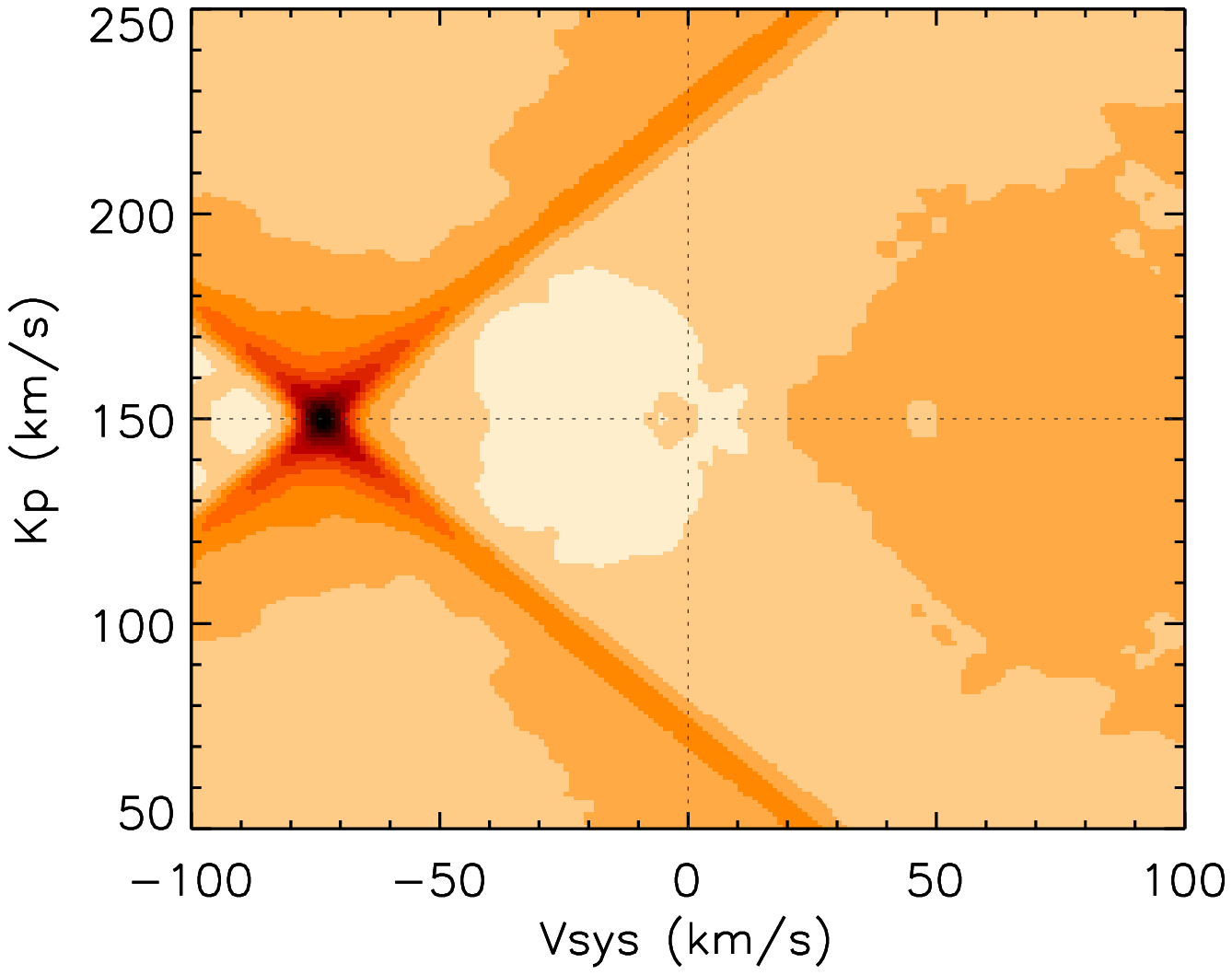} \\
Image I1: Isotopologue test $^{12}$C$^{17}$O Doppler map &
Image I2: Isotopologue test $^{12}$C$^{17}$O CCF
\vspace{0.4cm}\\
\end{tabular}
\parbox{17.6cm}{{\bf Figure~\ref{fig:simulations}.} -- continued. Image G shows the impact of
  a spurious phase-offset of +0.03 from the actual orbital phase. Images H
  and I show a more extreme case of Images D \& E, where the strength
  of $^{12}$C$^{16}$O is enhanced by a factor of 100 compared to
  $^{12}$C$^{17}$O. In the case of the Doppler tomogram maps, we
  search for $^{12}$C$^{16}$O and $^{12}$C$^{17}$O simultaneously,
  while for the CCFs we target each molecule separately. Image H show
  the results for $^{12}$C$^{16}$O and Image I for
  $^{12}$C$^{17}$O. As can be seen Doppler tomography recovers both
  species, while the CCF only clearly detects $^{12}$C$^{16}$O.
}  
\end{figure*}

\begin{figure*}
\begin{tabular}{ccc}
\includegraphics[width=5.5cm,keepaspectratio]{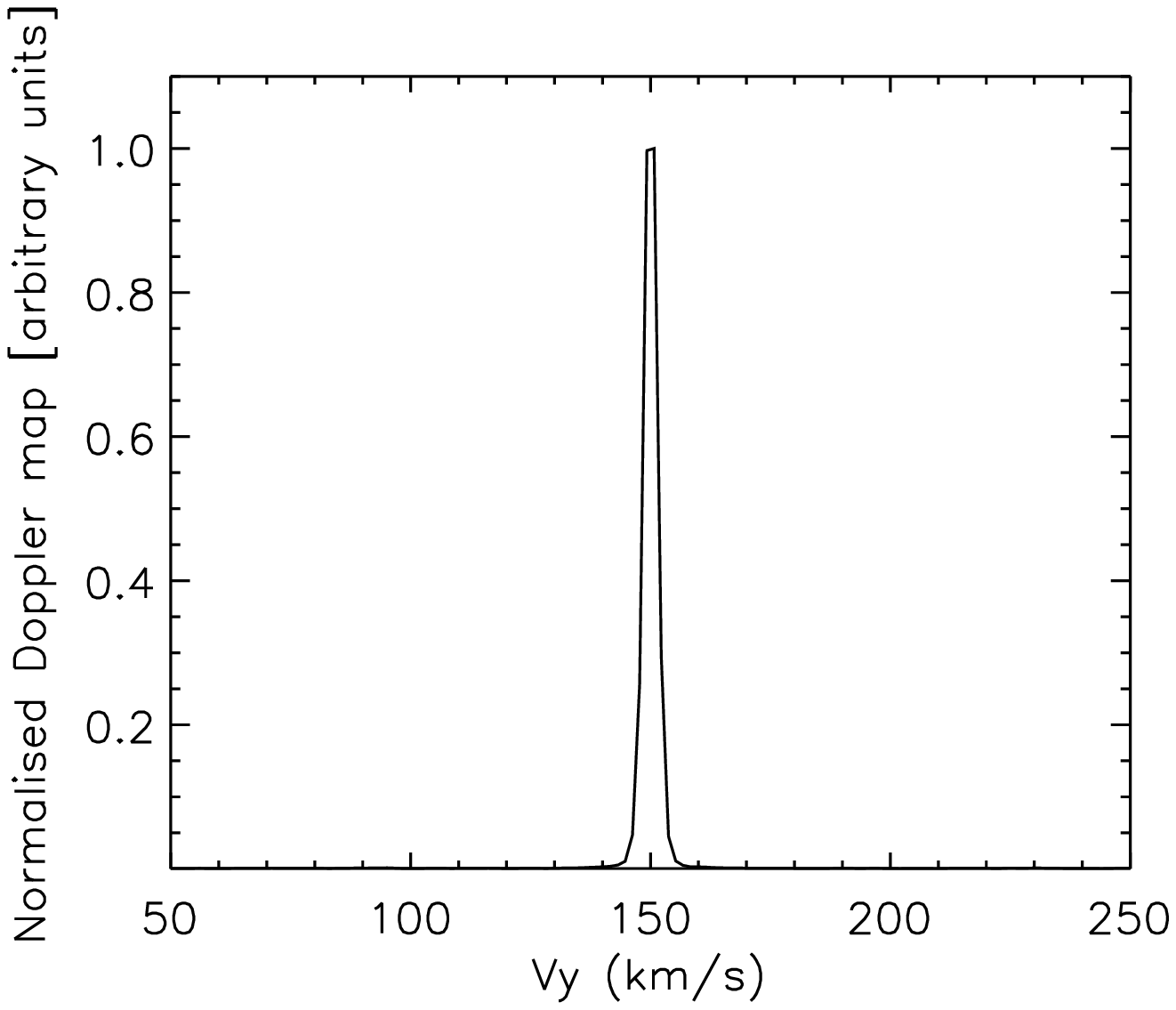} &
\includegraphics[width=5.5cm,keepaspectratio]{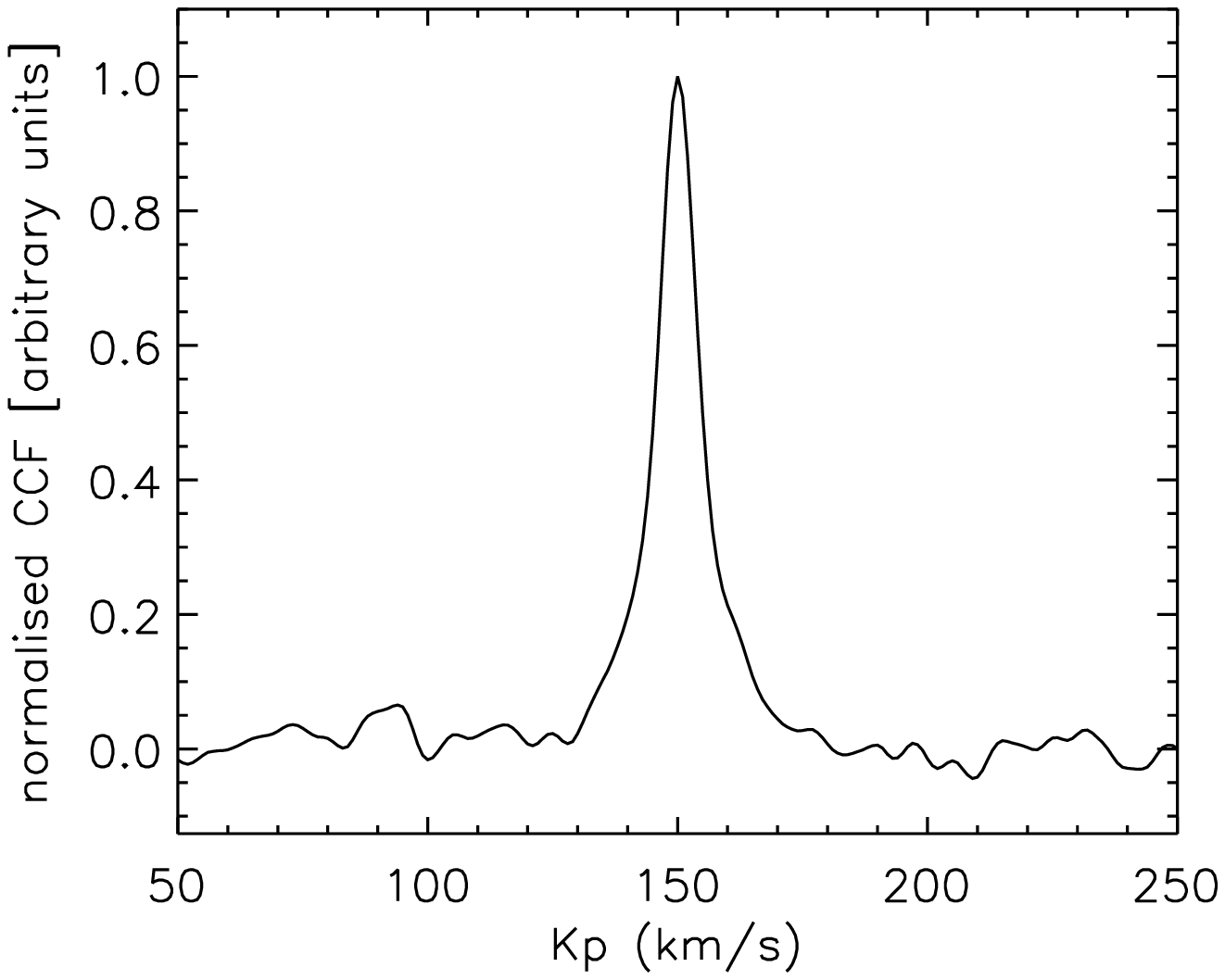} &
\includegraphics[width=5.5cm,keepaspectratio]{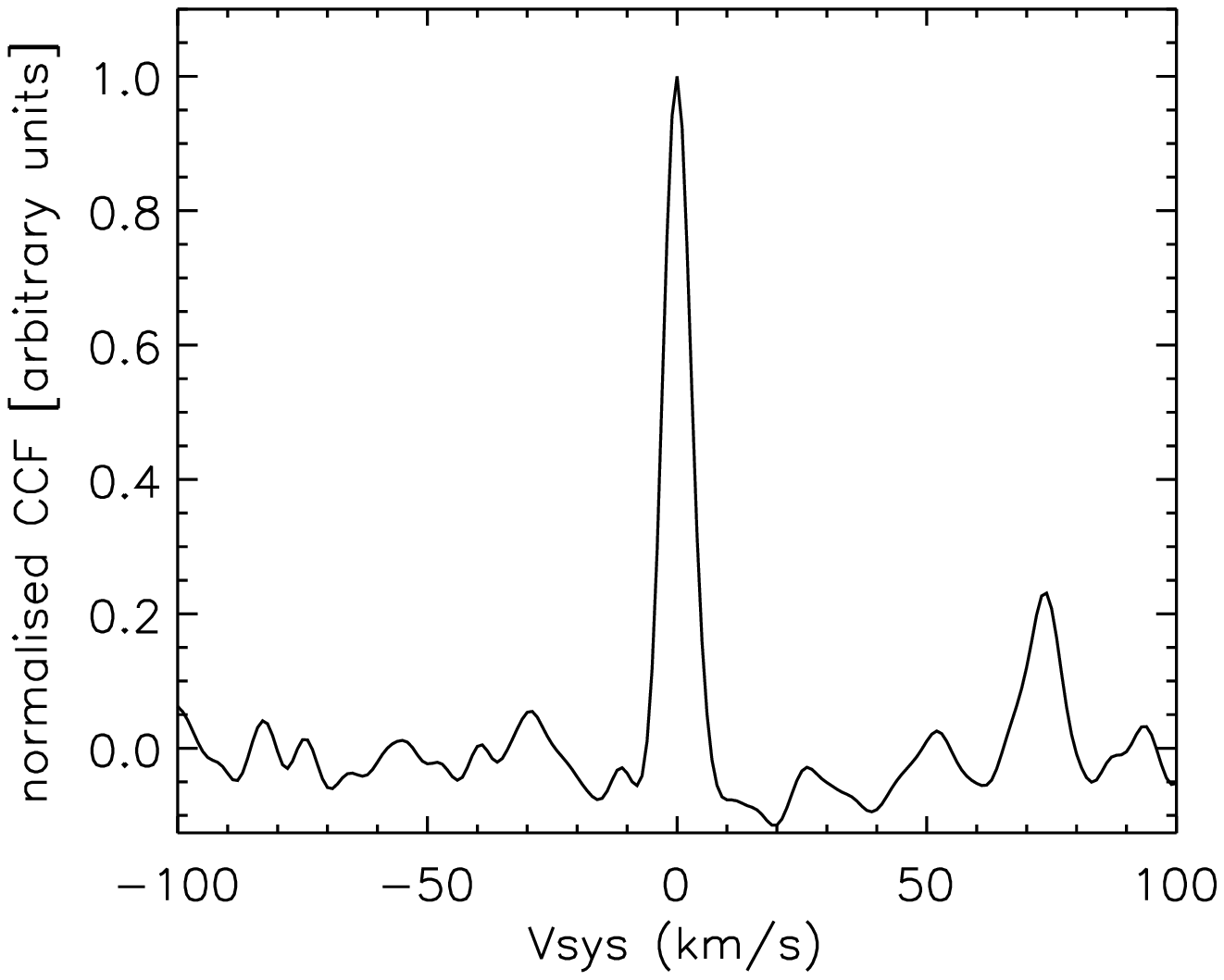} \\
Image E1 & Image E2 & Image E3
\vspace{0.4cm}\\
\includegraphics[width=5.5cm,keepaspectratio]{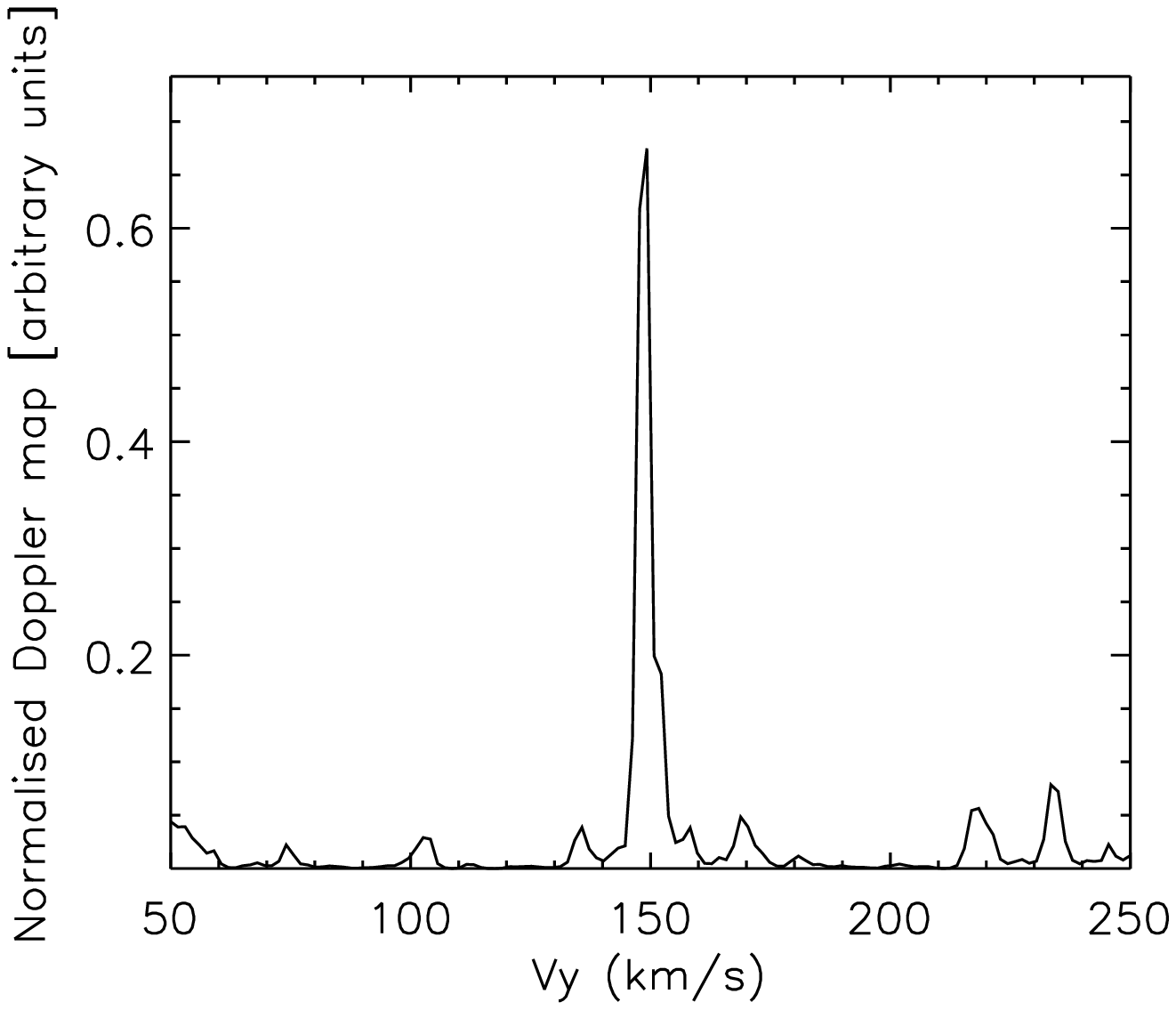} &
\includegraphics[width=5.5cm,keepaspectratio]{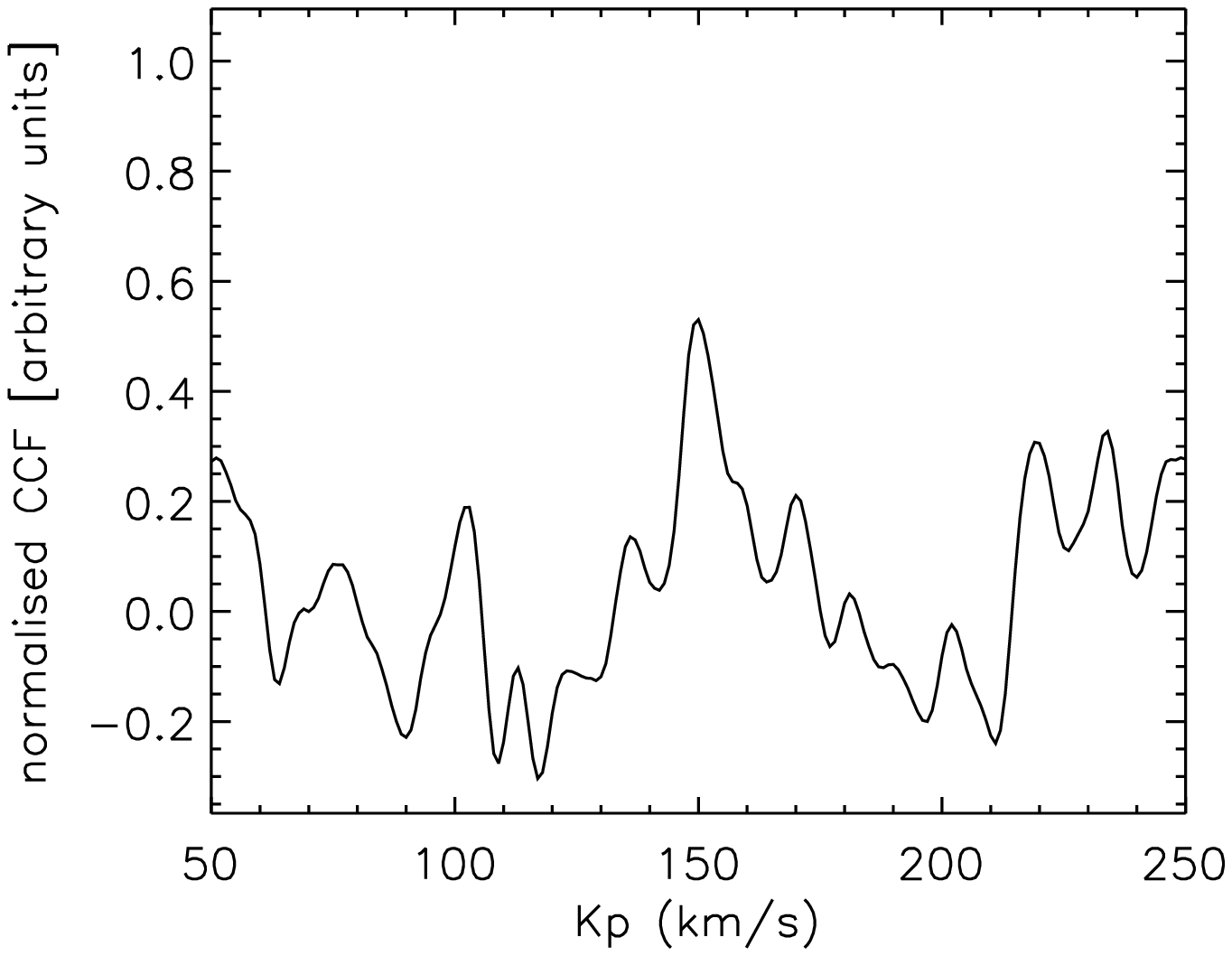} &
\includegraphics[width=5.5cm,keepaspectratio]{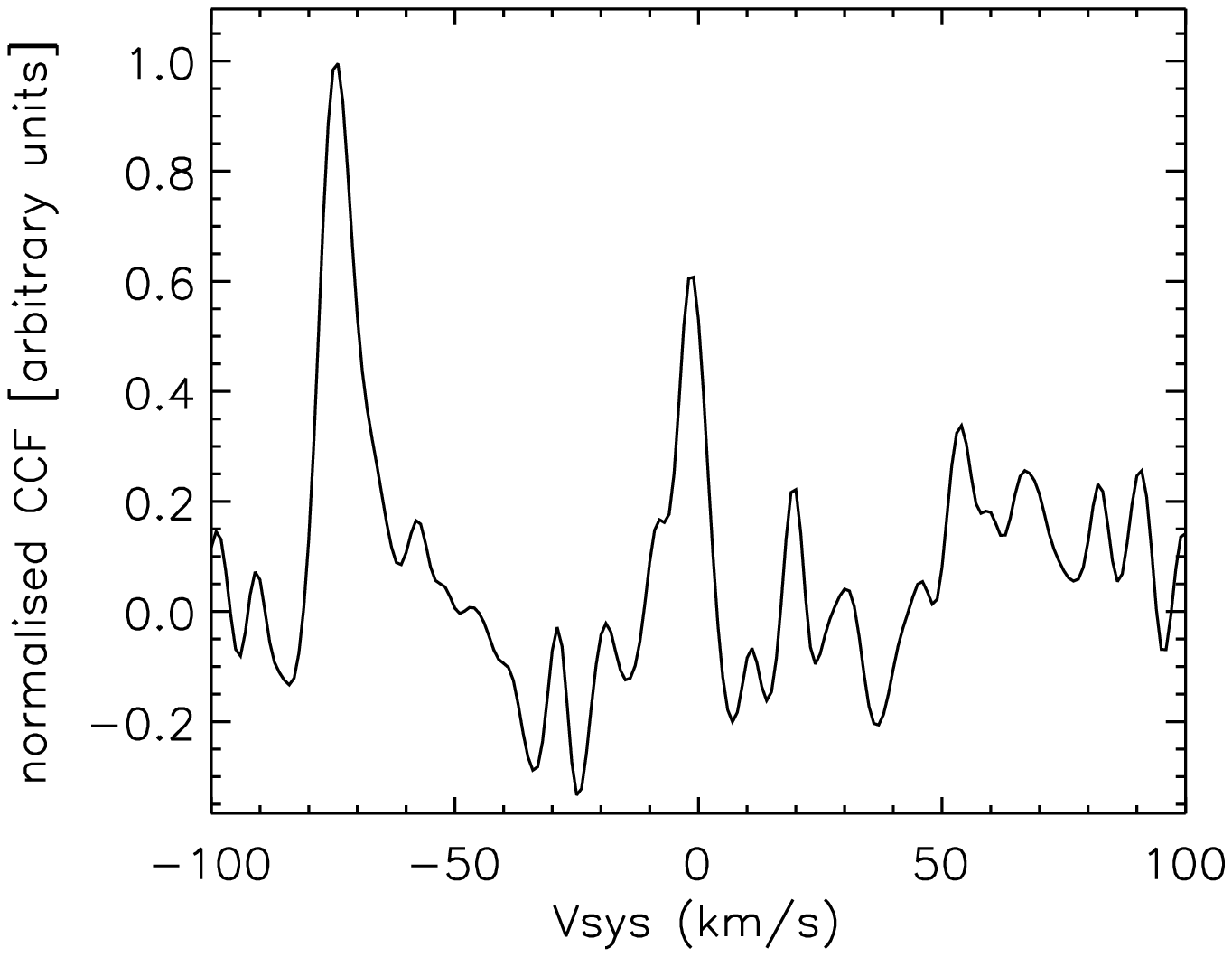} \\
Image F1 & Image F2 & Image F3
\vspace{0.4cm}\\
\includegraphics[width=5.5cm,keepaspectratio]{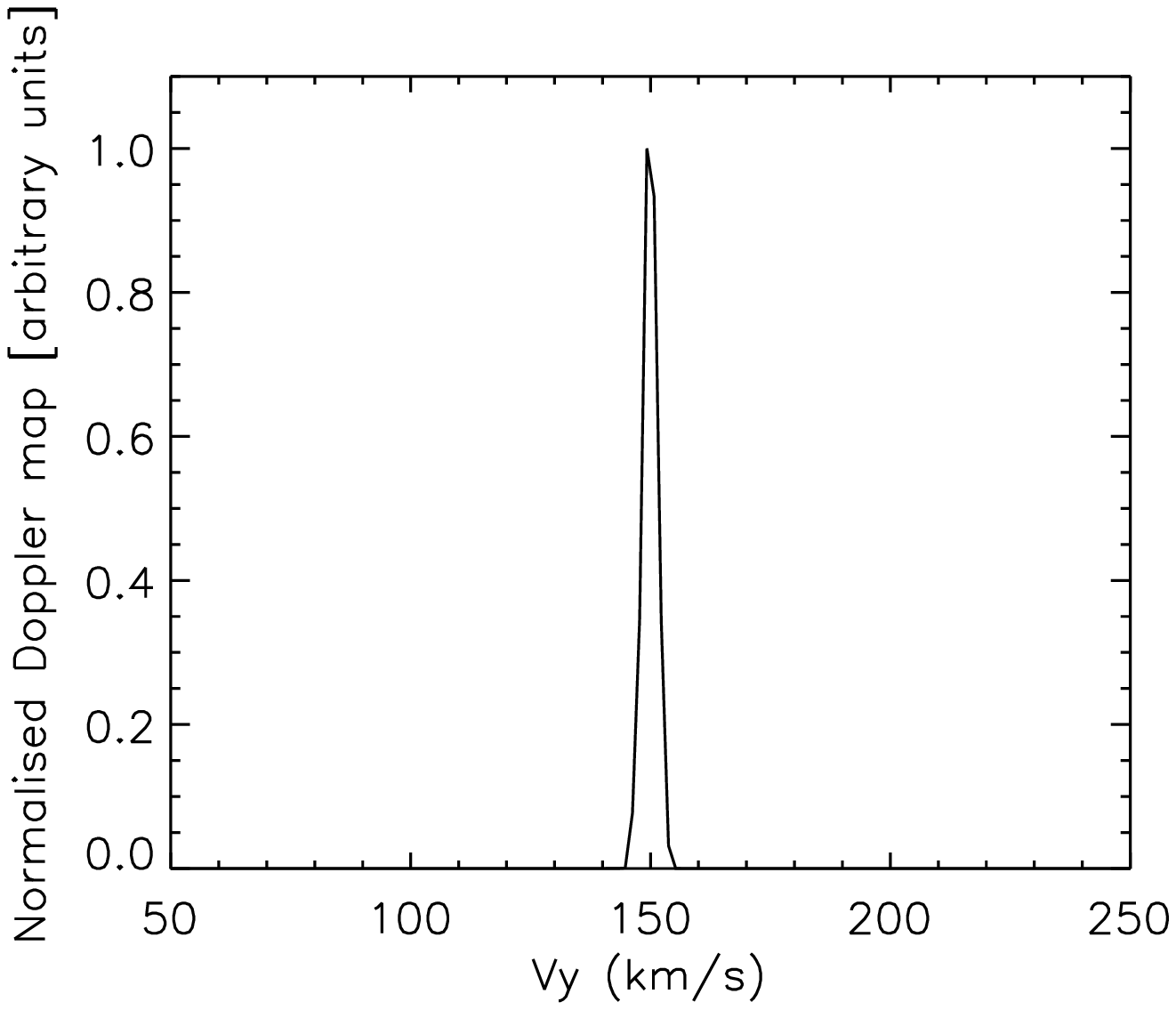} &
\includegraphics[width=5.5cm,keepaspectratio]{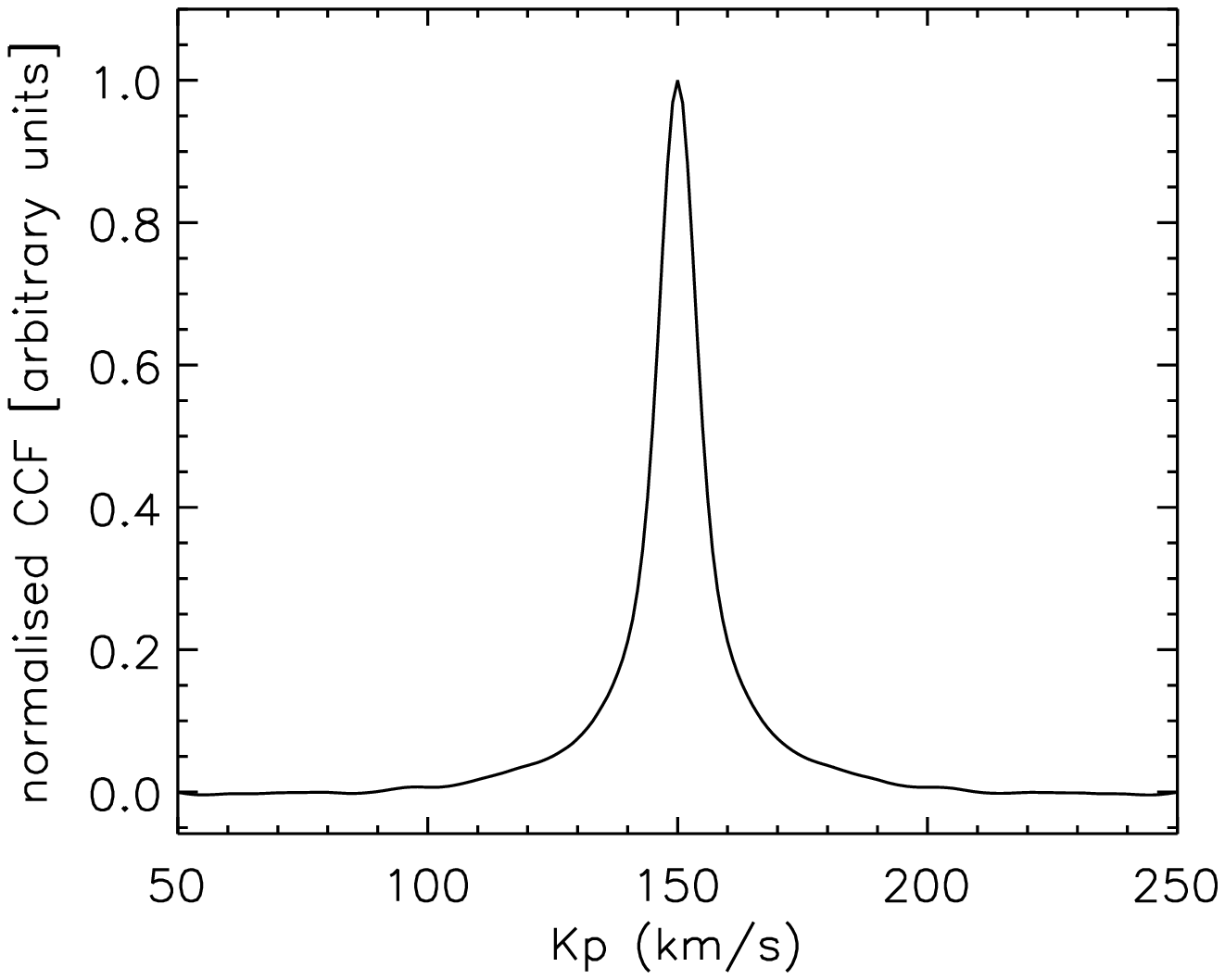} &
\includegraphics[width=5.5cm,keepaspectratio]{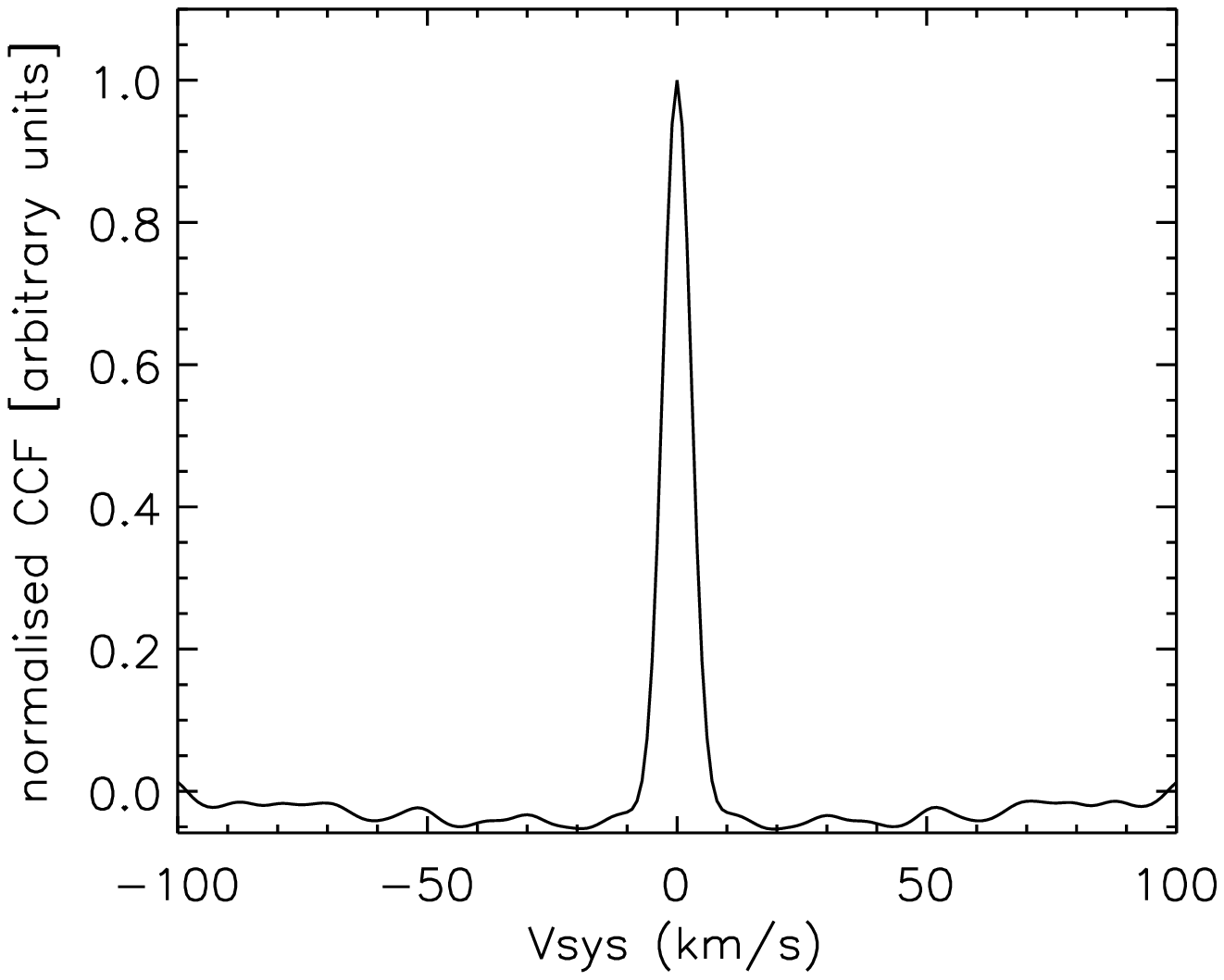} \\
Image H1 & Image H2 & Image H3
\vspace{0.4cm}\\
\includegraphics[width=5.5cm,keepaspectratio]{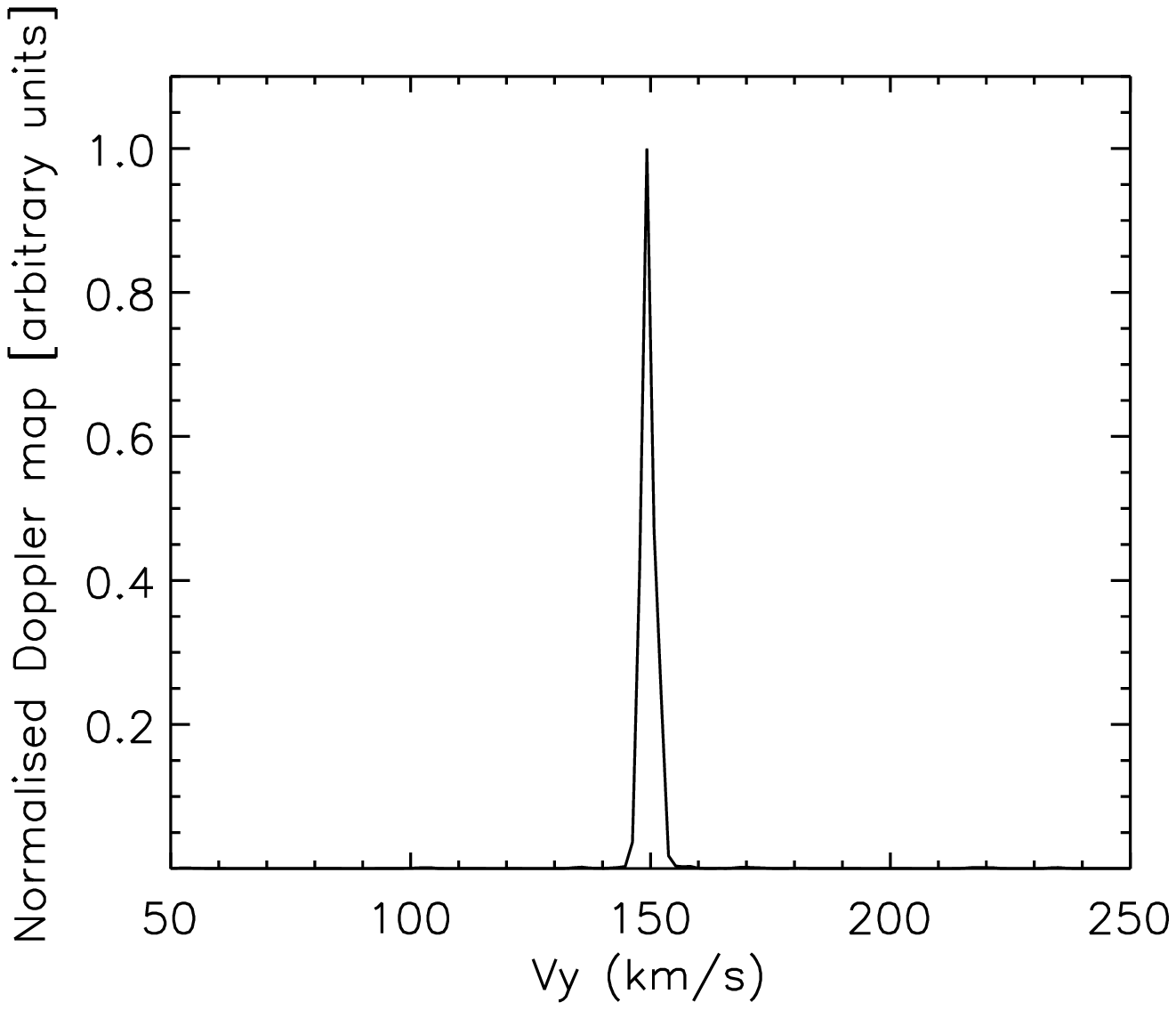} &
\includegraphics[width=5.5cm,keepaspectratio]{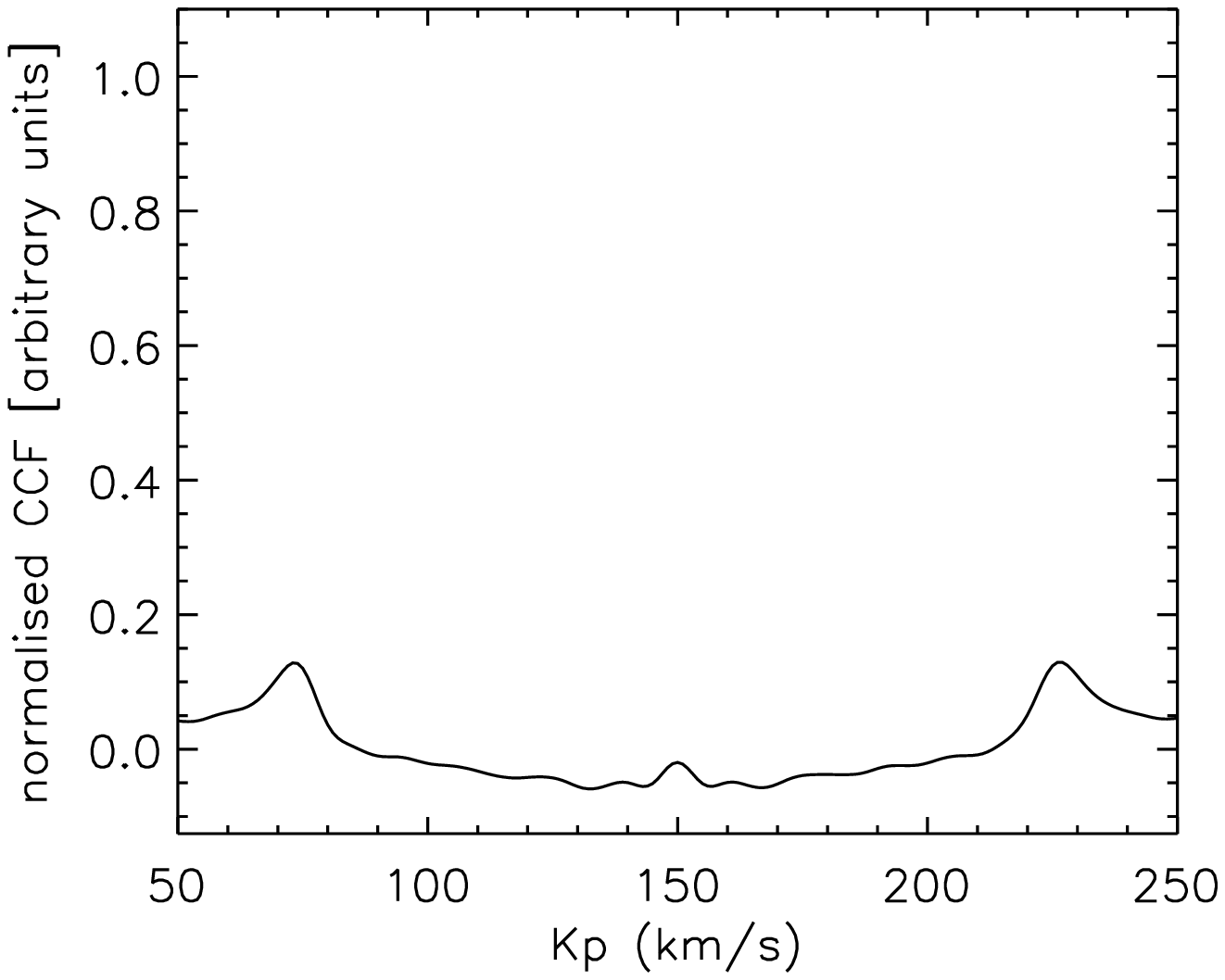} &
\includegraphics[width=5.5cm,keepaspectratio]{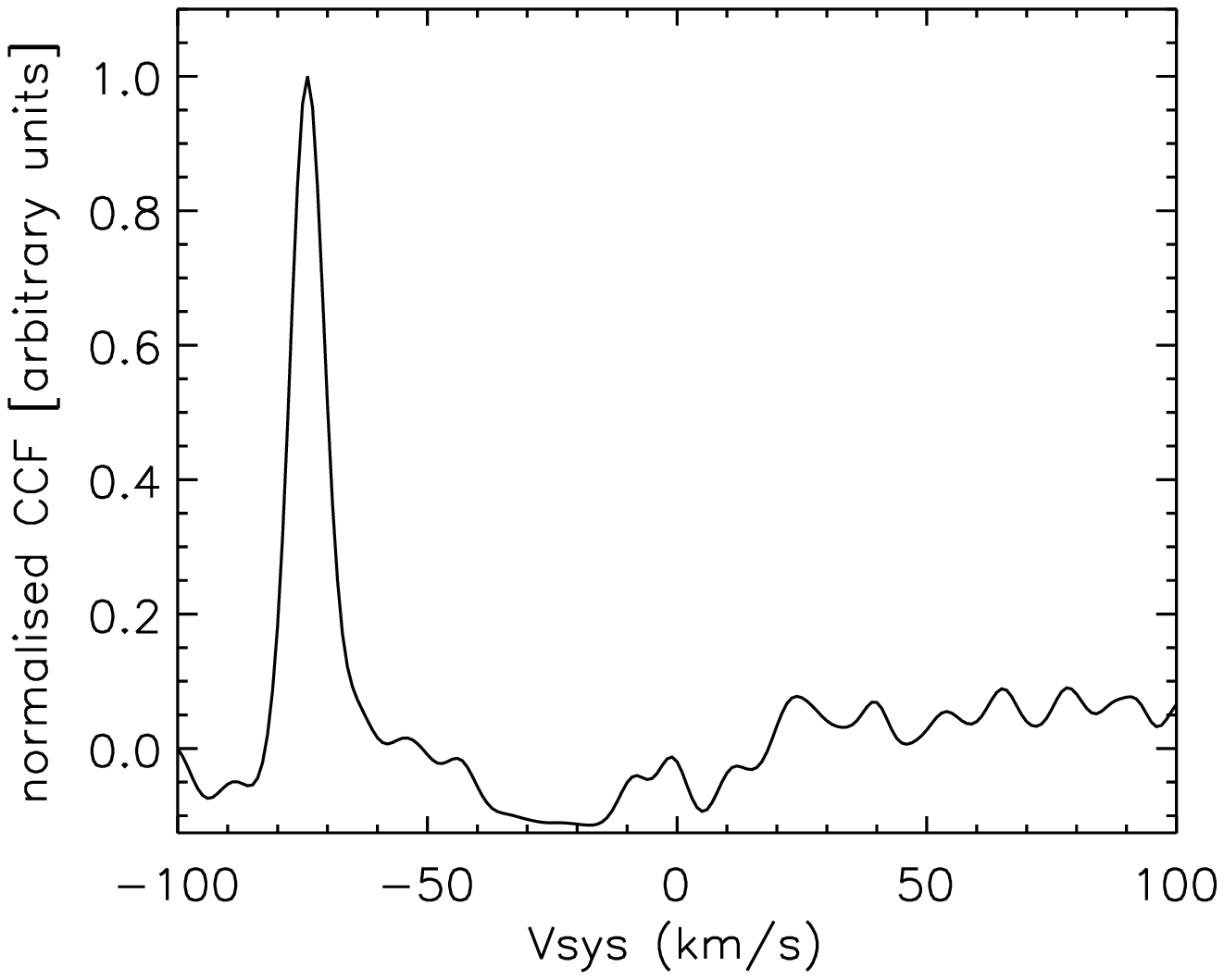} \\
Image I1 & Image I2 & Image I3
\vspace{0.4cm}\\
\end{tabular}
\centerline{{\bf Figure~\ref{fig:simulations2}.} -- continued}
\end{figure*}

\section*{Acknowledgements}

CAW would like to acknowledge support from UK Science Technology and
Facility Council grant ST/P000312/1. NPG gratefully acknowledges support
from the Royal Society in the form of a University Research Fellowship.
The authors would like to thank the referee, Prof Dr. Ignas Snellen, for very
helpful comments and suggestions.

\appendix

\section{Alternative CCF representations}
\label{sec:appendix}

\begin{figure*}
\begin{tabular}{cc}
\includegraphics[width=7.8cm,keepaspectratio]{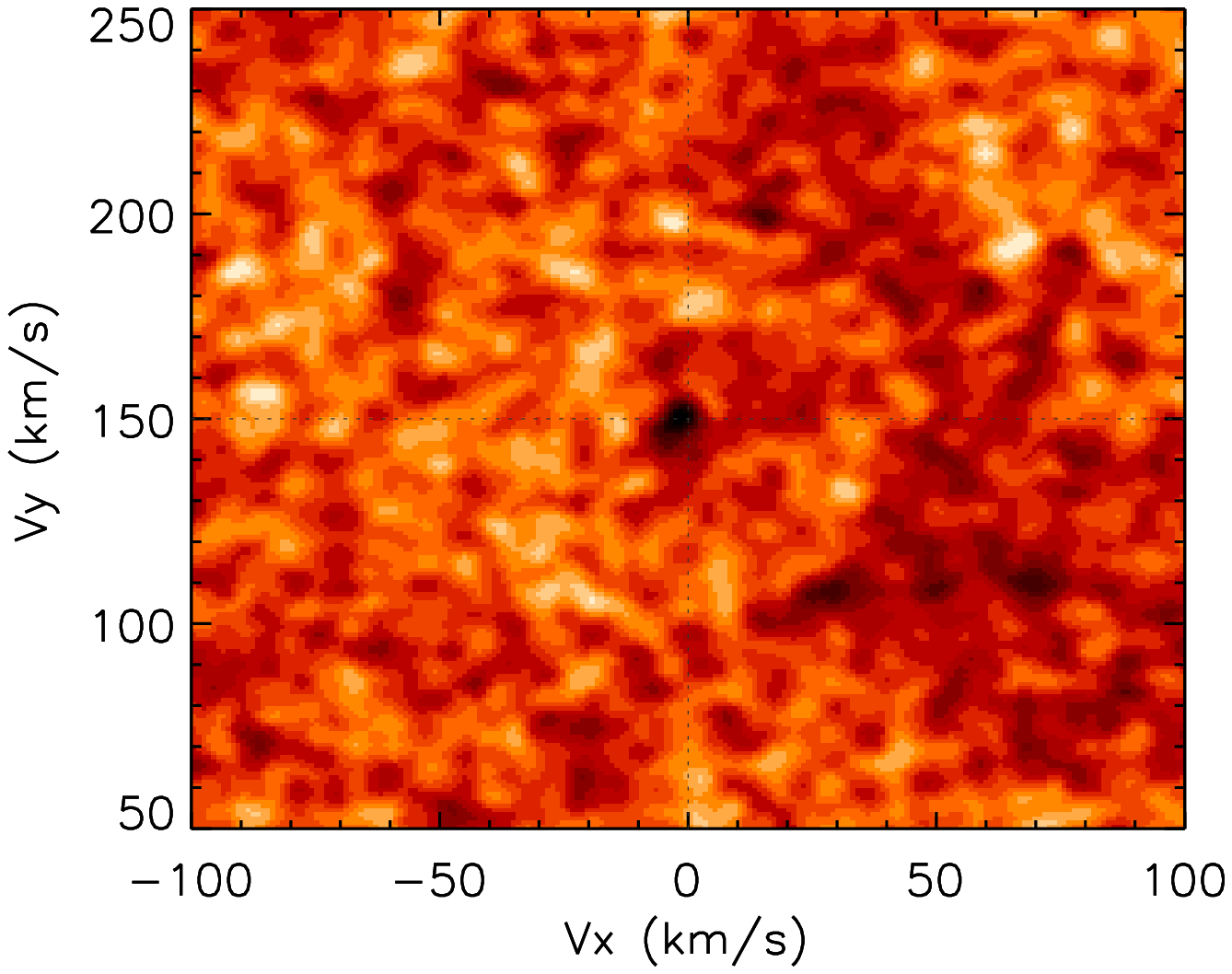} &
\includegraphics[width=7.8cm,keepaspectratio]{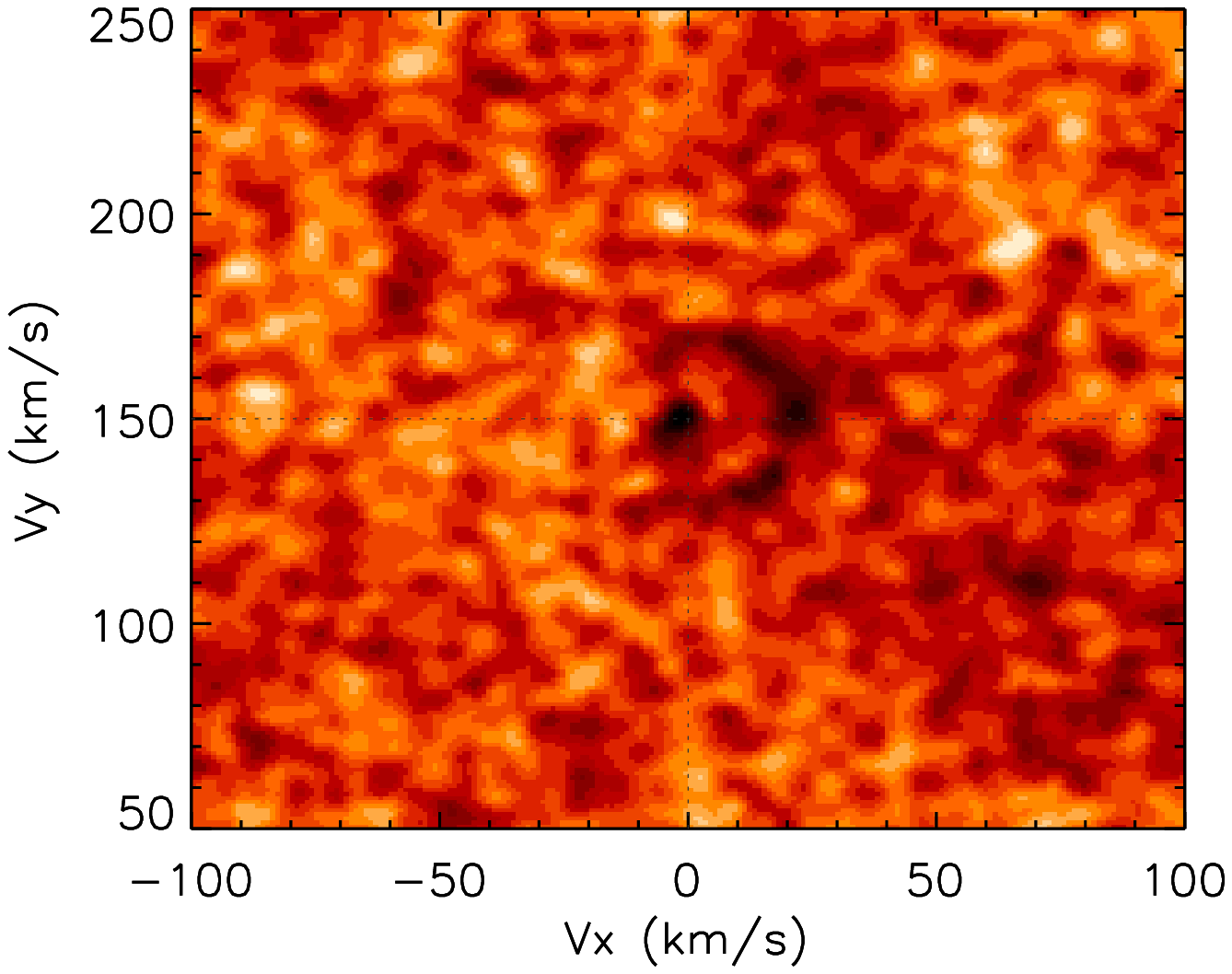} \\
Image A: CCF back-projection, contaminating line at +50 km s$^{-1}$ &
Image C: CCF back-projection, contaminating line at +20 km s$^{-1}$.
\vspace{0.4cm}\\
\includegraphics[width=7.8cm,keepaspectratio]{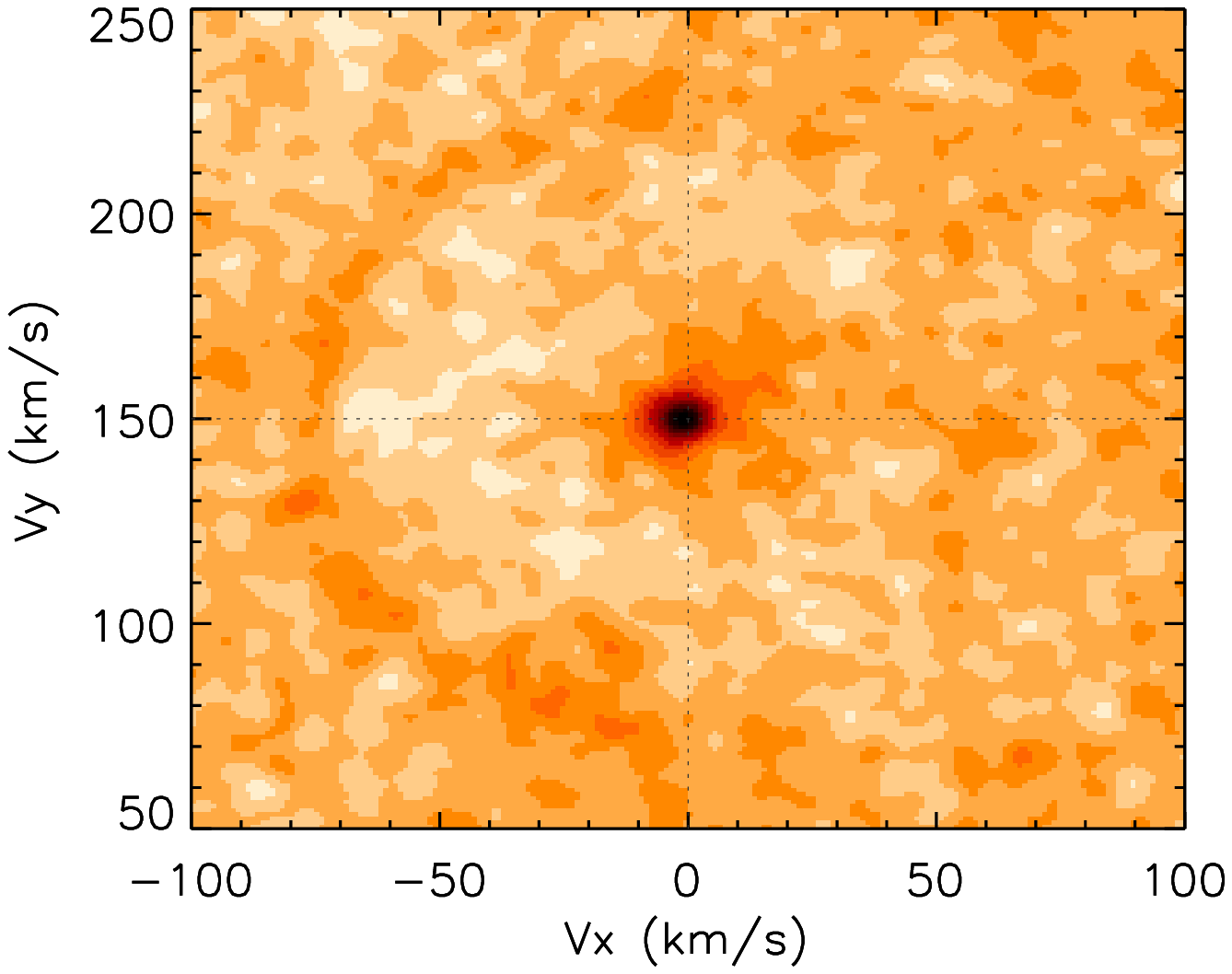} &
\includegraphics[width=7.8cm,keepaspectratio]{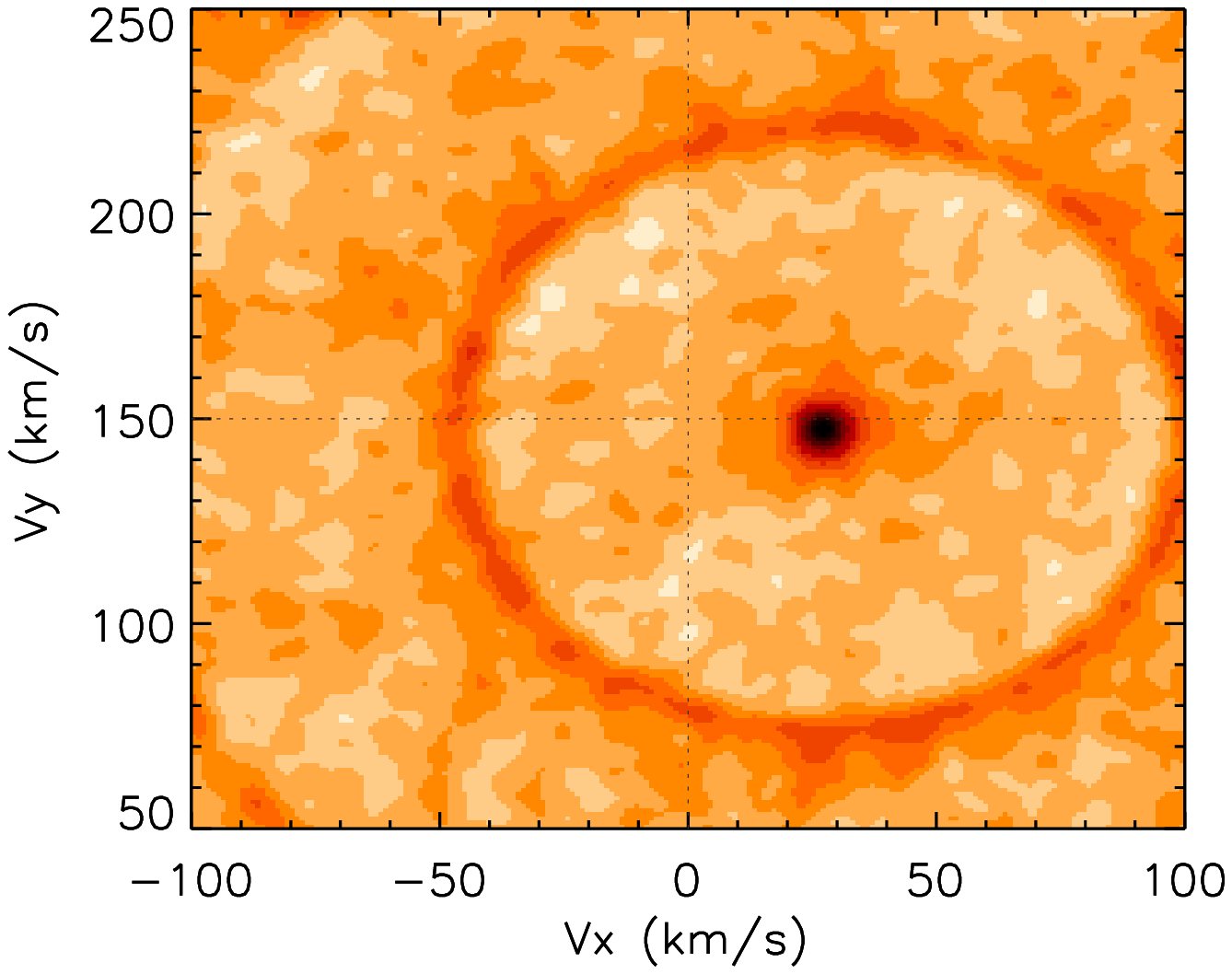} \\
Image D: CCF back-projection for $^{12}$C$^{17}$O. &
Image G: CCF back-projection for data with $\Delta\phi$=+0.03
\vspace{0.4cm}\\
\includegraphics[width=7.8cm,keepaspectratio]{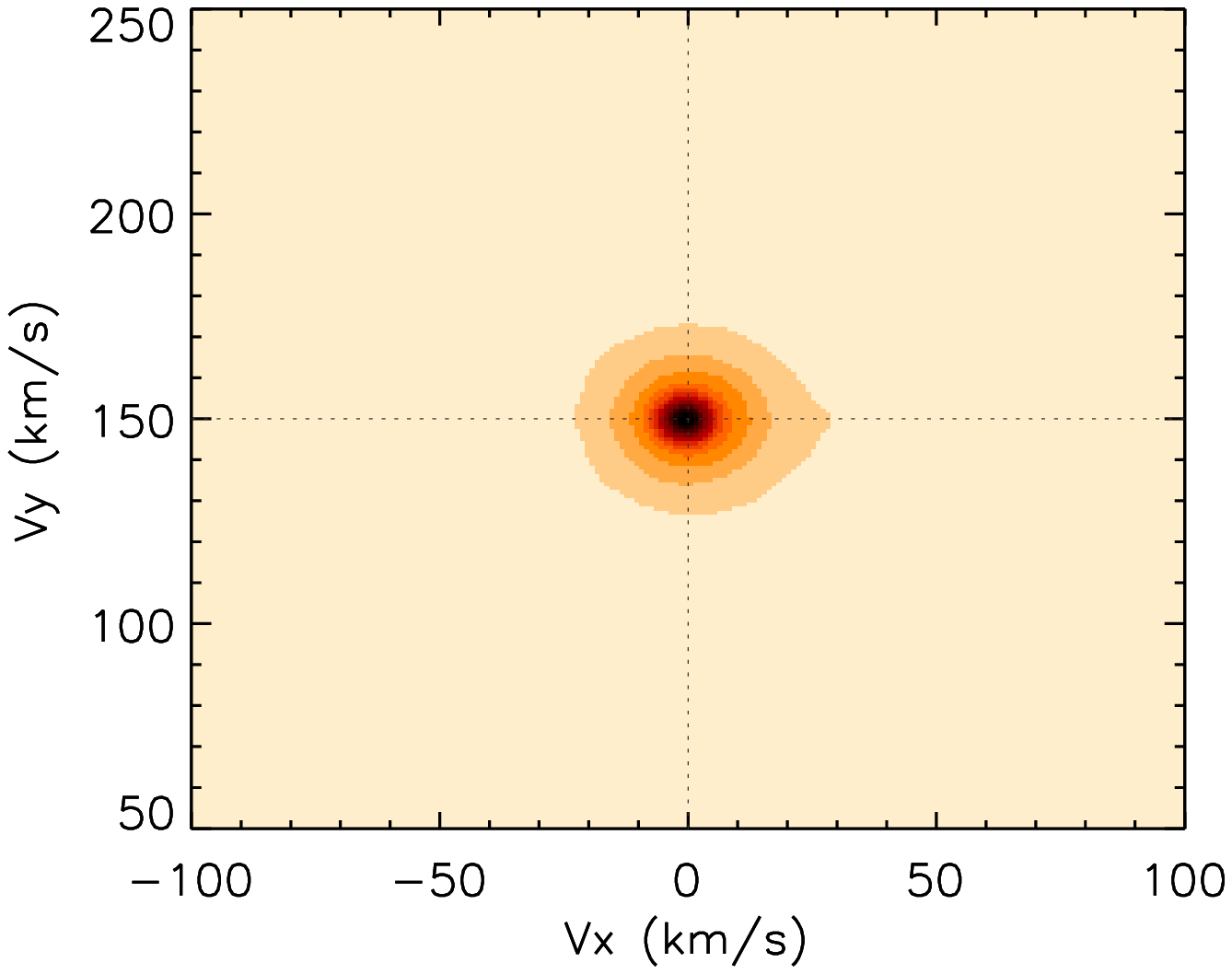} &
\includegraphics[width=7.8cm,keepaspectratio]{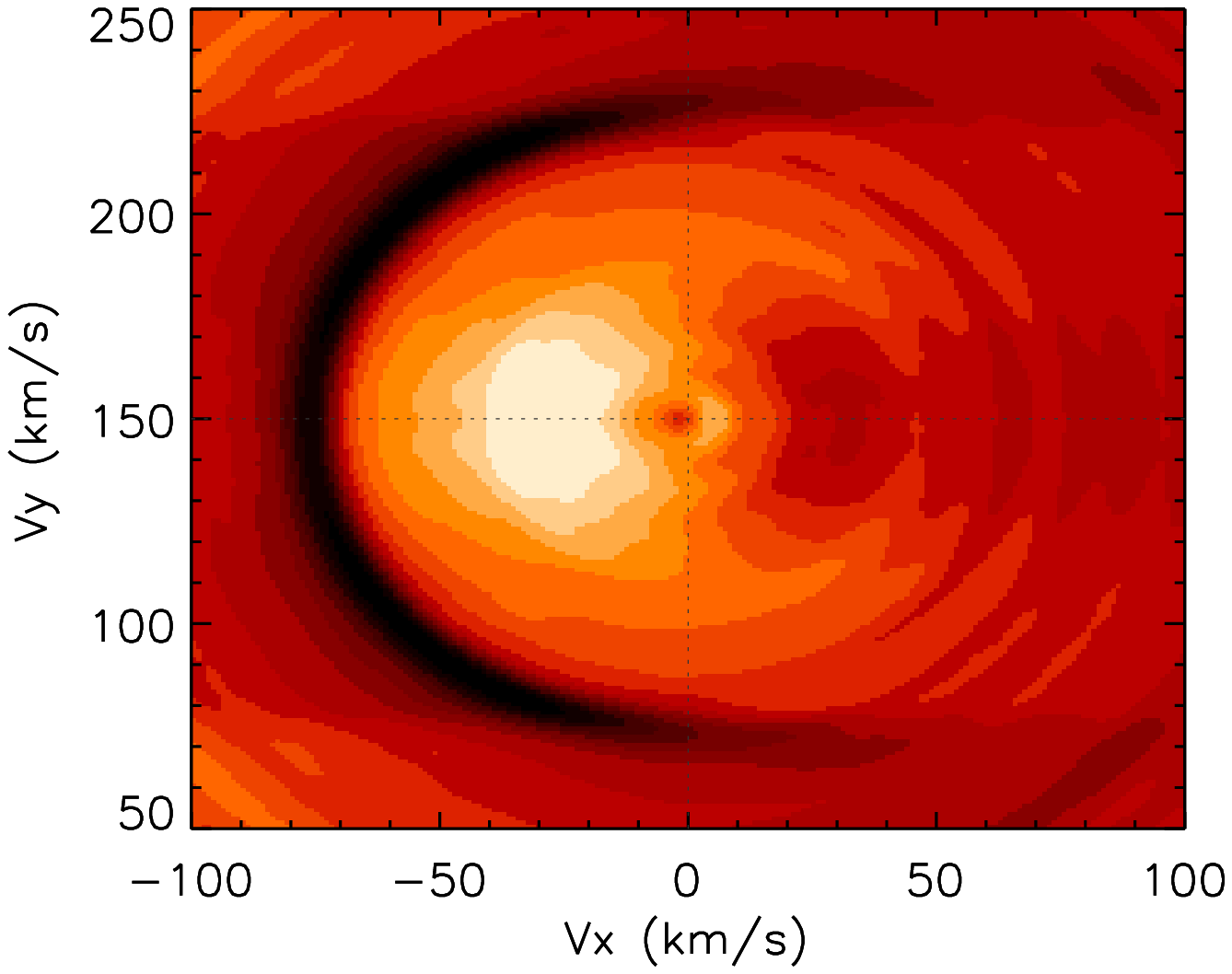} \\
Image H: Isotopologue test $^{12}$C$^{16}$O CCF back-projection &
Image I: Isotopologue test $^{12}$C$^{17}$O CCF back-projection
\vspace{0.4cm}\\
\end{tabular}
\caption{CCF back-projections for different simulations, the Image nomenclature mirrors that of the rest
  of the paper for ease of comparison. All images
  are linearly scaled from the minimum to the maximum values in the
  respective map. The intersection of the dashed lines indicate the location of
  the true injected signal.  Images A and C show the effects of a
  contaminating line (that is not included in the linelist) offset from
  the line of interest by +50 km s$^{-1}$ and +20 km s$^{-1}$, respectively.
  Image D shows the case where data containing $^{12}$C$^{16}$O and
  $^{12}$C$^{17}$O at equal strength is analysed using {\em only} the
  $^{12}$C$^{17}$O linelist. Image G shows the impact of
  a spurious phase-offset of +0.03 from the actual orbital phase. Images H
  and I show the case where the strength
  of $^{12}$C$^{16}$O is enhanced by a factor of 100 compared to
  $^{12}$C$^{17}$O. In the case of the Doppler tomogram maps presented in Figure~\ref{fig:simulations}, we
  search for $^{12}$C$^{16}$O and $^{12}$C$^{17}$O simultaneously,
  while for the CCFs we target each molecule separately. Image H show
  the results for $^{12}$C$^{16}$O and Image I for
  $^{12}$C$^{17}$O.}
\label{fig:appsims}
\end{figure*}

\begin{figure*}
\begin{tabular}{cc}
\includegraphics[width=7.8cm,keepaspectratio]{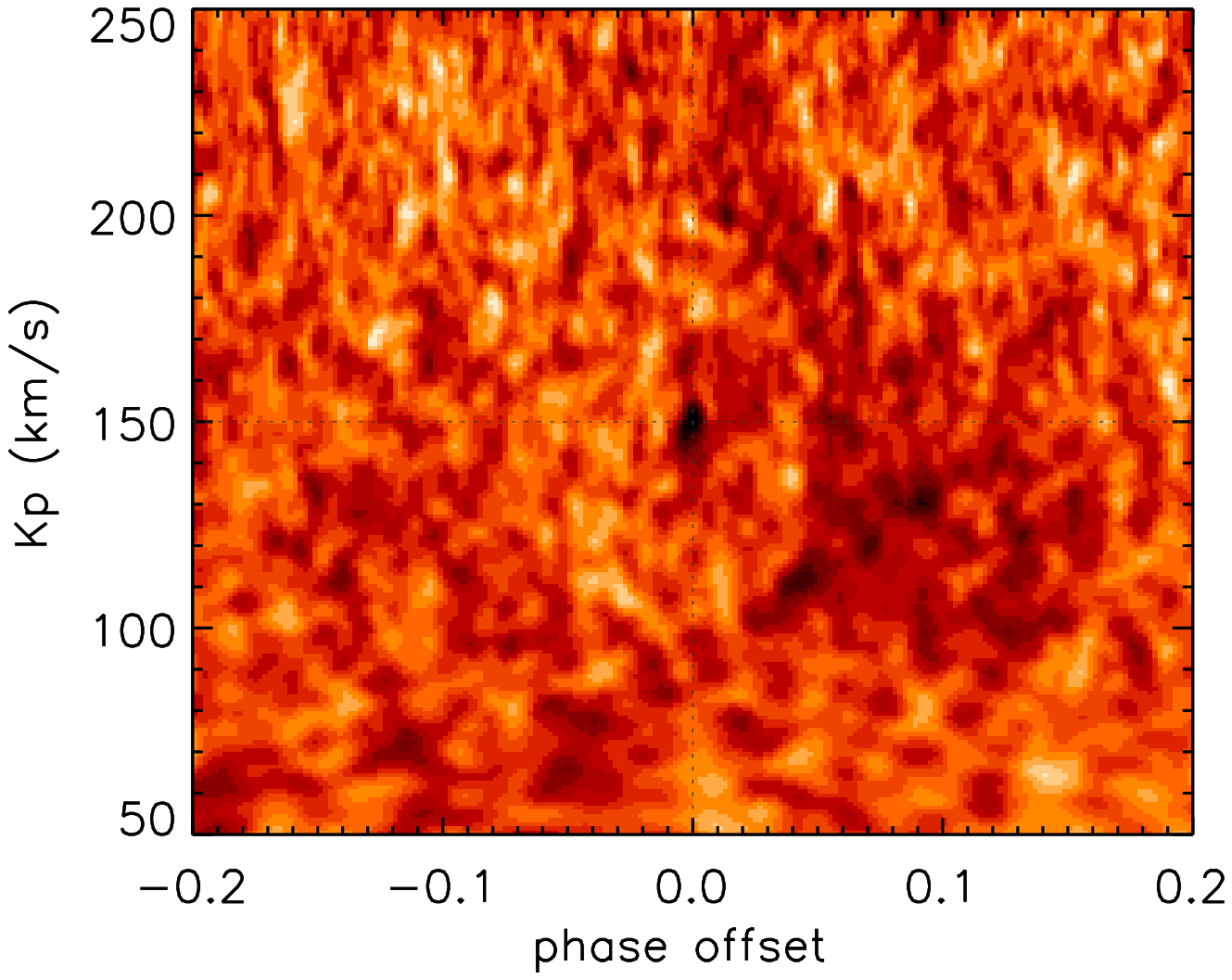} &
\includegraphics[width=7.8cm,keepaspectratio]{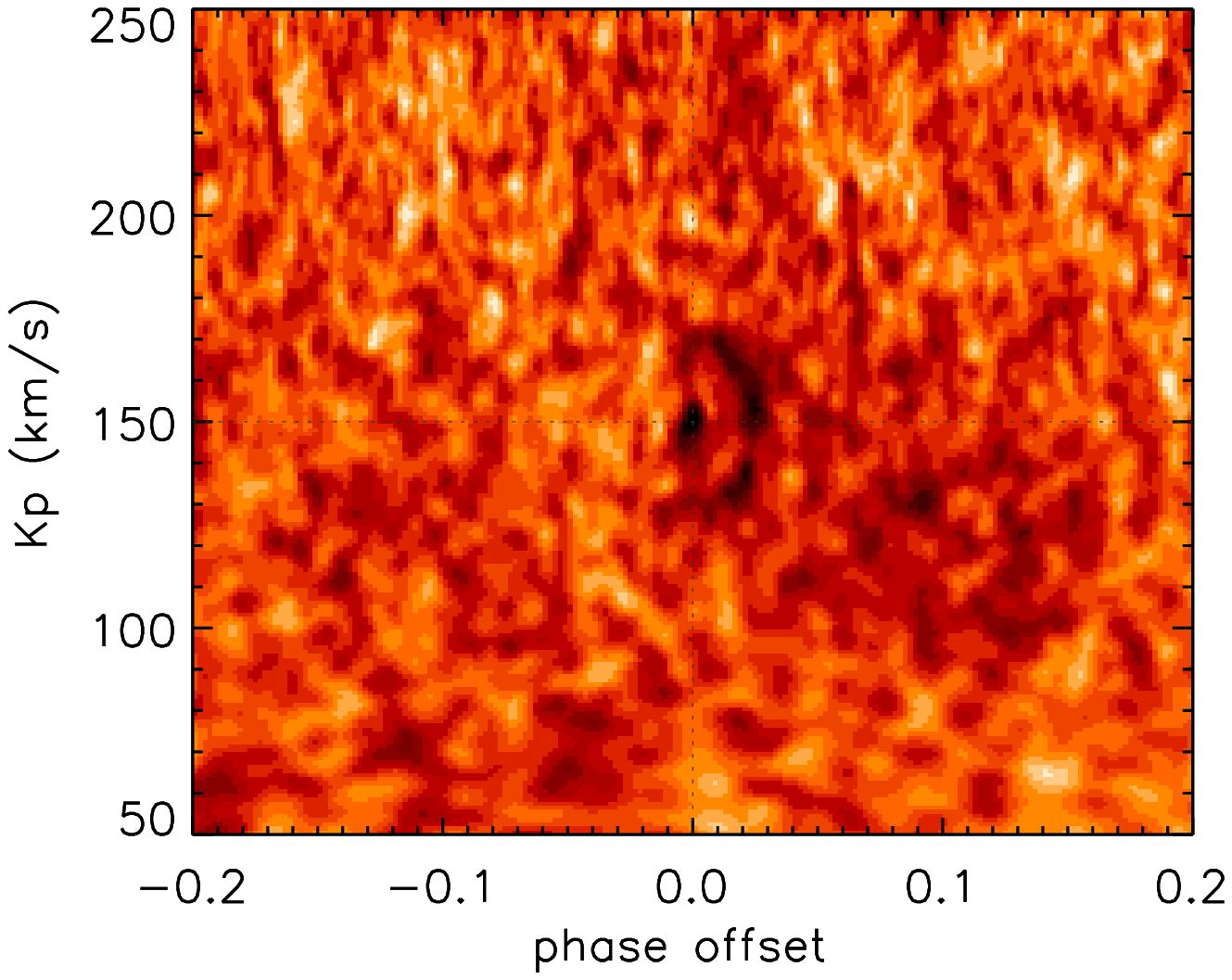} \\
Image A: Phase-folded CCF, contaminating line at +50 km s$^{-1}$ &
Image C: Phase-folded CCF, contaminating line at +20 km s$^{-1}$.
\vspace{0.4cm}\\
\includegraphics[width=7.8cm,keepaspectratio]{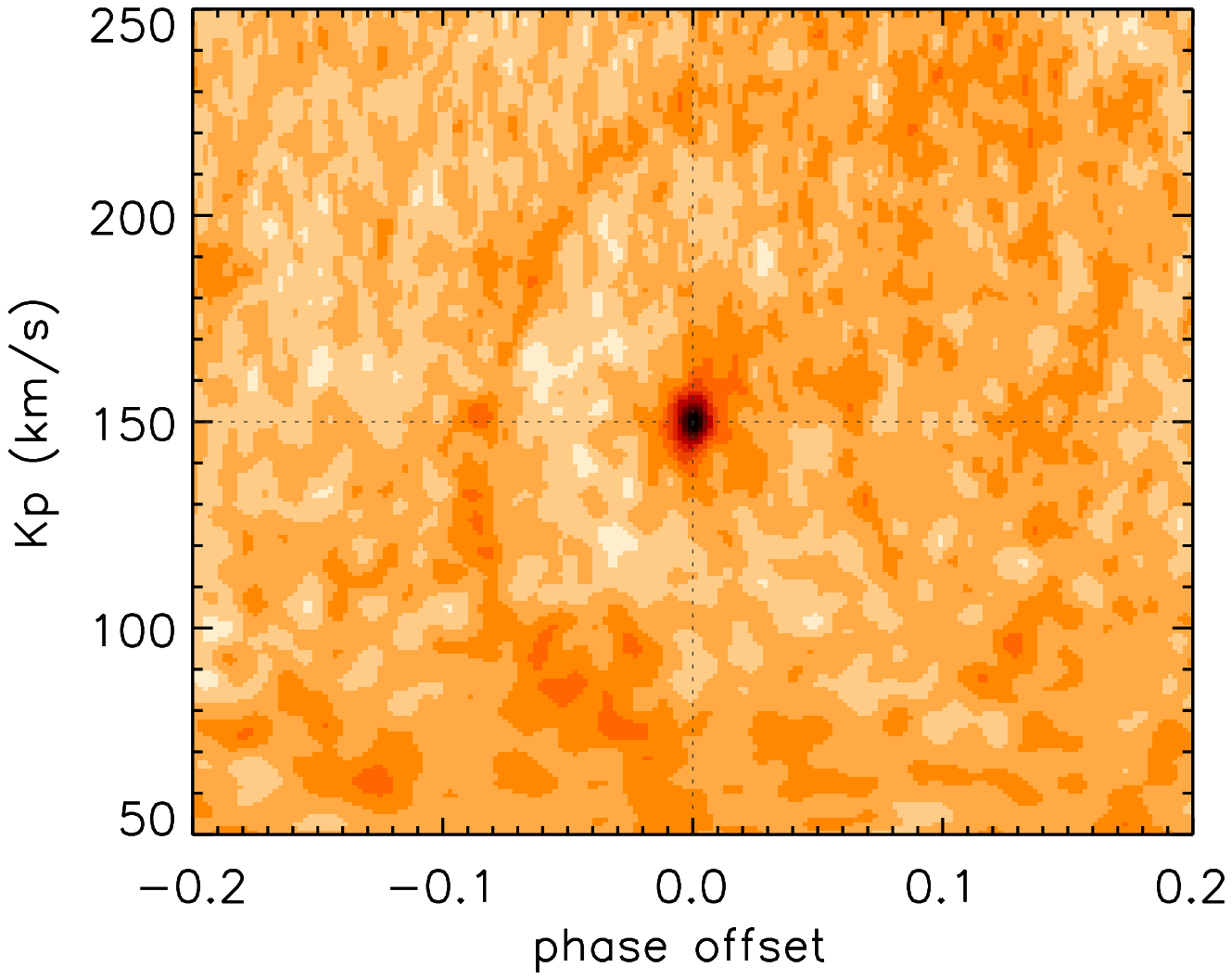} &
\includegraphics[width=7.8cm,keepaspectratio]{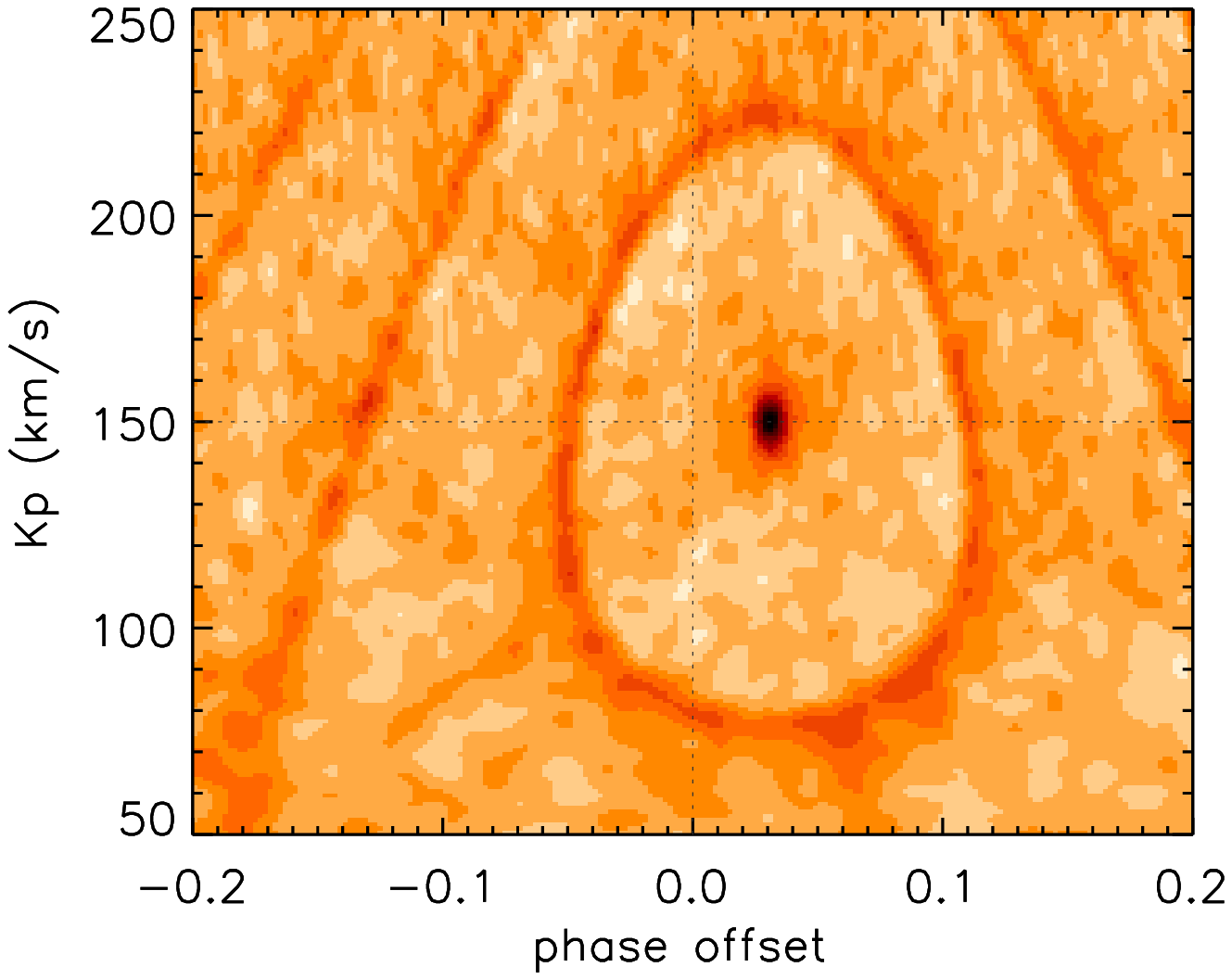} \\
Image D: Phase-folded CCF for $^{12}$C$^{17}$O. &
Image G: Phase-folded CCF for data with $\Delta\phi$=+0.03
\vspace{0.4cm}\\
\includegraphics[width=7.8cm,keepaspectratio]{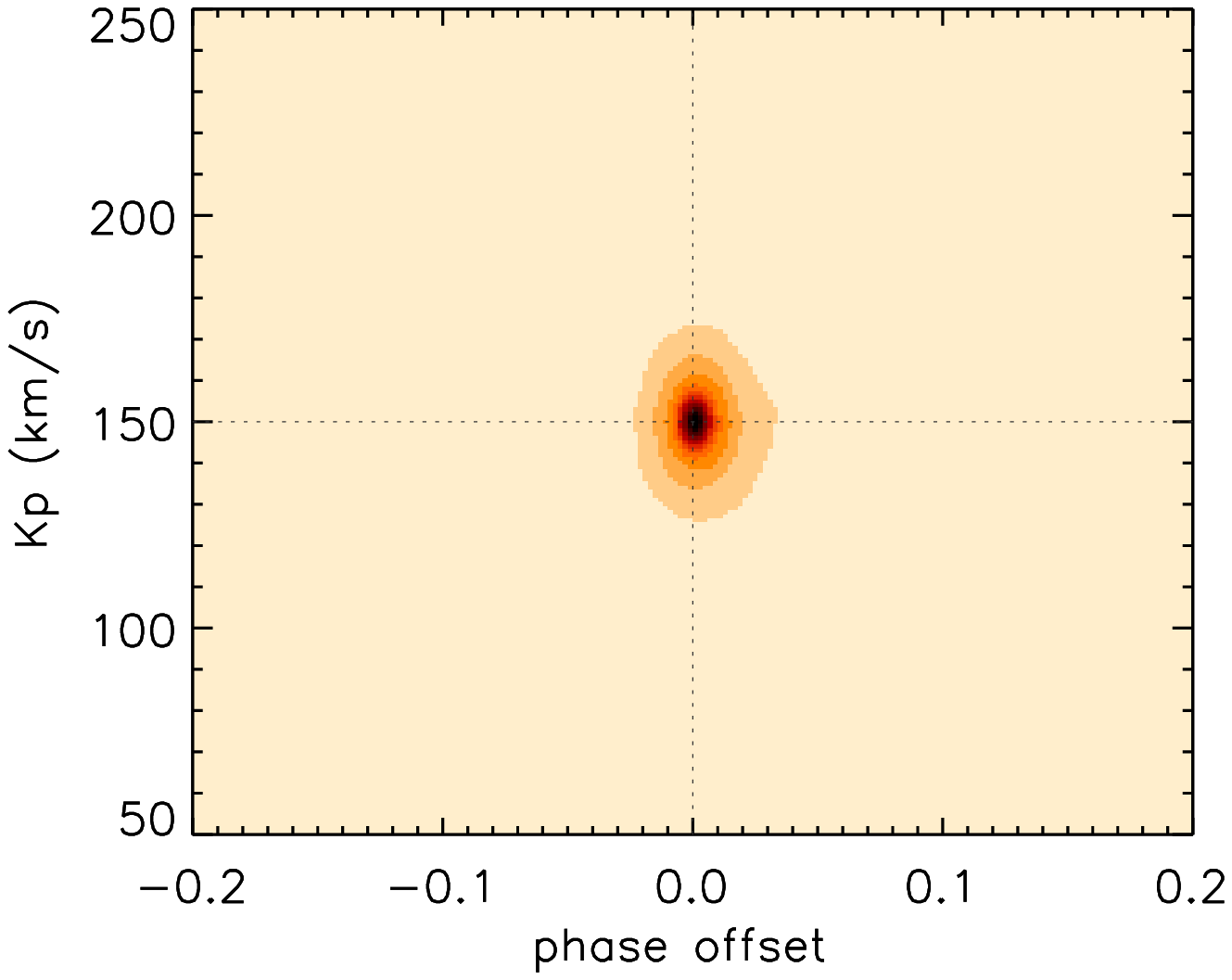} &
\includegraphics[width=7.8cm,keepaspectratio]{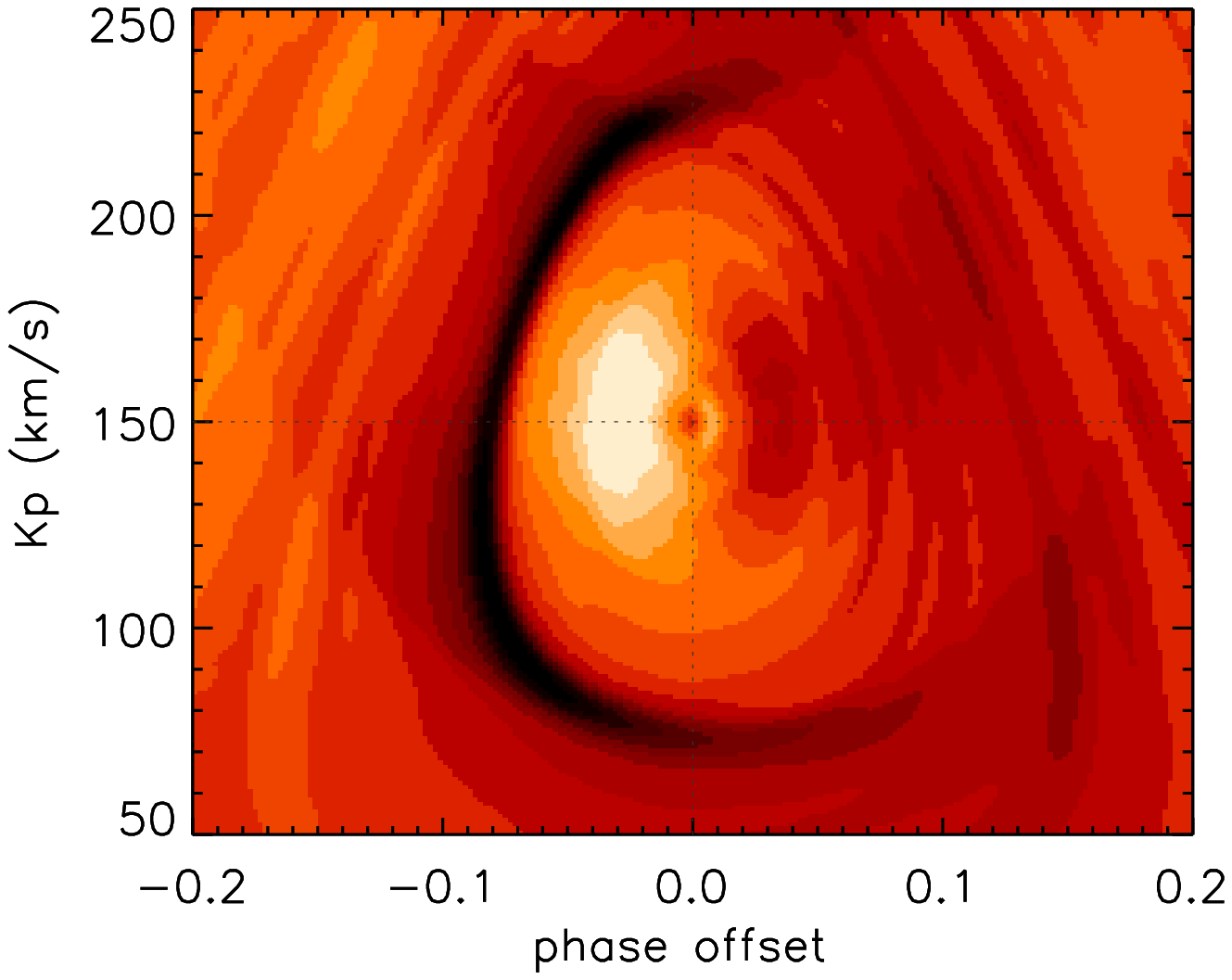} \\
Image H: Phase-folded CCF, isotopologue test $^{12}$C$^{16}$O &
Image I: Phase-folded CCF, isotopologue test $^{12}$C$^{17}$O
\vspace{0.4cm}\\
\end{tabular}
\caption{Alternative representation of the phase-folded CCFs for different simulations, in this case as a function of
  K$_P$ and phase offset, keeping v$_{sys}$ fixed at the correct value. Again, the Image
  nomenclature mirrors that of the rest of the paper for ease of comparison.}
\label{fig:appsims2}
\end{figure*}

In this paper we have compared the Doppler tomograms to the conventional that phase-folded
  CCFs have been presented in the literature. As pointed out by the referee, this is not entirely a one-to-one
  comparison, and therefore we have investigated two additional ways of representing the CCF analysis: back-projection,
  and phase-folding the CCFs as a function of the orbital phase-offset and K$_P$ (keeping v$_{sys}$ constant at the
  correct value). These are presented in Figures~\ref{fig:appsims} \&~\ref{fig:appsims2}, respectively.
  The back-projection approach follows that outlined in Section~\ref{sec:doptom} (see Figure~\ref{fig:doppler_plot}
  for a schematic), where a time-series of CCFs are integrated along different
  radial-velocity curves according to Equation~\ref{eq:vr}, and these can be directly compared to 
  the Doppler maps presented in the main body of this paper.

\bibliographystyle{mnras}
\bibliography{abbrev,refs}







\bsp	
\label{lastpage}
\end{document}